\documentclass{aa}  

\usepackage{graphicx}
\usepackage{txfonts}
\begin{document}

   \title{Galaxy populations in the Hydra I cluster from the VEGAS survey}
   \subtitle{I. Optical properties of a large sample of dwarf galaxies\thanks{Full table 5 is only available in electronic form at the CDS via anonymous ftp to cdsarc.u-strasbg.fr (130.79.128.5) or via http://cdsweb.u-strasbg.fr/cgi-bin/qcat?J/A+A/}}

   \author{Antonio La Marca \inst{1,2,3}\fnmsep\thanks{\email{antoniolamarca46@gmail.com}}
          \and Reynier Peletier\inst{3}
          \and Enrichetta Iodice \inst{1}
          \and Maurizio Paolillo \inst{1,2}
          \and Nelvy Choque Challapa \inst{3}
          \and Aku Venhola \inst{4}
          \and Duncan A. Forbes \inst{5}
          \and Michele Cantiello \inst{6}
          \and Michael Hilker  \inst{7}
          \and Marina Rejkuba \inst{7}
          \and Magda Arnaboldi \inst{7}
          \and Marilena Spavone \inst{1}
          \and Giuseppe D'Ago \inst{8}
          \and Maria Angela Raj \inst{9}
          \and Rossella Ragusa \inst{1,2}
          \and Marco Mirabile \inst{2}
          \and Roberto Rampazzo \inst{10}
          \and Chiara Spiniello \inst{11}
          \and Steffen Mieske \inst{12}
          \and Pietro Schipani\inst{1}
          }

   \institute{INAF $-$ Astronomical Observatory of Capodimonte, Salita Moiariello 16, I-80131, Naples, Italy 
         \and
             University of Naples ``Federico II'', C.U. Monte Sant'Angelo, Via Cinthia, 80126, Naples, Italy
        \and
            Kapteyn Institute, University of Groningen, Landleven 12, 9747 AD Groningen, the Netherlands
        \and
            Space Physics and Astronomy Research Unit, University of Oulu, P.O. Box 3000, FI$-$90014, Oulu, Finland
        \and
            Centre for Astrophysics \& Supercomputing, Swinburne University of Technology, Hawthorn VIC 3122, Australia
        \and
            INAF Osservatorio Astronomico d’Abruzzo, Via Maggini, 64100 Teramo, Italy
        \and
            European Southern Observatory, Karl$-$Schwarzschild-Strasse 2, 85748 Garching bei München, Germany
        \and
            Instituto de Astrofísica, Facultad de Física, Pontificia Universidad Católica de Chile, Av. Vicuña Mackenna 4860, 7820436 Macul, Santiago, Chile
        \and
            INAF $-$ Astronomical Observatory of Rome, Via Frascati, 33, 00078, Monte Porzio Catone, Rome, Italy
        \and 
            INAF $-$ Astronomical Observatory of Padova, Via dell’Osservatorio 8 , I-36012, Asiago (VI), Italy
        \and 
            Department of Physics, University of Oxford, Denys Wilkinson Building, Keble Road, Oxford OX1 3RH, UK
        \and 
            European Southern Observatory, Alonso de Cordova 3107, Vitacura, Santiago, Chile
             }

   \date{Received 29 July 2021 / Accepted 08 December 2021 }

 
  \abstract
   {Due to their relatively low stellar mass content and diffuse nature, the evolution of dwarf galaxies can be strongly affected by their environment. 
   Analyzing the properties of the dwarf galaxies over a wide range of luminosities, sizes, morphological types, and environments, we can obtain insights about their evolution.
   At $\sim50\;Mpc$, the Hydra I cluster of galaxies is among the closest cluster in the $z\simeq0$ Universe, and an ideal environment to study dwarf galaxy properties in a cluster environment. 
    }
   {We exploit deep imaging data of the Hydra I cluster to construct a new photometric catalog of dwarf galaxies in the cluster core, which is then used to derive properties of the Hydra I cluster dwarf galaxy population as well as to compare it with other clusters.
   Moreover, we investigate the dependency of dwarf galaxy properties on their surrounding environment.
    }
   {The new wide-field $g$- and $r$-band images of the Hydra I cluster obtained with the OmegaCAM camera on the VLT Survey Telescope (VST) in the context of the VST Early-type GAlaxy Survey (VEGAS) were used to study the dwarf galaxy population in the Hydra I cluster core down to r-band magnitude $M_r=-11.5\;mag$. 
   We used an automatic detection tool to identify dwarf galaxies from a $\sim1\;deg^2$ field centered on the Hydra I core, covering almost half of the cluster virial radius. 
   The photometric pipeline was used to estimate the principal photometric parameters for all targets. 
   Scaling relations and visual inspection were used to assess the cluster membership and construct a new dwarf galaxy catalog. 
   Finally, based on the new catalog, we studied the structural (Sérsic index $n$, effective radius $R_e$, and axis ratio) and photometric (colors and surface brightness) properties of the dwarf galaxies, also investigating how they vary as a function of clustercentric distance.
   }
   {The new Hydra I dwarf catalog contains 317 galaxies with a luminosity between $-18.5<M_r<-11.5\;mag$, a semi-major axis larger than $\sim200\;pc$ ($a=0.84\arcsec$), of which 202 are new detections, and previously unknown dwarf galaxies in the Hydra I central region.
   We estimate that our detection efficiency reaches $50\%$ at the limiting magnitude $M_r=-11.5\;mag$, and at the mean effective surface brightness $\overline{\mu}_{e,r}=26.5\;mag/arcsec^2$. 
   We present the standard scaling relations for dwarf galaxies, which are color-magnitude, size-luminosity, and Sérsic $n$-magnitude relations, and compare them with other nearby clusters.
   We find that there are no observational differences for dwarfs scaling relations in clusters of different sizes.
   We study the spatial distribution of galaxies, finding evidence for the presence of substructures within half the virial radius.
   We also find that mid- and high-luminosity dwarfs ($M_r<-14.5\;mag$) become, on average, redder toward the cluster center, and that they have a mild increase in $R_e$ with increasing clustercentric distance, similar to what is observed for the Fornax cluster.  
   No clear clustercentric trends are reported for surface brightness and Sérsic index.
   Considering galaxies in the same magnitude bins, we find that for high and mid-luminosity dwarfs ($M_r<-13.5\;mag$), the $g-r$ color is redder for the brighter surface brightness and higher Sérsic $n$ index objects.
   This finding is consistent with the effects of harassment and/or partial gas stripping. 
   }
   {}

   \keywords{Galaxy clusters: Abell 1060 (Hydra I) -- galaxies: low surface brightness galaxies -- galaxies: dwarf : evolution -- galaxies: photometry }

   \maketitle
%

\section{Introduction}\label{sec:Intro}

Dwarf galaxies are the most abundant type of galaxies in the Universe, and they play a crucial role in testing the current $\Lambda$ Cold Dark Matter cosmological model \citep[$\Lambda$CDM;][]{Bullock2017LCDM}, and galaxy formation and evolution theories.
Indeed, dwarf-sized proto-galactic fragments built up the bulges of massive galaxies very rapidly, which are their building blocks, while other (metal-poor) dwarf galaxies were accreted later to form the outer galactic halos \citep[growth through bottom-up assembly processes][]{springel2005simul,mancillas2019probing}.
Dwarfs are an important resource to study the physics of galaxies because they have a low stellar mass and, they are usually diffuse and, therefore, are more affected by environmental processes.   
Furthermore, dwarfs are found both in isolation, in groups of galaxies, and in dense, cluster environments.
Different samples can be used to investigate intrinsic (internal) processes versus extrinsic (environmental) processes that shape their evolution.
Some of the relatively recent examples of low density environment studies are \citet{Leaman2014}, \citet{Gallart2015}, \citet{Kacharov2017}, \citet{Hermosa-munoz2020}, and in cluster environment: \citet{sanchez2008properties}, \citet{peng2010mass,peng2012mass,peng2014mass} investigated processes shaping dwarf galaxies in cluster environments.
Hence, in conclusion, the combination of their abundance and their vulnerability makes them a suitable instrument to investigate the effects of environment on galaxy evolution.

Recent deep, wide-field surveys provided homogeneous and complete samples of galaxy populations, down to the low luminosity regime.
Among these, the Next Generation Virgo cluster Survey \citep[NGVS; ][]{Ferrarese2012}, the Next Generation Fornax Survey \citep[NGFS; ][]{Munoz2015ngfs,Eigenthaler2018ngfs}, the VST Early-type GAlaxy Survey \citep[VEGAS; ][]{Capaccioli2015,iodice2021vegas}, the Fornax Deep Survey \citep[FDS; ][]{Peletier2020}, and the Mass Assembly of early Type gaLAxies with their fine Structures \citep[MATLAS; ][]{duc2015atlas3d,habas2020newly} revealed large samples of faint galaxies in various environments,  which were previously unknown, constituting a powerful resource to analyze how galaxies change as a function of different environments.
On the theoretical side, the current large-scale cosmological simulations \citep[such as IllustrisTNG, ][]{pillepich2018simulating} have reached sufficient high resolution, making a direct comparison with dwarf galaxies with a stellar mass of $10^{8-9}M_{\odot}$ now possible. 
However, this stellar mass range is typical of larger dwarf galaxies, and simulations are still not good at reproducing galaxies below that stellar mass limit \citep[for example, look at the EAGLE simulation lower mass limit,][]{Schaye2015eagle}.

The faintest galaxies discovered in the new imaging surveys usually lack accurate distance information, and many of them have such a low surface brightness that obtaining spectroscopic redshifts for a complete sample is not possible with the available instruments. 
Thus, to assess the cluster memberships of very faint objects, one needs to exploit their photometric properties.
Surface brightness fluctuations (SBFs) enable us to determine the distance of a galaxy measuring the variance in its light distribution, arising from fluctuations in the numbers of and luminosities of individual stars per resolution element \citep{tonry1988new}.
However, measuring SBFs for large samples of faint galaxies at distances D>20 Mpc is really challenging with observing facilities \citep[see][to get an idea of the potential of the SBFs method, and its application to the Hydra I cluster]{mieske2003potential, mieske2005sbf,Carlsten2019SBF,Greco2021,blakeslee2018sbf}.
Another way is to use the photometric redshifts, often used to obtain distances for a large samples of galaxies \citep[see for example ][]{bilicki2018photometric}.
Finally, it is possible to address cluster memberships using the known scaling relations for galaxies. 
In fact, in clusters there are up to thousands of galaxies located at a similar distance, and many of their properties scale with each other.
On the other hand, background galaxies are spread over a wide range of distances, so their apparent parameters do not follow any relation.
Useful relations commonly used for identifying cluster members are the color-magnitude and the magnitude-surface brightness relations \citep[see for instance ][]{Misgeld2009cent}.
Already the work by \cite{binggeli1985studies} have used colors, the luminosity-surface brightness relation and galaxy morphology to define the membership status in the Virgo cluster.
More recently, \cite{venhola2018,venhola2019optical} used the mentioned scaling relations to construct the FDS dwarf galaxy catalog for the Fornax cluster. 

Correlations among global parameters of galaxies are not only useful to assess their membership, but they also provide insight into the physical processes that impact the formation and evolution mechanisms of these galaxies. 
For example, luminosity, color, size, surface brightness, light concentration, central velocity dispersion are related to each other,  \citep[such as ][]{visvanathan1977color,kormendy1977brightness,faber1976velocity,djorgovski1987fundamental}
Specifically, the fundamental plane, the color-magnitude relation (CMR) and the luminosity-surface brightness relation link the physical properties of the underlying stellar populations and the global structural properties with the galaxy mass.
Investigating these scaling relations in multiple environments sets constraints for galaxy evolutionary models.
Interestingly, since the observed properties of galaxies reflect their evolution, it is possible to infer how they reached the actual configuration, and what were the main processes responsible for their evolution. 
The evolutionary processes can be divided into two main categories: internal and environmental. 
The set of the former processes is usually called mass quenching because they are all related to the mass of the galaxy.
These internal mechanisms usually affect mostly massive early-type galaxies \citep{Su2021}. 
Studies focused on isolated early-type galaxies, for which external influences can be excluded, have shown that mass quenching is effective only for galaxies more massive than $10^9\;M_{\odot}$ \citep{geha2012stellar}.
Concerning dense environments, such as galaxy clusters, dwarf galaxies can experience various environmental processes. 
Ram-pressure stripping is a process affecting the interstellar matter in galaxies: when a galaxy falls into a cluster, its cold gas component interacts with the hot gas of the cluster, and the pressure between the two gas components removes the cold gas from the galaxy's potential well \citep{Gunn1972stripping}.
Because the stars are not removed, galaxy evolution after the gas-removal happens via fading of its stellar populations (id est, through internal processes).
If not all the gas of the galaxy is removed at once, it is likely that a small gas fraction is retained in the center of the potential well, where it may have become overpressurized leading to star formation. 
Indeed, there are dwarf galaxies in clusters with blue centers, which are indicative of a recent star formation episode \citep[for example ][]{Lisker2006dwarf, hamraz2019young}.
These cases are often referred to as partial ram-pressure stripping.
In dense environments, cluster galaxies go through high speed encounters, and the cumulative effect of multiple high-speed collisions is the stripping of stellar, gas and dark matter components from galaxies, the material of which will build up the intra-cluster component. 
This process is called harassment \citep{Moore1996harassment,Moore1998morph}.
Harassment is not efficient in removing a large number of stars \citep{Smith2015sensitivity}, and generally the material is stripped off firstly from galaxy's outskirts, increasing its light concentration.
Moreover, the star formation activity is quenched \citep{Moore1998morph,Mastropietro2005}.
Another drastic consequence of the gravitational interactions is that they can be so strong in the cluster center that they may cause complete disruption of dwarf galaxies \citep{mcglynn1990remnants,Koch2012,Sasaki2007}.
The ripped off material then ends up in the intra-cluster medium, in which it piles up. 

On this paper we focus on the cluster core, where vivid examples of tidal dissolution of galaxies have already been observed \citep{arnaboldi2012,Koch2012}. 
However, observationally there is still a lack of systematic studies analyzing the environmental effects, and the properties of dwarf galaxies for a wide portion of the Hydra I cluster.
Recently, the FDS approached this problem from an observational point of view for the Fornax cluster dwarf galaxies, using imaging data from the ESO VLT Survey Telescope \citep{venhola2019optical}, and so we aim to analyze Hydra I cluster similarly in this paper, with images from the same instrument.


\object{Hydra I}, also known as \object{Abell 1060}, is among the closest and largest galaxy clusters in the nearby Universe.
It is the prototype of an evolved and dynamically relaxed cluster, dominated by early-type galaxies and having a regular core shape.
The cluster center is inhabited by its two brightest members, \object{NGC 3309} and \object{NGC 3311}, of which the latter is the central Dominant (cD) galaxy of the cluster. 
Both are surrounded by an extended diffuse stellar halo \citep{arnaboldi2012}.

In Table \ref{tab:hydra} we summarize the adopted values for the different parameters of \object{Hydra I}. 
Throughout the paper we assume the average cluster distance of $51\pm 4\;Mpc$ ($(m-M)=33.55\;mag$), assuming $D=\overline{cz}/H_0$\footnote{We adopt a $\Lambda$CDM cosmology with $\Omega_m= 0.3$, $\Omega_{\Lambda}= 0.7$ and $H_0 = 72\;km\;s^{-1}\;Mpc^{-1}$, \cite{freedman2001final}.}.
\citet[][]{Richter1982} and \citet{Richter1987} found evidence for an empty space of $40-50\;Mpc$ extending in front and behind the \object{Hydra I} cluster along the line of sight. 
This implies that background objects appear smaller on the images than \object{Hydra I} galaxies (at least a factor $\times0.5$), and that the cluster is relatively isolated in redshift space.

\begin{table*}[ht]
    \caption{Hydra I galaxy cluster properties. }
    \centering
    \begin{tabular}{cccc}
    \hline \hline
        Parameter & Estimate & Reference & Notes \\
        \hline
         $\overline{cz}$ & $3683\pm46\;km/s$ & \cite{Christlein2003} & - \\
         $\sigma$ & $724\pm31\;km/s$ & \cite{Christlein2003} & - \\
         Core radius, $r_c$ & $170h^{-1}kpc$ & \cite{Girardi1995} & - \\
         $R_{200}$ & $\sim1.6\;Mpc$ & - & Derived in this work from the other parameters \\
         $M_{200}$ & $1.9\times10^{14}h^{-1}M_{\odot}$ & \cite{Girardi1998} & Virial mass from optical studies \\
         $M_{200}$ & $2.1\times10^{14}h^{-1}M_{\odot}$ & \cite{Tamura2000} & Virial mass from X-ray study \\
        \hline
    \end{tabular}
    \label{tab:hydra}
\end{table*}

The center of the cluster-wide X-ray emission looks to be centered near \object{NGC 3311} \citep[slightly displaced toward northwest][]{hayakawa2004inhomogeneity}.
The total mass estimated from X-rays measurements is relatively high compared to the number of known galaxies in the cluster \citep{Tamura2000,loewenstein1996mass}. 
This could mean either that the fraction of dark matter (DM) content in \object{Hydra I} is somehow higher than seen in similar clusters, or that there is a number of still undetected faint galaxies. 
The work of \cite{misgeld2008early}, as well as the spectroscopic catalog by \cite{Christlein2003} show a rich environment, with a large population of dwarf galaxies in its core. 
However, the first work is focused on a small area around the cluster core, while the latter is limited to mid-luminosity dwarfs ($M_R \lesssim -14\;mag$).

Several other works, studying the light distribution and the kinematics of the \object{Hydra I} cluster core, demonstrated that the mass assembly is still active around \object{NGC 3311}, and showed the presence of ongoing interactions between galaxies \citep{arnaboldi2012,ventimiglia2011,hilker2018,barbosa2018,barbosa2021,Koch2012}. 
Hereafter, we assume that the cluster center is represented by \object{NGC 3311}: $R.A.=159.17842\;deg$, and $Dec. = -27.528339\;deg$ (respectively 10:36:42.82 and -27:31:42.02).

In this paper we present a new catalog of dwarf galaxy ($M_r>-18.5\;mag$) members of \object{Hydra I} galaxy cluster. 
The aim of the work is to use this new catalog to study the structural (effective radius $R_e$, Sérsic index $n$, and axis ratio) and photometrical (colors, surface brightness) properties of the galaxies, as well as the differences between the populations of galaxies of different morphological types in the \object{Hydra I} cluster. 
In Section \ref{Observations} we describe the observations used in this work;
then we explain all the steps necessary to construct the Hydra cluster dwarf catalog in Sect. \ref{Extraction} and \ref{Membership}.
The final catalog of dwarf galaxies is presented in Sect. \ref{Final catalog}.
In Section \ref{Results} we present the dwarfs spatial distribution, their colors, their structural scaling relations, and how these properties vary as a function of the clustercentric distance and of their total luminosity.
Finally, in Sections \ref{Discussion} and \ref{Summary} we discuss and summarize the results.

\section{Deep images of the Hydra I cluster from VEGAS}\label{Observations}

The \object{Hydra I} galaxy cluster is one of the targets of the VEGAS research project. 
VEGAS is a deep multiband ($u, g, r, i$) imaging survey \citep{Capaccioli2015,Iodice2020,iodice2021vegas}, carried out at the European Southern Observatory (ESO) VLT Survey Telescope \citep[VST,][]{Schipani2012}.
The optical camera OmegaCAM \citep{kujiken2011} on the VST, used to record wide field images spanning $1\times1 deg^2$ field of view, has a mosaic of 32 CCDs with a resolution of $0.21\;arcsec\;pixel^{-1}$.

The deep $g$ and $r$ bands images of the \object{Hydra I} cluster have been already presented in \citet{Iodice2020c}. 
The observations have been acquired with the step-dither observing strategy, which guarantees an accurate estimate of the  sky  background \citep[see for instance ][]{iodice2016fornax,venhola2018}.
The data reduction has been performed using \emph{VST-Tube}, one of the dedicated pipelines to process OmegaCAM images \citep{Grado2012vsttube,Capaccioli2015}.
The sky field has been acquired on the west side of the cluster, adopting an overlap of $\sim0.3$ deg with the 1 deg field centered on the core of the cluster.
Therefore, the final sky-subtracted reduced mosaic for the \object{Hydra I} cluster extends over $1^{\circ}\times2^{\circ}$, corresponding to $\sim 0.9\times1.8\;Mpc$ at Hydra's distance, which means that the cluster is symmetrically covered out to $\approx 0.4$ virial radius. 
During the data acquisition, we have taken special care to put the bright (7th-magnitude) foreground star, which is on the NE side of the cluster core, always in one of the two wide OmegaCam  gaps, thereby reducing the scattered light. 
The residual light from this bright star has been modeled and subtracted from the mosaic, in both bands.
The light distribution of the second brightest star in the field, located SE the core, is also modeled and subtracted from the parent image \citep[see Fig.1 in][]{Iodice2020c}.

Thanks to the long integration times ($2.8$ hours in the $g$-band and $3.22$ hours in the $r$-band), and the specific observing strategy, the final stacked images reach surface brightness depths of $\mu_g=28.6\pm0.2\;mag/arcsec^2$, and $\mu_r=28.1\pm0.2\;mag/arcsec^2$.
They are derived as the flux corresponding to $5\sigma$, with $\sigma$ averaged over $1\;arcsec^2$ from an empty area.
The median FWHM of the point spread function are $0.81\arcsec$ in the $r$-band, and $0.85\arcsec$ in the $g$-band. 

\begin{figure*}
    \centering
    \includegraphics[width=0.92\textwidth]{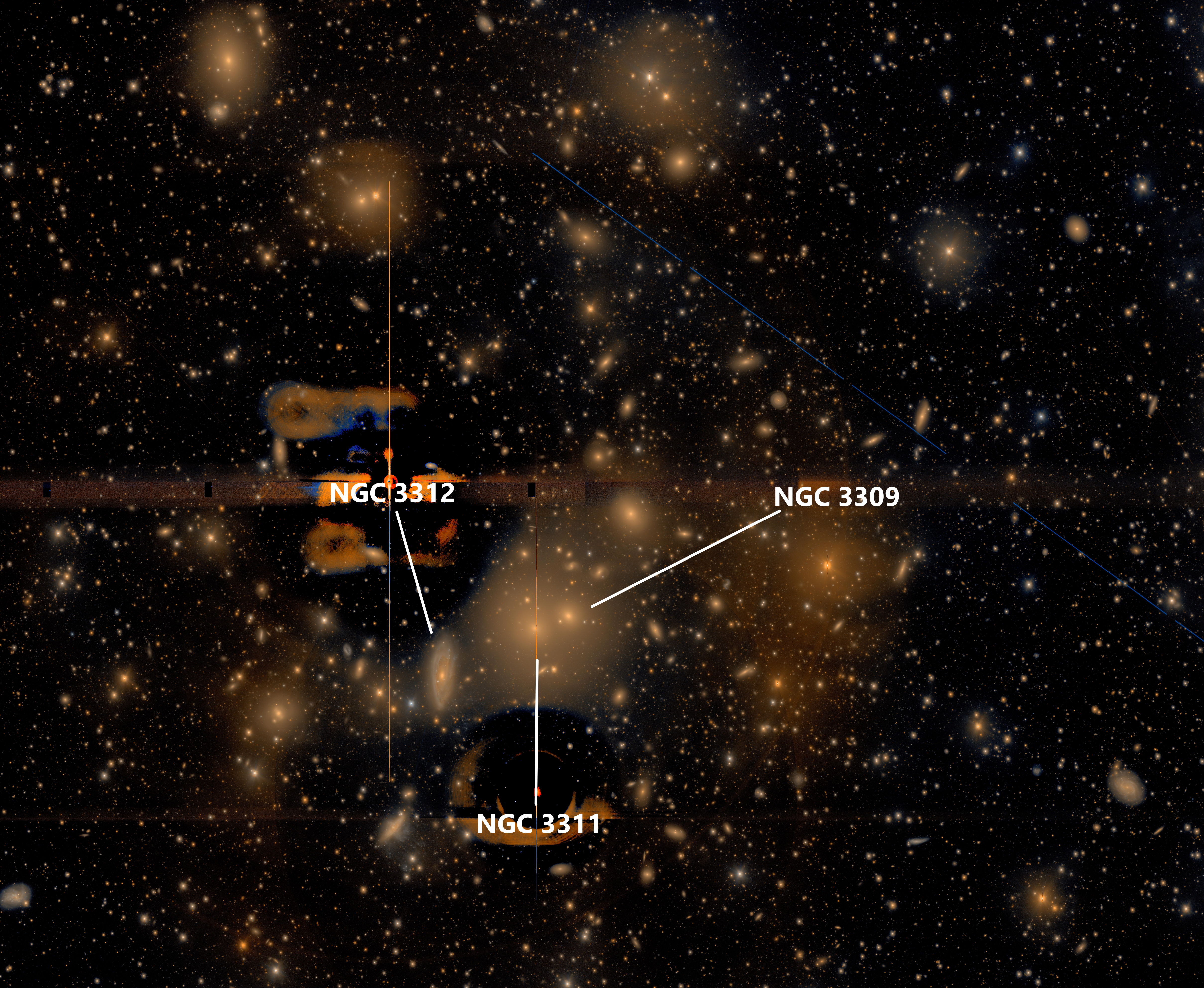}
    \caption{Color composite image ($g$ and $r$ bands) of the VST mosaic's portion used in this paper ($56.7\times46.6\;arcmin^2$). The two cD galaxies \object{NGC 3311} and \object{NGC 3309}, which lie in the cluster center, are labeled in the figure. Close to them, to the left is marked \object{NGC 3312}, a large spiral galaxy. Residual light from the two subtracted stars is still visible. North is up, and east is on the left.
    The color RGB image is created using the \cite{Lupton2004rgb} scheme, using the $r$-band image as R filter input, the $g$-band as B filter input, and the average image between $r$ and $g$ bands as G filter input.}
    \label{fig:mosaic}
\end{figure*}

In this work we focus on a $56.7\times 46.6\; arcmin^2$ wide portion of the VST/OmegaCAM mosaic that maps a symmetric area around the cluster core (Fig. \ref{fig:mosaic}).
This portion, at the assumed \object{Hydra I}'s distance, corresponds to $\sim 0.8\times0.7\;Mpc$, enabling us to study the cluster in a systematic way up to a considerable fraction of the virial radius ($\approx 0.4R_{vir}$).

\section{Extraction of the sample}\label{Extraction}

In this section we describe the detection algorithm for the Low Surface Brightness (LSB) galaxies, which is based on the method presented in \citet{venhola2018} and successfully applied to the VST data of the FDS.

\subsection{Detection of the sources}\label{Detection}

We used the automatic detection algorithm {\sc SExtractor} \citep{bertin1996} to identify dwarf and LSB galaxies in the \object{Hydra I} cluster.
{\sc SExtractor} works detecting objects with a certain number of connected pixels, brighter than a chosen threshold. 
Hence, the first parameters to configure are the detection threshold and the minimum number of pixels above that value, \texttt{DETECT\_TRESH} and \texttt{DETECT\_MINAREA} respectively. Both parameters have to be set according to the type of galaxy one wants to detect. 

The software creates a background model by defining a grid of image pixels, and then estimating the background level in each grid box. 
The background grid size is therefore another critical parameters of {\sc SExtractor}. 
For the more extended galaxies the grid size should be set to be large enough to prevent false detections. 
For the smaller galaxies, which often are spatially blended with brighter ones, the grid size should be set to a smaller value, so that the larger galaxies can be included into the background model.

Following \citet{venhola2018}, we use different parameters for three different classes of objects: Large, LSB, Small sources. 
We ran {\sc SExtractor} in the so-called double-image mode, using the $r$-band image (after subtracting the light from the two brightest stars) for the detection, while the $g$-band was used for measurements only. 
The final combination of the main detection parameters is shown in Table \ref{tab:parameters}.

\begin{table}[ht]
    \caption{Main configuration parameters used in {\sc SExtractor} for the three different classes of sources.}
    \begin{center}
        \begin{tabular}{c|c|c|c}
        \hline \hline
         List & Thresh. ($\sigma_{sky}$) & Min. area (pix) & Back. size (pix)  \\
         \hline
         Large & 50 & 10000 & 21000$\times$21000 \\
         LSB & 5 & 25 & 256$\times$256 \\
         Small & 1 & 10 & 64$\times$64 \\
         \hline
        \end{tabular}
    \end{center}
    \
    \label{tab:parameters}
\end{table}

\begin{figure}
    \centering
    \includegraphics[width=0.49\textwidth]{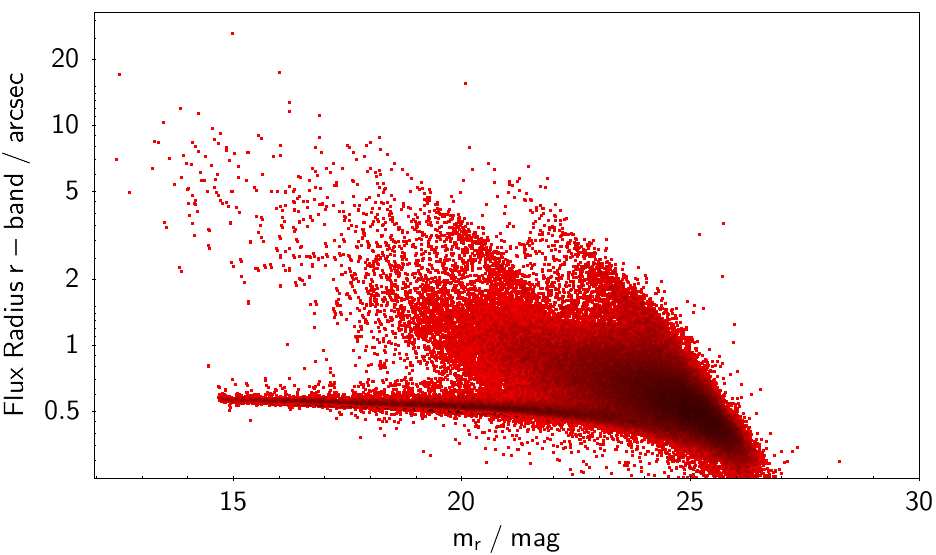}
    \caption{Flux radius in $arcsec$ vs. $r$-band magnitude. A tight band extends horizontally from $m_r\sim14\;mag$ to $\sim22\;mag$, and it is inhabited by stars. Extended objects are arranged in a more diffuse way.}
    \label{fig:flux_radius}
\end{figure}

\begin{figure}
    \centering
    \includegraphics[width=0.49\textwidth]{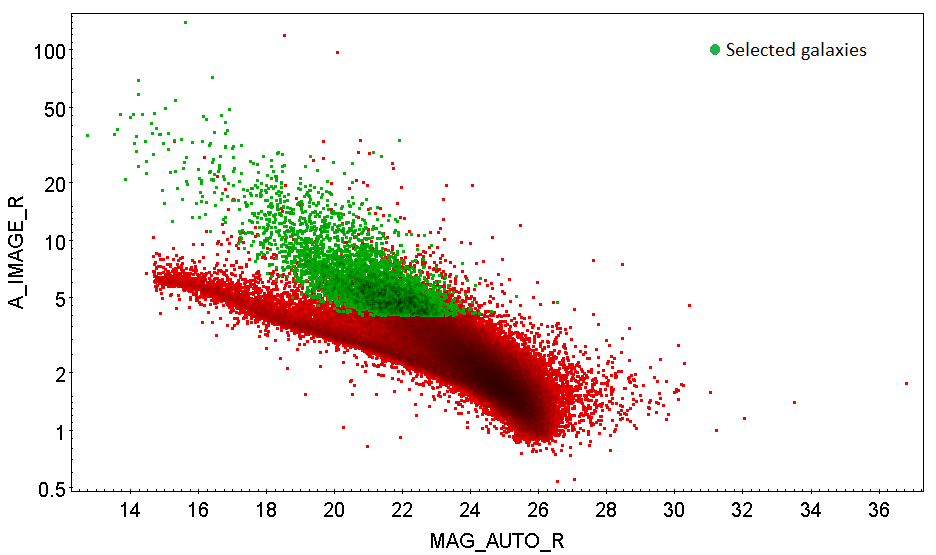}
    \caption{Size-magnitude diagram for detected sources, with size in pixels. Green dots represents the galaxies selected for the final clean catalog, i.e. have semi-major axis larger than 4 pixels, and have passed all the previous steps.}
    \label{fig:selected galaxies}
\end{figure}

Given the goal of this work, we focused only on the \emph{Small} and \emph{LSB} tables, which both consist of a huge number of sources, over $30,000$ and $150,000$ objects, respectively. 
{\sc SExtractor} gives as output object lists with a certain number of parameters associated with each detection. 
Most of these detections are foreground stars, globular clusters hosted by \object{Hydra I} galaxies, false detections or unresolved background galaxies, which must be removed from the list.
To this aim, we proceeded as follows: 
firstly, we removed sources with magnitude $m_r=99.0\;mag$ or $m_g=99.0\;mag$, which are clearly software errors;
secondly, we removed all the objects which contain the {\sc SExtractor} flag $\ge 4$\footnote{For a full reference of all {\sc SExtractor}'s flags see \url{https://sextractor.readthedocs.io/en/latest/Flagging.html}};
thirdly, we divided the sample into "stars" and "not-stars" using a criterion based on the \texttt{FLUX RADIUS}\footnote{This {\sc SExtractor}'s parameter estimates the radius enclosing a certain quantity of the total object's flux. In this case, we consider half of the total light.}. 
Stars are likely to have smaller flux radii since they are point sources, while galaxies are much less concentrated objects. 
In a flux radius vs magnitude plot (Fig. \ref{fig:flux_radius}) this is particularly visible, with stars inhabiting a tight sequence around $1\arcsec$. 
This band merges with diffuse objects for magnitude fainter than $\sim 22\;mag$.
Given that, we have marked all objects with flux radius lower than $0.9\arcsec$, brighter than $m_r=22\;mag$, as "stars", and therefore excluded them from the catalogs.
We used the same criteria also for the \textit{g}-band.
Finally, we excluded the unresolved galaxies and globular clusters.
This is done by removing all the objects with semi-major axis smaller than $4\; pixels$ (or $0.84\;arcsec$), as measured by {\sc SExtractor}'s parameter \texttt{A\_IMAGE}, see Fig. \ref{fig:selected galaxies}.
This size limit corresponds to $\sim 200\;pc$ at Hydra's distance, therefore we are aware that we are excluding also some intrinsically small, unresolved \object{Hydra I} cluster galaxies, like for example the smallest Ultra Compact Dwarfs \citep[UCDs, e.g.][]{misgeld2011}.
Importantly, we are not removing from the sample of \object{Hydra I} cluster galaxies similar to the Local Group dwarf spheroids (dSphs), that usually have effective radii between $200\; pc <R_e<1000\; pc$.

These selection criteria exclude $\sim 98\%$ of the initial {\sc SExtractor}'s detections in the field. 
The remaining objects in the two lists were joined, avoiding to insert duplicates: objects within $2\;arcsec$ from each other are considered the same galaxy. 
For duplicates, parameters and coordinates were taken from the \emph{Small} list.
As a result, a cleaned catalog in all bands is obtained, containing $4535$ candidate galaxies for \object{Hydra I}.
Coordinates, magnitudes and morphological parameters from {\sc SExtractor} are used as initial guesses for the photometric pipeline (see next Section).
We did not filter any further since, as described in the next Section, we use more robust photometric measurements, in order to divide background galaxies from cluster members. 

\subsection{Efficiency of the detection} \label{mock-gal}

In order to determine the typical detection limits, we test the completeness of our detection algorithm. 
We use a python script to generate iteratively 3000 mock galaxies, divided in bunches of 250 individuals, and added them to 12 template images, that have the same size and seeing conditions of the $r$-band images. 
We use the seeing of the r-band because the detection is done on this filter. 
We also include 30000 foreground stars to better resemble the observed field
\footnote{Stars are injected according to an exponential law in the magnitude range $15\leq m_r \leq 23\;mag$. The total number of injected stars is approximately normalized by estimating -for the appropriate magnitude range- the total number of stellar detections from GAIA \citep{Gaia2016} over the same area.}.
To simulate the light profiles of artificial galaxies, we use 2D-Sérsic profiles as artificial galaxies, convolved with the PSF of OmegaCAM.
Moreover, poissonian noise is added into each pixel.
The galaxies are embedded into the template images with random locations and position angles.
We chose a wide range of input parameters to cover the expected parameter space of dwarf galaxies and UDGs in the \object{Hydra I} cluster, which are summarized in the following:
\begin{itemize}
    \item  $\bar{\mu}_{e,r} = 22-30\;mag\;arcsec^{-2}$
    \item $R_e= 0.4-8\;arcsec$ ($\sim 0.1-2\;kpc$)
    \item Sérsic $n=0.5-3.0$
    \item Axis ratio $b/a=0.2-1.0$
\end{itemize}
The only shortcoming of this procedure is that we are not considering galaxies with 2 or more components.
However, since LSB galaxies usually do not show many features,  this should not affect the detection \citep{Su2021}.

\begin{figure*}
        \centering
        \includegraphics[width=.32\textwidth]{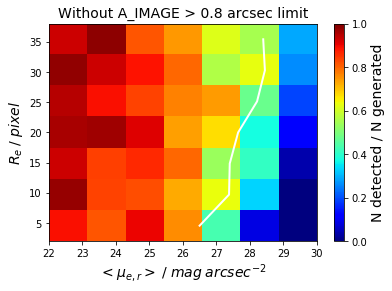}
        \hfill
        \includegraphics[width=.32\textwidth]{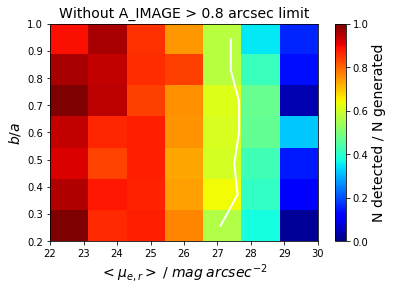}
        \hfill
        \includegraphics[width=.32\textwidth]{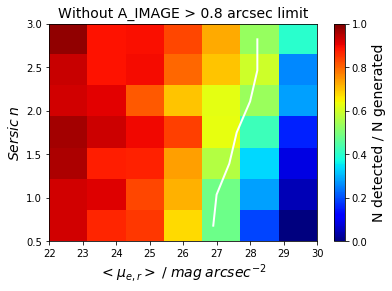}
        \\
        \centering
        \includegraphics[width=.32\textwidth]{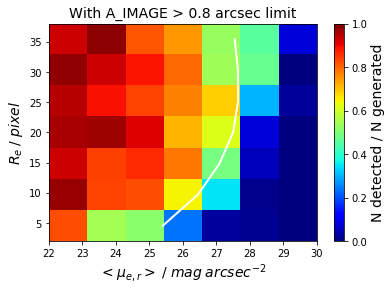}
        \hfill
        \includegraphics[width=.32\textwidth]{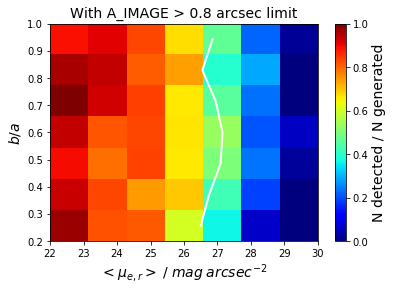}
        \hfill
        \includegraphics[width=.32\textwidth]{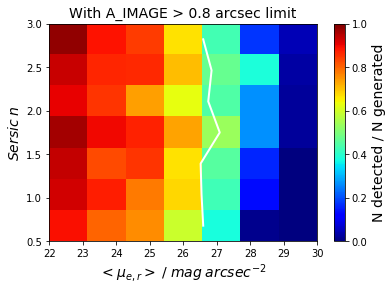}

        \caption{Detection efficiency (number of detected galaxies over number of generated ones in the same bin) of our detection algorithm is shown color-coded such that red means more efficient, while blue less efficient. The efficiency is plotted for effective radius ($R_e$, the pixel scale is $0.21\,arcsec/pixel$), axis ratio ($b/a$), and Sérsic index $n$, as function of mean effective surface brightness, without (top row) and with (bottom row) minimum size limit. The white lines show the $50\%$ detection efficiency limit.}
        \label{fig:Detection_efficiency}
\end{figure*}

We run {\sc SExtractor} on the simulated images, with the same combination of configuration parameters used for science frames.
We look at how many objects it is able to identify. 
For each mock galaxy we require the center to be detected within $2.5\;arcsec$ from the true central coordinates.
To understand the effect of the minimum size cut carried out, we remove the objects with $A\_IMAGES<0.84\;arcsec$ from the detections. 
Fig. \ref{fig:Detection_efficiency} shows how the detection efficiency changes as function of different structural parameters, with and without the minimum size limit. 
We find that the detection efficiency slightly depends on the Sérsic index so that more peaked galaxies are more efficiently detected. 
More extended sources are more easily recognized, while there is no clear dependence on the axis ratio, as expected.

The detection efficiency reaches $50\%$ at a $r$-band magnitude value of $m_r=23.0\;mag$ and at mean effective surface brightness value of $\bar{\mu}_{e,r} = 27.5\; mag\;arcsec^{-2}$. 
When the size cut is applied, the $50\%$ efficiency is reached at $m_r=22.0\;mag$ and $\bar{\mu}_{e,r} = 26.5\; mag\;arcsec^{-2}$.
We take FDS data for comparison: there, the exposure times in $g$- and $r$-band, in a single field, were of $2.3$ h, with typical seeing of $1.1$ and $1.0$ arcsec, and surface brightness depths of $28.4$ and $27.8\;mag/arcsec^2$, respectively \citep{iodice2016fornax,venhola2018}. 
Data used in this work are deeper, and our detection efficiency reaches the $50\%$ level slightly deeper than in FDS.
Indeed, \cite{venhola2018} detection has the limiting $r$-band magnitude with $50\%$ detection efficiency of $m_r=21\;mag$ and $\bar{\mu}_{e,r} = 26\; mag\;arcsec^{-2}$.
However, since Fornax has less than half the distance of \object{Hydra I}, their dwarf catalog goes fainter than ours by $1\;mag$ (up to $m_r\simeq-10.5\;mag$).

\subsection{Photometric measurements}

In order to identify \object{Hydra I} cluster members and to study their optical properties, we derive the photometric parameters for all galaxies in the catalog of candidates as obtained in Section \ref{Detection}, namely all galaxies that have {\sc SExtractor} semi-major axis larger than $0.84\;arcsec$ ($\sim200\;pc$). 
They are derived by fitting the 2D light distributions of all targets using {\sc GALFIT} \citep{Peng2002,Peng2010}, assuming a Sérsic profile to obtain the effective radii ($R_e$), magnitudes and colors, axis ratios and position angles, and Sérsic indices.

To this aim, we run a semi-automatic photometric pipeline over all objects in the final candidates catalog. 
The pipeline was originally developed to work with VST data of the FDS \citep{Peletier2020} data, therefore we do not need to modify it.
A full description of how the photometric pipeline works is given by \citet{venhola2017fornax} and \citet{venhola2018}.
Here we briefly list the main operations:

\renewcommand{\labelitemi}{\textendash}
\begin{itemize}
    \item Post-stamp images of the galaxies are made in both bands, using as center the coordinates of the object as detected by {\sc SExtractor}.
    The semi-width of images is limited to 10 \texttt{A\_IMAGE}, again, as measured by {\sc SExtractor}. 
    These semi-major axis lengths are not always accurate, and sometimes the cutouts can be too small for the selected galaxy candidates. 
    In such cases we increase the size manually.
    
    \item For the stamps of each candidate, other objects than the main galaxy are masked, like faint stars or nearby galaxies. 
    This is done with an automatic masking routine\footnote{For further details see Section 7.1 \citet{venhola2018}.}, followed by a manual masking when needed. 
    When two galaxies are overlapping or are so close that masking is not feasible, both are modeled simultaneously by {\sc GALFIT}. 
    
    \item The estimate of the initial input parameters of {\sc GALFIT} is done by making an azimuthally averaged radial profile of the galaxy, using two pixels wide circular bins. 
    Then the clipped average of each bin is taken to make a cumulative profile up to three times the semi-major axis value (where the semi-major axis is taken from {\sc SExtractor}).
    From the growth curve, effective radius and magnitude values are estimated, and used as input parameters for {\sc GALFIT}. 
    Also the center is measured at this step: if the object has a clear center, a 2D-parabola is fitted and the peak is considered the center of the galaxy. For galaxies with flat center, {\sc SExtractor}'s coordinates are used as center. 
    
    \item At this point the post stamps are ready for the {\sc GALFIT} run. 
    The objects are fitted in both $g$ and $r$ band with a single Sérsic function, or with a combination of a Sérsic function and a point-source for the nucleus, if required (it is up to the user to choose, based on the visual appearance of the galaxy). 
    The pipeline allows the Sérsic component to have a different center than the nucleus, because it is possible to have off-centered nuclei\footnote{Some examples are given in \citet{bender2005hst} and \citet{Cote2006acs}.}.
    
    \item The fits are inspected to see if they resemble real galaxies. 
    The inspection is done by looking at the radial profile with the model overlapped, at the residuals, and the original image with an aperture of radius $1R_e$ overlaid. 
    The latter is done in order to understand if the estimated $R_e$ is reasonable. 
    
\end{itemize}
\renewcommand{\labelitemi}{\textbullet}
The colors of the galaxies are derived separately, the $g-r$ colors are calculated from the $g$ and $r$ band magnitudes within the effective aperture, as calculated from {\sc GALFIT} models. 
In particular, we measure the flux in the elliptical effective aperture and calculate the corresponding magnitude. 
For this task, we use $r$-band parameters also for $g$-band measurement, in order to obtain consistent magnitudes.
Colors calculated in this way are more stable than using total magnitudes, because in the outermost part of the galaxies the determination of the background can severely affect the colors.

In conclusion, the pipeline returns a catalog containing photometric and structural parameters of the galaxies: the effective radius $R_e$, axis ratio $b/a$, position angle $\theta$, Sérsic index $n$, magnitudes and aperture $g-r$ color
\footnote{All magnitudes and colors are corrected for Galactic extinction using correcting values from \citet{schlegel1998maps} and \citet{Schlafly2011}}.
It is important to highlight that we consider only the structural parameters in the $r$-band, because of the higher S/N. 

Inferring accurate errors for the {\sc GALFIT} models is crucial. 
As pointed out by \cite{haussler2007gems}, {\sc GALFIT} tends to underestimate errors on the fitted parameters. 
Full details of how we infer errors for the photometric parameters are given in Appendix \ref{App:A}.


\section{Selection of the cluster galaxies}\label{Membership}

From the source catalog we aim at distinguishing between the cluster and the background galaxies.
The most recent and complete spectroscopic sample for the \object{Hydra I} cluster is given by \cite{Christlein2003}, which fully covers the area studied in this work. 
They consider \object{Hydra I}'s member galaxies with recession velocity within the range $2292 - 5723\;km/s$.
However, the spectroscopic redshift sample of \cite{Christlein2003} is limited to bright galaxies, therefore calibrated selection limits should be carefully extended to low luminosity. 

We exploit the effects of greater distance on galaxy properties to remove background galaxies from our sample.
Firstly, with the increasing distance, galaxies appear to have smaller angular sizes, while their  surface brightness stays (almost\footnote{The increasing redshift dims the surface brightness too, but the contribution is relatively small for low and mid redshifts. The dimming factor is $1/(1+z)^4$, that means that at \object{Hydra I}'s redshift $z\sim0.01$ there is $\sim3\%$ dimming.}) constant. 
Hence intrinsically bright galaxies at large distance will show a low apparent total magnitude, but high surface brightness. 
On the other hand, cluster galaxies follow the surface brightness vs luminosity relation \citep{binggeli1984studies}, thus the fainter the galaxy, the lower the surface brightness.
Therefore, we can use this relationship to separate background and foreground galaxies. 
Galaxies in clusters follow also a color-magnitude relation (CMR), that is they become bluer going to lower luminosities \citep{visvanathan1977color,roediger2017ngvs,venhola2019optical}.

Additionally, large background galaxies can have structures, such as spiral arms or bars, while their angular size is comparable to the size of Hydra’s dwarfs. 
These features are unlikely shown by small dwarf galaxies \citep{Janz2014}, although they can be present in the brightest dwarfs. 
The latter, anyway, are bright enough to have measured spectroscopic redshifts.
Hence, there is no degeneracy in that sense, and a visual classification based on galaxies morphological appearance is helpful to further distinguish background galaxies from cluster members.

Optical photometry alone is not sufficient for defining the cluster membership, and some degree of degeneracy exists in the apparent structural and color properties of the cluster galaxies and those at higher redshift.
However, calibrating the selection limits using available archival spectroscopic redshifts, it is possible to minimize the degeneracy problem.
As shown for example by \cite{venhola2018}, this approach gives satisfying results when supported with a visual classification.

In the following paragraphs we explain in detail the adopted procedure to isolate the \object{Hydra I} galaxies in our sample. 

\subsection{Photometric selection criteria}

\begin{figure}
    \centering
    \includegraphics[width=.45\textwidth]{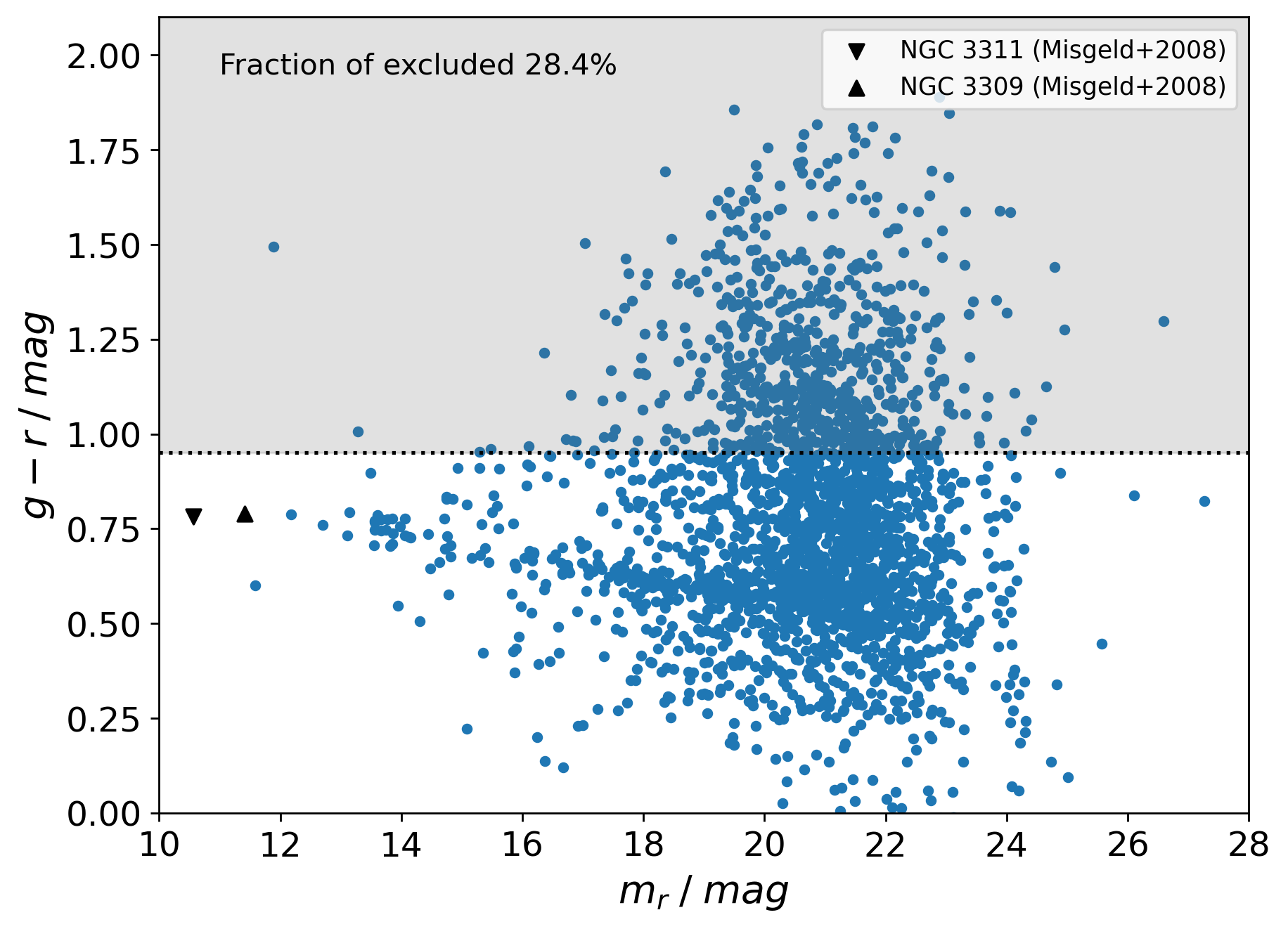}
    \caption{Color selection criteria used to separate the cluster members from the background galaxies. The horizontal dotted line represents the upper color limit we applied, $g-r=0.95\;mag$: galaxies in the shaded gray area are excluded. The two BCGs are marked with black triangles.}
    \label{fig:color_cut}
\end{figure}

We proceed in the following order with the selection cuts based on photometry: initially we perform an upper color cut, secondly a cut based on the surface brightness vs luminosity relation, and finally we use a CMR from a previous work on \object{Hydra I} \citep{misgeld2008early}. 
For the upper color cut, we consider the two brightest cluster members, NGC~3311 and NGC~3309, which have $g-r$ color $\simeq 0.8$, as reported by \cite{misgeld2008early}. 
Given that limit, we exclude all the galaxies that are at least $0.15\;mag$ redder than the $0.8\;mag$, that is $g-r > 0.95$~mag.
In Fig. \ref{fig:color_cut} there is an illustration of the color selection we apply here. 

Doing so, we are mostly ruling out the large background early-type galaxies (ETGs) and spirals, as their intrinsic color could be similar to the largest cluster members, but their apparent color is significantly redder due to the higher redshift. 
\cite{price2009hst} demonstrated that is possible to find in galaxy clusters compact dwarf galaxies that are significantly redder than the sequence defined by normal cluster members. 
However, usually these systems are at most only marginally redder than the brightest cluster galaxies \citep{hamraz2019young}.
We are therefore confident that our limits are sufficiently broad that none of the likely cluster members is excluded.
However, some background spirals with a small or no bulge will be still left in the sample.

\begin{figure*}
    \centering
    \includegraphics[width=0.45\textwidth]{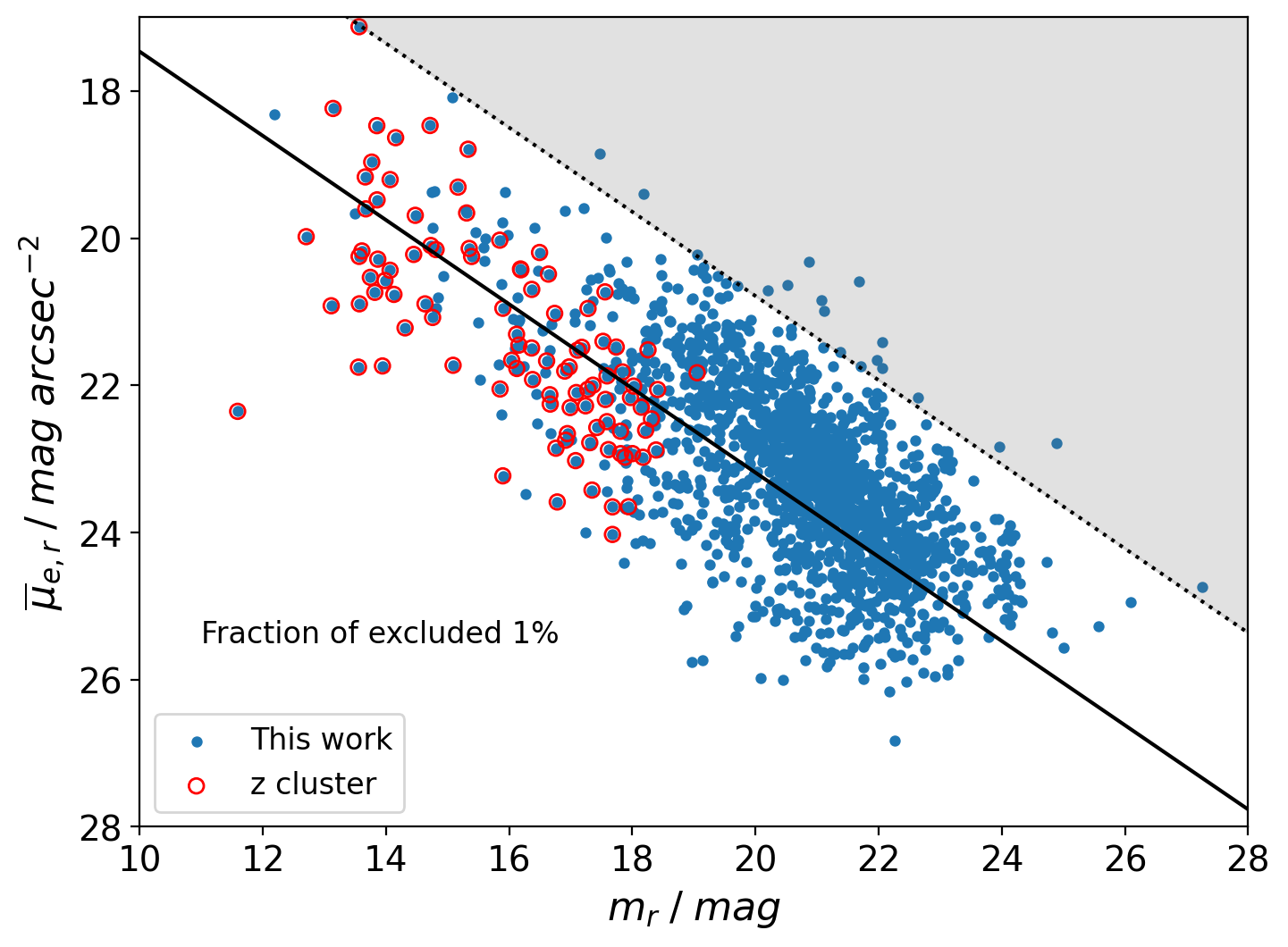}
     \includegraphics[width=0.45\textwidth]{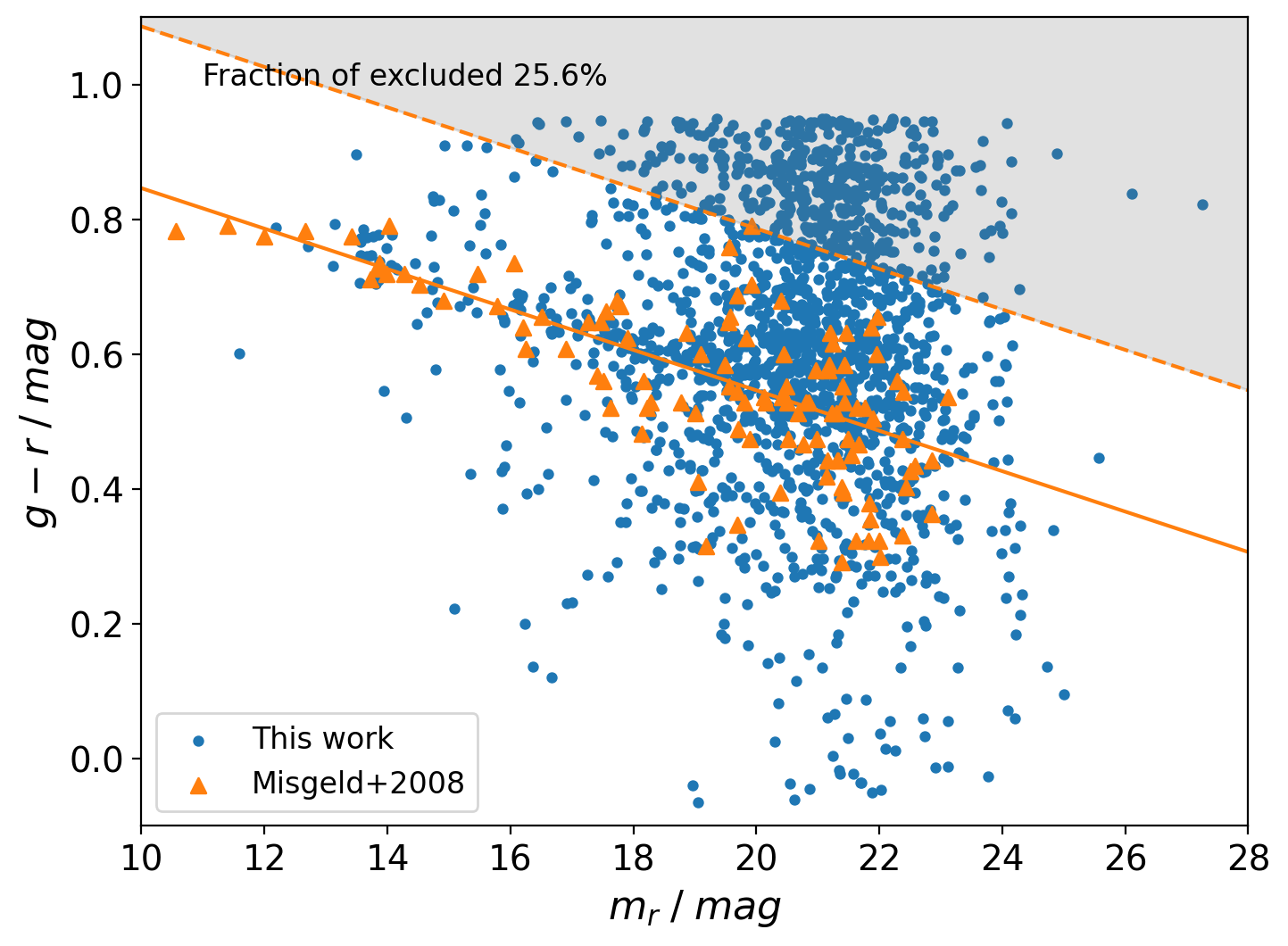}
   \caption{Scaling relations used for the galaxy selection.
   \emph{Left panel} - Mean effective surface brightness vs total magnitude ($r$-band) plane for catalog objects that passed the first selection cut. 
    The black solid line is the linear fit of the redshift confirmed cluster members. The data of \cite{Christlein2003} are represented by red open circles. 
    The dotted line indicates $2$ standard deviations from the cluster sequence: the objects in the shaded gray area are removed from the sample.
    \emph{Right panel} - Color-magnitude diagram for galaxies that passed the first two cuts. Orange triangles are early-type galaxies found by \cite{misgeld2008early}, and the orange solid line is their CMR. The dashed line is the 2 rms level we use to define galaxies as members of the Hydra I cluster.}
    \label{fig:selection_cuts}
\end{figure*}

As second criterion we use the surface brightness vs. magnitude relation.
We take confirmed cluster members \citep{Christlein2003} and make a linear fit of the $\overline{\mu}_{e,r}-m_r$ relation. 
As mentioned by \cite{binggeli1984studies}, the slope of this relationship appears to flatten for brighter galaxies.
Our study is focused on the dwarf population, therefore we fit only the galaxies fainter than $m_r=13\;mag$ (corresponding to $M_r=-20.5\;mag$).
We then take two standard deviations as confidence level, and exclude all the galaxies that passed the first cut with a $\overline{\mu}_{e,r}$ $2\sigma$ brighter than the cluster sequence. 
The left panel of Fig. \ref{fig:selection_cuts} shows the effect of this selection.
With this selection criterion, 17 galaxies from the sample are removed.

To further improve the selection, we use the CMR for the early-type dwarf galaxy population of the Hydra I cluster, provided by Eq. (1) in \cite{misgeld2008early}.
To be sure not to exclude cluster members, we consider an upper confidence interval of $0.24\;mag$, which is $2$ times larger than the root mean square error the authors reported, and remove all galaxies $0.24\;mag$ redder than this sequence. 
This prevents removing galaxies which are already confirmed cluster members:  all dwarf galaxies from \cite{misgeld2008early} sample are included, as shown in the right panel of Fig. \ref{fig:selection_cuts}.
We note that even with this cut, the red compact dwarfs are retained within our sample, because $g-r\sim0.8\;mag$ for  $m_r\;\leq\;24\;mag$.

Even after the last cut, the sample of selected galaxies still includes some of the spectroscopically confirmed background galaxies from \citet{Christlein2003}. 
Considering a tolerance of 5 arcsec for the position of the galaxies, we excluded 69 more galaxies from the sample.

\subsection{Visual classification}

As a next step, a visual inspection is performed for all galaxies in the selected sample, in order to make 
morphological classification of each of them.
This step is only used to identify more background galaxies and clean the Hydra I sample in this way. 
It is not used for detailed morphological analysis, and no new galaxy is added at this stage. 

\begin{figure*}
    \centering
    \textbf{Background galaxies}
    \\
    \includegraphics[width=.19\textwidth]{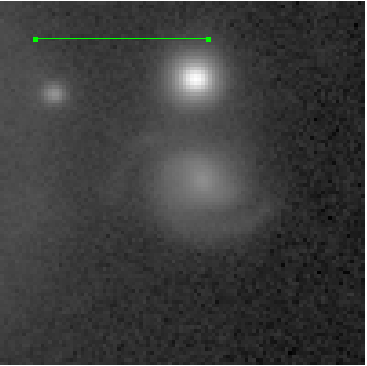}
    \includegraphics[width=.19\textwidth]{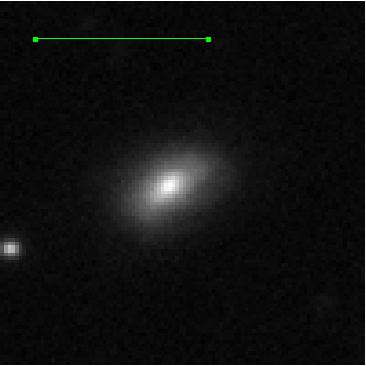}
    \includegraphics[width=.19\textwidth]{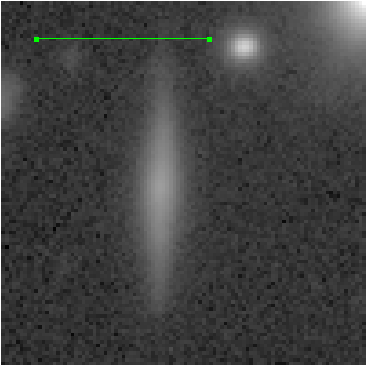}
    \includegraphics[width=.19\textwidth]{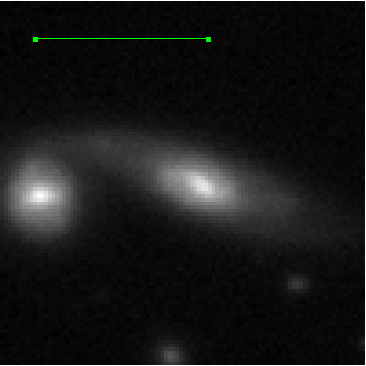}
    \includegraphics[width=.19\textwidth]{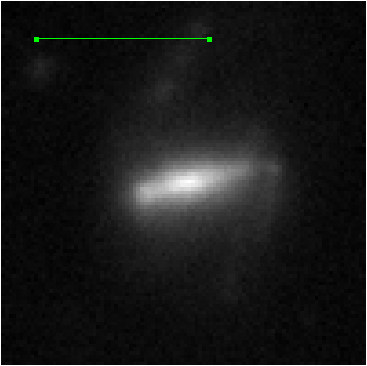}
    \\
    \textbf{Late-type galaxies}
    \\
    \includegraphics[width=.19\textwidth]{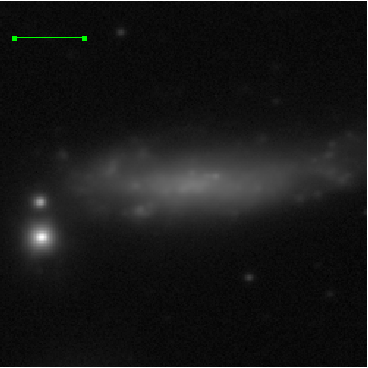}
    \includegraphics[width=.19\textwidth]{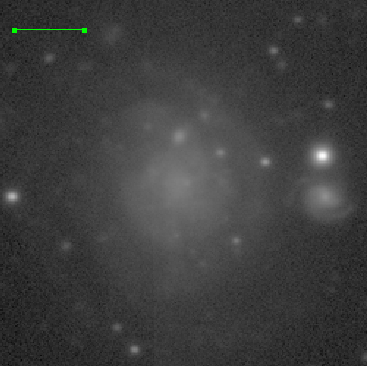}
    \includegraphics[width=.19\textwidth]{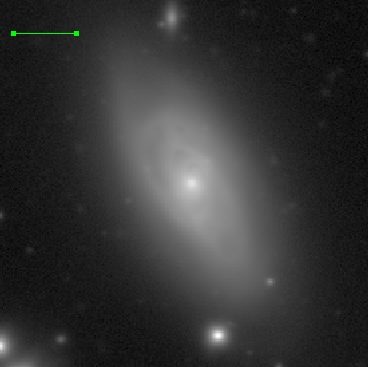}
    \includegraphics[width=.19\textwidth]{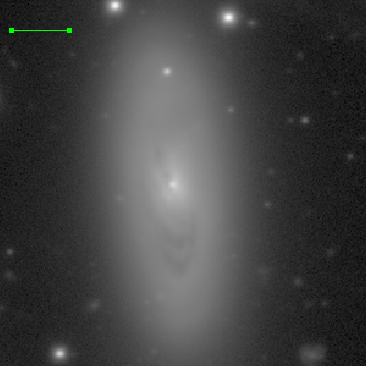}
    \includegraphics[width=.19\textwidth]{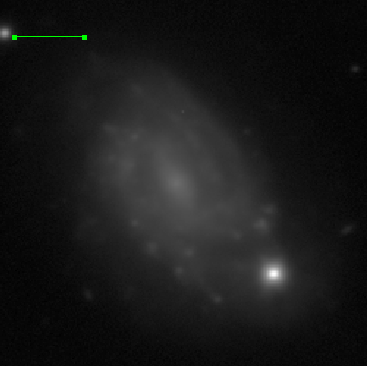}
    \\
    \textbf{Early-type galaxies with structures}
    \\
    \includegraphics[width=.19\textwidth]{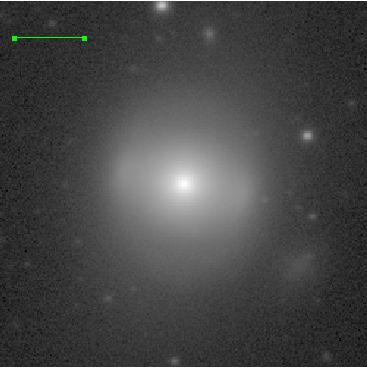}
    \includegraphics[width=.19\textwidth]{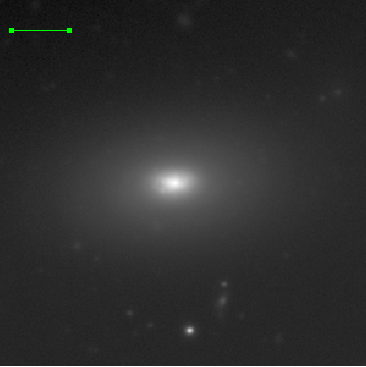}
    \includegraphics[width=.19\textwidth]{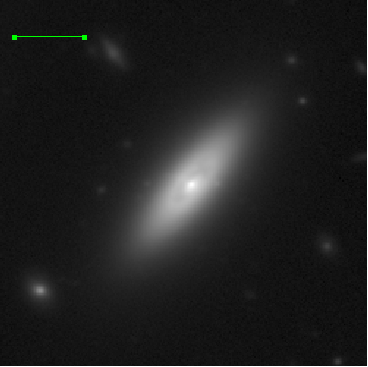}
    \includegraphics[width=.19\textwidth]{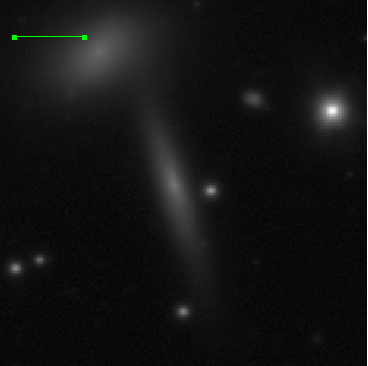}
    \includegraphics[width=.19\textwidth]{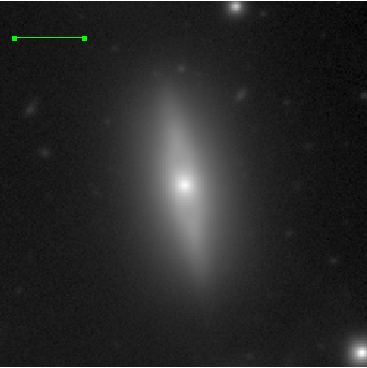}
    \\
    \textbf{Smooth early-type galaxies}
    \\
    \includegraphics[width=.19\textwidth]{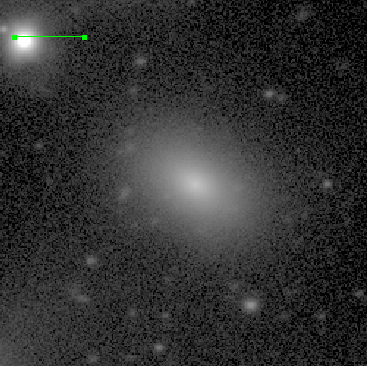}
    \includegraphics[width=.19\textwidth]{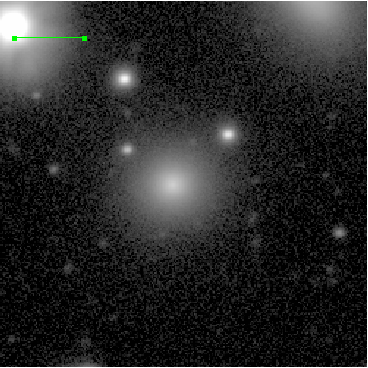}
    \includegraphics[width=.19\textwidth]{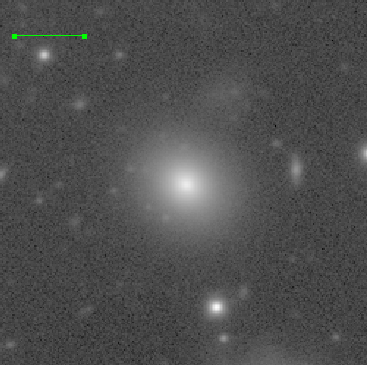}
    \includegraphics[width=.19\textwidth]{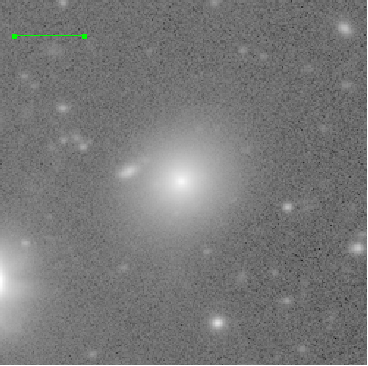}
    \includegraphics[width=.19\textwidth]{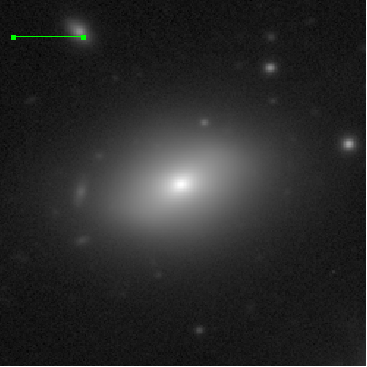}
        
    \caption{$r$-band images of the galaxies as they are classified in the visual inspection phase. 
    Green bars are of equal length in each cutout, $10\;arcsec$.
    Background galaxies show spiral features, but they are considerably smaller than the Late-type galaxies that inhabit the cluster. 
    The difference between the LTGs and the ETGs with structures in our classification scheme is mainly given by the color, hence is not evident in gray-scale images.}
    \label{fig:Visual_insp}
\end{figure*}

To visual classify galaxies, in practice we need extended objects to be sharp and well resolved, thus we decided to exclude all the galaxies with an effective radius smaller than $1\;arcsec$. 
This is motivated by considering that the average seeing of the data is $0.8\;arcsec$. 
Taking into account that we reach the $50\%$ completeness at  $m_r\;=\;22.0\;mag$, we limit this analysis to all objects brighter than this limit.
The objects that satisfy these two constraints are visual classified, looking at their color and residual images.
We divided the galaxies into the following four groups according to their morphology:
\renewcommand{\labelitemi}{\textperiodcentered}
\begin{itemize}
    \item \textit{Late-type}: Galaxies that have blue color and have blue star-forming clumps. 
    In this category spirals, blue compact dwarfs and irregular galaxies are included. 
    \item \textit{Early-type with structure}: Red galaxies with no blue star-forming clumps, but with structures such as a bar, spiral arms, or a bulge. 
    In this group we include S0s galaxies and dEs with an evident disk component.
    \item \textit{Smooth early-type}: Galaxies that appear red and smooth, with no clear structures. 
    The dwarf ellipticals, and the giant ETGs without structures are included here. 
    We do not distinguish between nucleated and non-nucleated dwarfs; both are added to this group.
    \item \textit{Background galaxies}: Small galaxies that show features like spiral arms or bars. 
    These properties are not likely to appear in low-mass dwarfs \citep{Janz2014,Su2021}, thus we conclude that they are background galaxies. 
    We are aware that some massive dwarf galaxies can show prominent disk-like features, with star forming regions, resembling spiral galaxies \citep{Lisker2006dwarf,michea2021brought}. 
    However, the fraction of light in the disk, compared to the smooth spherical component is low and there is no risk to confuse these dwarfs with the background spirals.
\end{itemize}
\renewcommand{\labelitemi}{\textbullet}
In Figure \ref{fig:Visual_insp} we show some examples of galaxies for each morphological category, as we classified them.
In total $916$ identified galaxies were clear enough to be classified. 
569 of them were classified as \emph{background} systems. The remaining $347$ are likely cluster galaxies. 
Specifically, $288$ galaxies were marked as "Smooth early-type galaxies", $36$ as "Late-type galaxies", and the remaining $23$ as "Early-type galaxies with structures". 

\section{Final catalog and comparison with previous Hydra I catalogs } \label{Final catalog}

The final sample of Hydra I cluster members consists of $347$ galaxies. 
We adopt a limit of $M_r\; \ge\;-18.5\;mag$ to define our sample of Hydra I dwarf galaxies, which gives a sample of 317.
In Teb. \ref{table} we present an extract of the complete catalog containing photometric and structural parameters of dwarfs.
The catalog covers the Hydra I cluster core, up to $0.4$ virial radius, and has a minimum semi-major axis limit of $0.84\;arcsec$. 
It reaches a $50\%$ completeness level at the mean effective surface brightness ($r$-band) value of $\overline{\mu}_{e,r}=26.5\;mag\;arcsec^{-2}$, and at $r$-band magnitude $M_r=-11.5\;mag$. 
70 galaxies in our sample are spectroscopically confirmed members: we matched our sample with \cite{Christlein2003} catalog for Hydra I, considering a range of $5\;arcsec$.
Of the 317 galaxies, 61 have already been presented by \cite{misgeld2008early} (in this case considering a tolerance of $3\;arcsec$ on the position). 
Hence, in this paper we report the discovery of 202 dwarf galaxies, previously unknown, in the central region of Hydra I. 

From the sample of \cite{Christlein2003} we miss 20 galaxies, of which 8 are in the area of the residual light of the two saturated stars, 7 overlap with large galaxies, or bright stars, so that they cannot be distinguished, 1 is on the edge of our image, and only the remaining 4 are the ones we truly missed. 
It is important to point out that none of these 4 missing galaxies have been excluded during the selection cuts phase.

The Hydra I cluster catalog (HCC) presented in \cite{misgeld2008early} consists of 111 galaxies, down to the magnitude limit of $M_V \simeq-10\;mag$.
The catalog covers an \emph{L}-shaped area, made of seven $7'\times 7'$ fields over the cluster core.
Their imaging data were obtained with VLT/FORS1, in Johnson V and I filters, with seeing conditions better than ours, averaging between $0.5$ and $0.7\;arcsec$.
Matching our sample with the \cite{misgeld2008early} catalog, and considering our magnitude range ($-18.5\:mag \leq\;M_r\;\leq\;-11.5\;mag$), we miss 30 galaxies. 
In Tab. \ref{tab:comparison} we show their median photometric parameters.
These systems are quite faint and small. 
\begin{table}[ht]
    \begin{center}
        \begin{tabular}{cc}
        \hline \hline
         Parameter & Median \\
         \hline
         $m_V$ & $20.72\;mag$ \\
         $V-I$ & $1.05\;mag$ \\
         $R_e$ & $2.93\;arcsec$ \\
         $\mu_e$ & $24.90\;mag/arcsec^2$ \\
         \hline
        \end{tabular}
    \end{center}
    \caption{Median values of the principal photometric parameters for the 30 missed objects from the \cite{misgeld2008early} Hydra I catalog. }
    \label{tab:comparison}
\end{table}
As shown in Fig. \ref{fig:selection_cuts}, our photometric selection cuts are not excluding any dwarf galaxy from \cite{misgeld2008early}. 
Rather, given their faint nature, we miss them during the detection phase. 
\cite{misgeld2008early} detection strategy is a combination of visual inspection of the images and the use of {\sc SExtractor} detection routines.

In conclusion, our catalog extends by almost a factor 3 the sample of known Hydra I galaxies within its $0.4$ virial radius, making it a more complete tool to exploit the photometric properties of dwarf galaxies in this environment.


\section{Results}\label{Results}

In this section we present the analysis carried out on the new photometric catalog of dwarf galaxies in the Hydra I cluster.
We exploit all the derived photometrical measurements, studying the spatial distribution of the galaxies and deriving several scaling relations, with the final goal of fully characterize the dwarf galaxy population over the central area of the Hydra I cluster.
In the following subsections we focus on the main properties derived for the dwarf galaxies in this sample. 
They represent $\sim 88\%$ of the galaxy population\footnote{The dwarf ratio is calculated as:
\begin{equation*}
    \frac{Number\;of\;dwarf\;candidates}{Number\;of\;dwarfs+Number\;of\;giants}=0.876
\end{equation*}
As number of giant galaxies we considered galaxies brighter than $M_r=-18.5\;mag$ which are spectroscopically confirmed Hydra I's galaxies, taking them from \cite{Christlein2003}.} 
in the studied area of the cluster.

\subsection{Spatial distribution of dwarfs and giants}

\begin{figure*}[t]
    \centering
    \includegraphics[width=.9\textwidth]{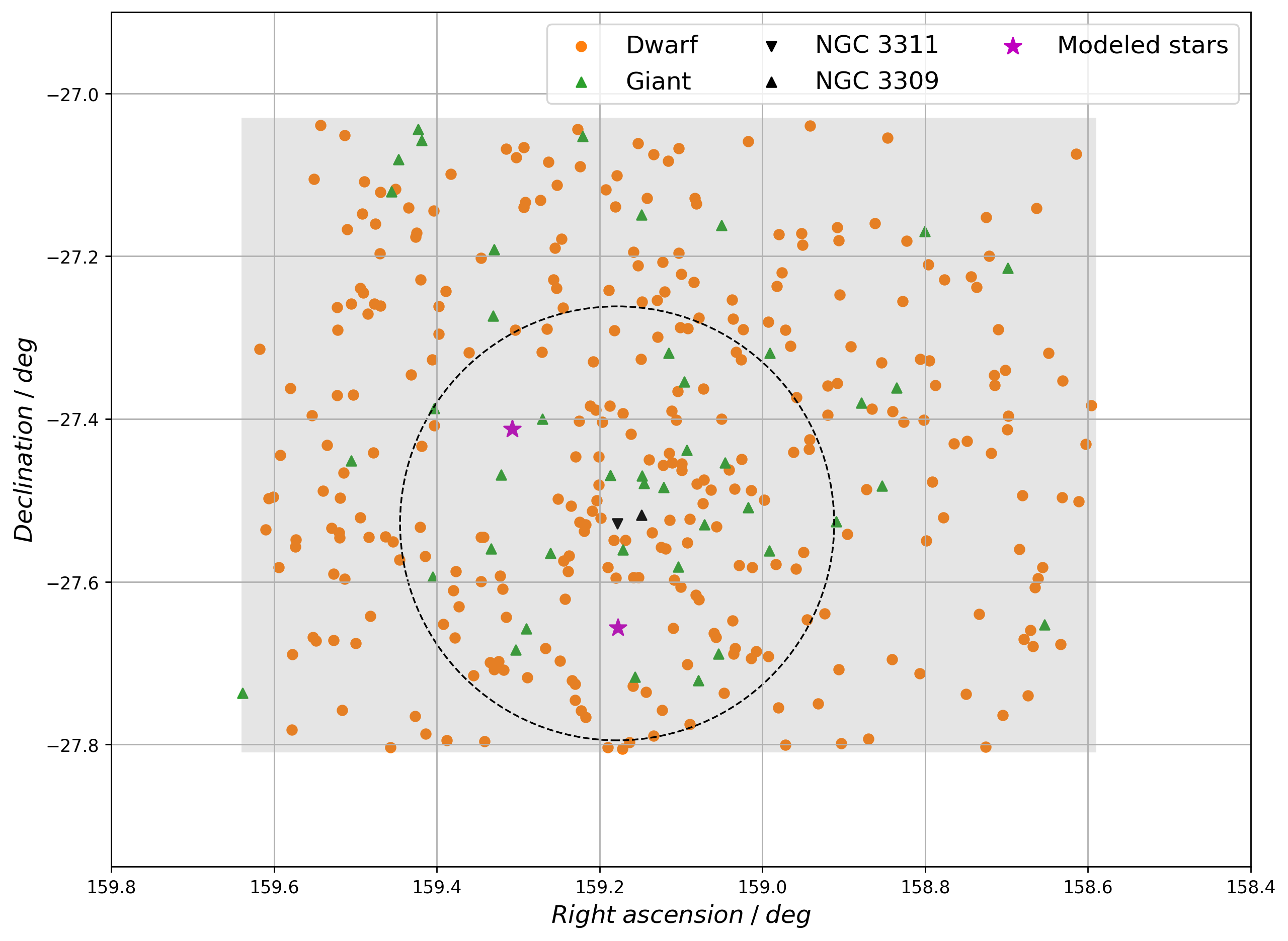}
    \caption{Spatial distribution of the likely Hydra I cluster dwarfs in our sample (orange circles), and giant galaxies from \cite{Christlein2003}, green triangles. The dividing magnitude is at $M_r=-18.5\;mag$.
    The black circle indicates the cluster core radius $r_c=170h^{-1}kpc$ \citep{Girardi1995}, assuming $h=0.72$, and the gray area indicates our studied field. The two BCGs, NGC~3311 \& NGC~3309 are marked separately as black triangles. The two magenta stars show the position of the two major saturated stars that we masked. In the figure north is up and east is left.}
    \label{fig:scattermap}
\end{figure*}

In Figure \ref{fig:scattermap} we show the locations of the dwarf galaxies in our sample. 
As expected, the galaxies are strongly concentrated around the center of the Hydra I cluster, dominated by the cD galaxy NGC~3311. 
The densest region is located within the cluster core \citep[black circle of radius $r_c=170\;h^{-1}kpc$][]{Girardi1995}.
Moreover, even considering the empty area due to the presence of the two heavily saturated stars, which have prevented the detection of some of the faint sources there, the galaxy distribution seems to be approximately symmetric around the two BCGs.

\begin{figure*}[ht]
    \includegraphics[width=0.49\textwidth]{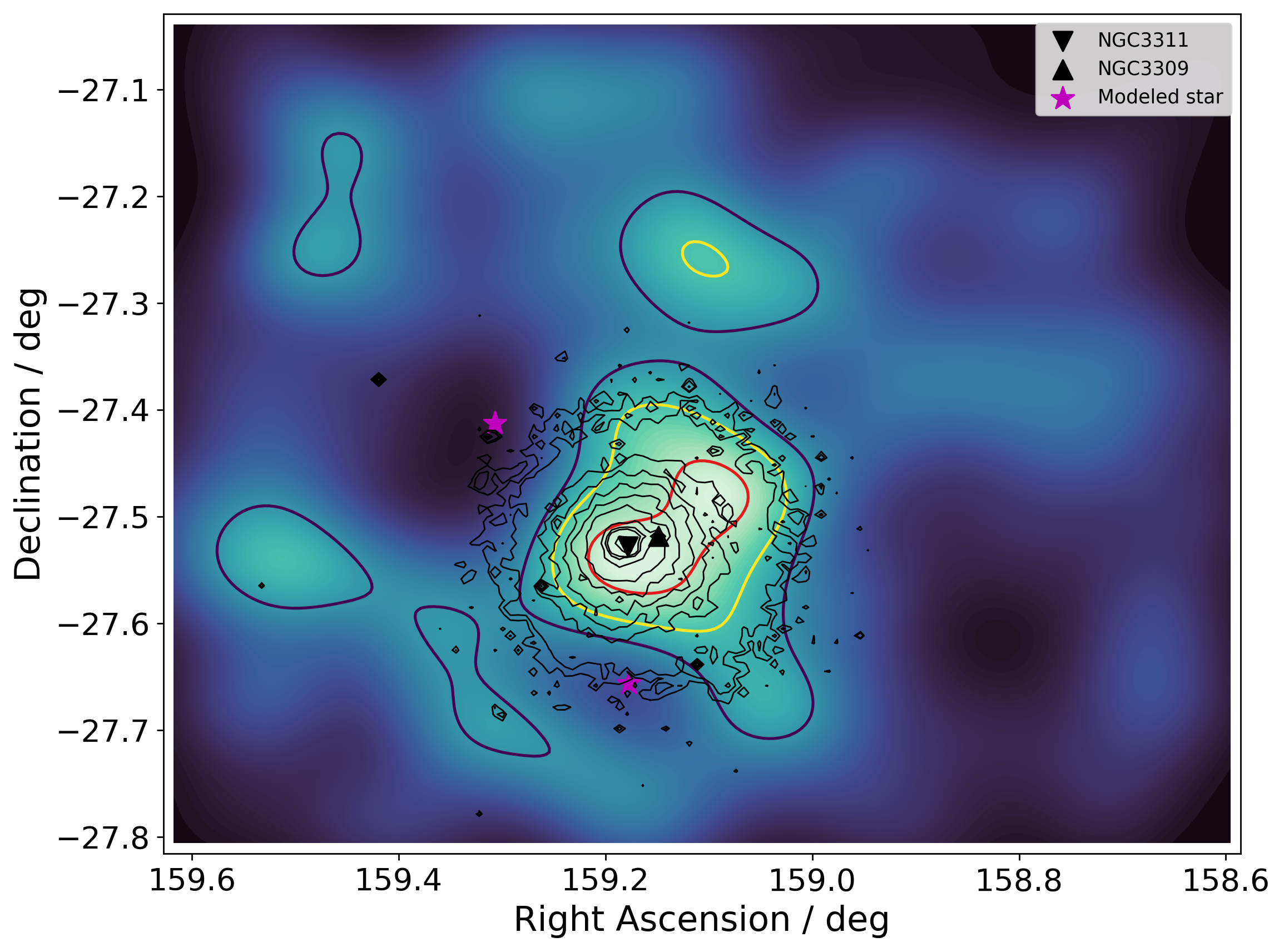}
    \includegraphics[width=0.49\textwidth]{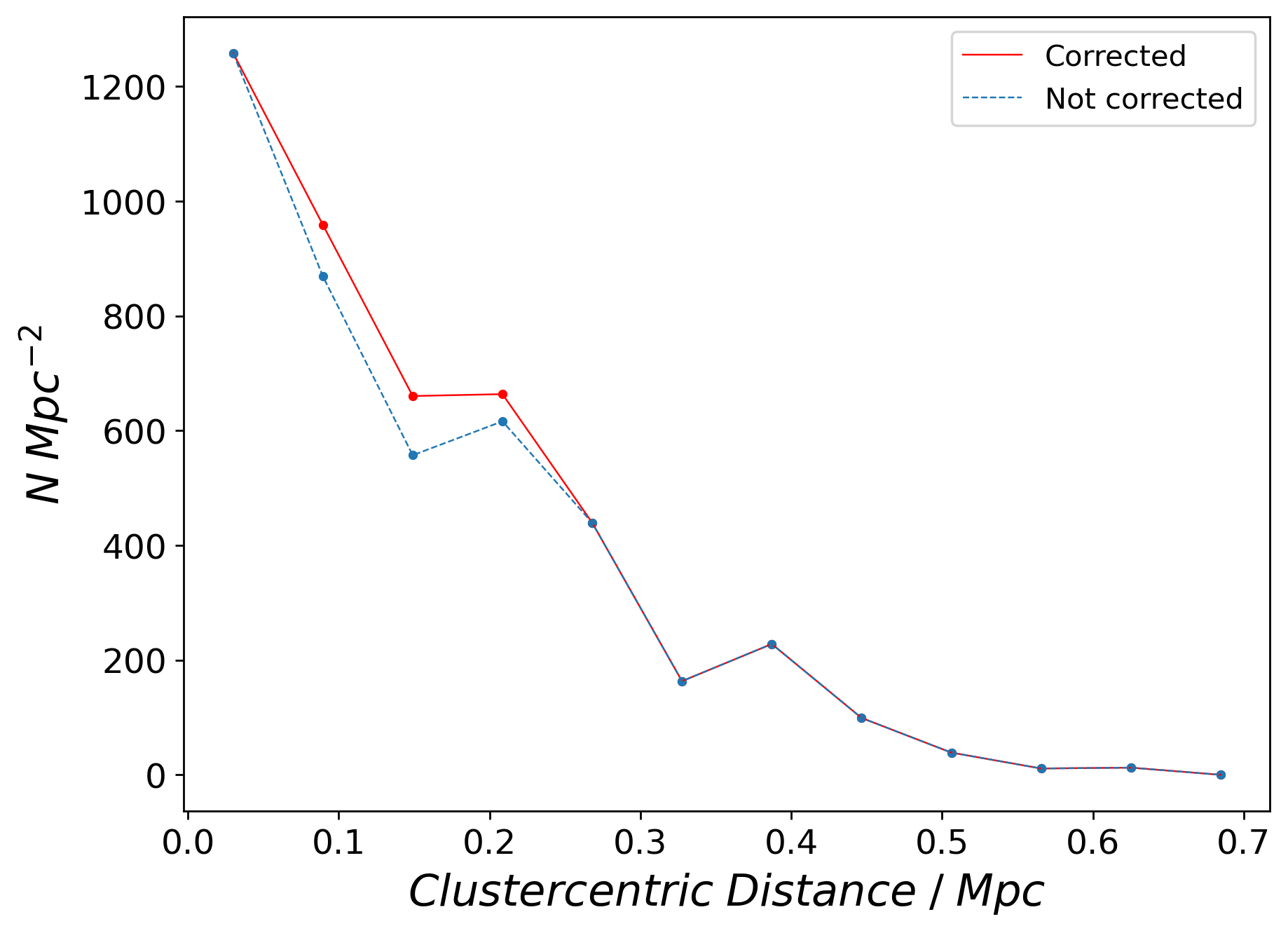}
    \caption{Projected galaxy density distribution inside the Hydra I cluster. {\it Left panel-} Smoothed number density distribution of the galaxies in the Hydra I galaxy cluster. The observed galaxy distribution is convolved with a Gaussian kernel with $\sigma=10\arcmin$.
    The colored contours highlight the iso-density curves (red, yellow and purple). 
    X-ray contours (thin black lines) are also overlaid to the map.
    The two giant galaxies at the center of the cluster are represented by the two black triangles, while the magenta stars are the two modeled stars in the field. North is up, east is left.
    {\it Right panel-} Projected galaxy surface density, as number of galaxies per square $Mpc$, plotted against the projected clustercentric distance. Red solid line is the density profile accounting for the area of the two saturated stars, while blue dashed line is the profile without correcting for that area.}
    \label{fig:smooth_map}
\end{figure*}

We derive a smoothed density map of the galaxies, by convolving the galaxy distribution with a Gaussian kernel with a standard deviation of $\sigma=10\arcmin$.
This is shown in Figure \ref{fig:smooth_map}, where also X-ray isophotes are plotted\footnote{X-ray data for Hydra I are taken from the XMM-Newton archive \url{http://nxsa.esac.esa.int/nxsa-web/#search}. The observation used are presented in \citet{hayakawa2006detailed}.} 
From the smoothed distribution we find a pronounced over-density in the northwest of the cluster center.
In this region, also the X-ray contours are a bit more broadened, just outside the core radius, and a structure on the southeast of the cluster center. 
The smoothed distribution also shows that overall, the western side of the cluster is less populated than the eastern side.

To better understand how the projected galaxy density changes going outward from the cluster center, we derive the radial clustercentric surface density for the dwarf galaxies in our sample. 
We count the number of galaxies in circular bins, with a step of $4\arcmin$ in radius ($\simeq60\;kpc$), and then divide that number by the bin area to get the surface density of the galaxies.
We assume a fixed distance for all the cluster members, equal to the distance of the cluster.
We plot the projected surface density as function of the clustercentric distance in the right panel of Figure \ref{fig:smooth_map}. 
The plot shows the number density both accounting for the incompleteness in the areas of the two brightest stars, and without correcting for it.
We point out that this is the projected surface density, and that we have considered a fix projected clustercentric distance. 
Therefore a realistic volume density of galaxies could give substantially different results.
As expected, on average, a decreasing galaxy density is found with increasing clustercentric distance.
The curve is also consistent with the 2D smoothed distribution of galaxies in Fig. \ref{fig:smooth_map},as we find some over-densities, instead of a monotonic decreasing profile.
A first overdensity at $d\sim0.2\;Mpc$ corresponds to the southeast group of galaxies, while the second peak at $\sim0.35\;Mpc$ traces the north group.

\begin{figure}[ht]
    \centering
    \includegraphics[width=0.49\textwidth]{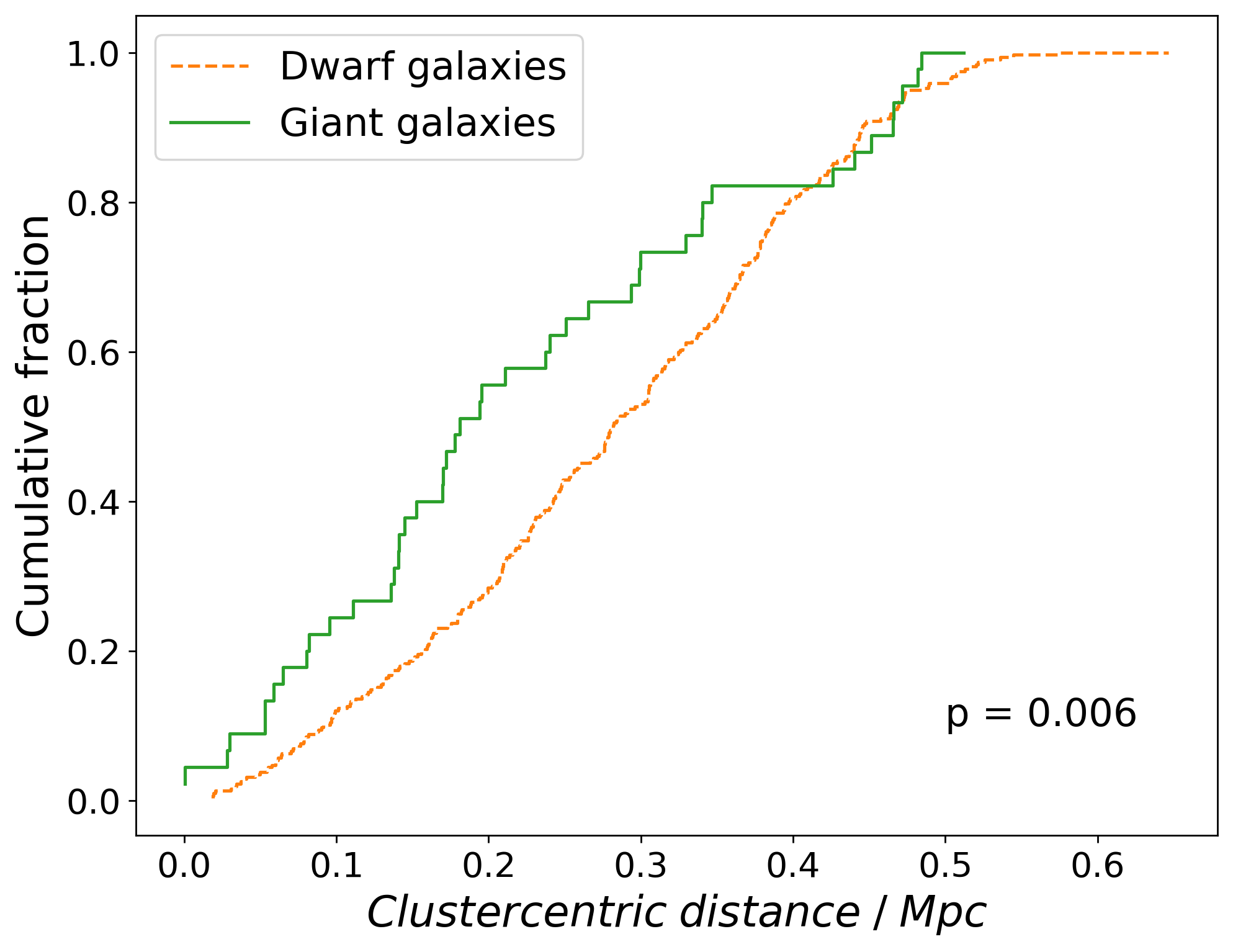}
    \caption{Cumulative functions of the clustercentric distances of giants \citep{Christlein2003}, green solid line, and of dwarfs, orange dashed line. The label refers to the p-value from the Kolmogorov-Smirnov test. The KS-test result states that the two radial distributions are statistically different.}
    \label{fig:cdist_cum}
\end{figure}

In addition, we compare the spatial distribution of giants and dwarfs, by studying their cumulative distribution as a function of their clustercentric distance, see Figure \ref{fig:cdist_cum}. 
We analyze the statistical differences between the two populations using a Kolmogorov-Smirnov (KS) test \citep{hodges1958significance}. 
The \textit{p-value} of the KS test gives the probability that giant and dwarf galaxies are taken from the same radial distribution, and this resulted to be $0.006$.
This means that giant and dwarf radial distributions are statistically different.
Specifically, Fig. \ref{fig:cdist_cum} highlights that giant galaxies are more concentrated toward the center of the cluster, with respect to the dwarfs. 

\subsection{Color-magnitude relation}\label{cmr_sec}

\begin{figure}[ht]
    \centering
    \includegraphics[width=0.49\textwidth]{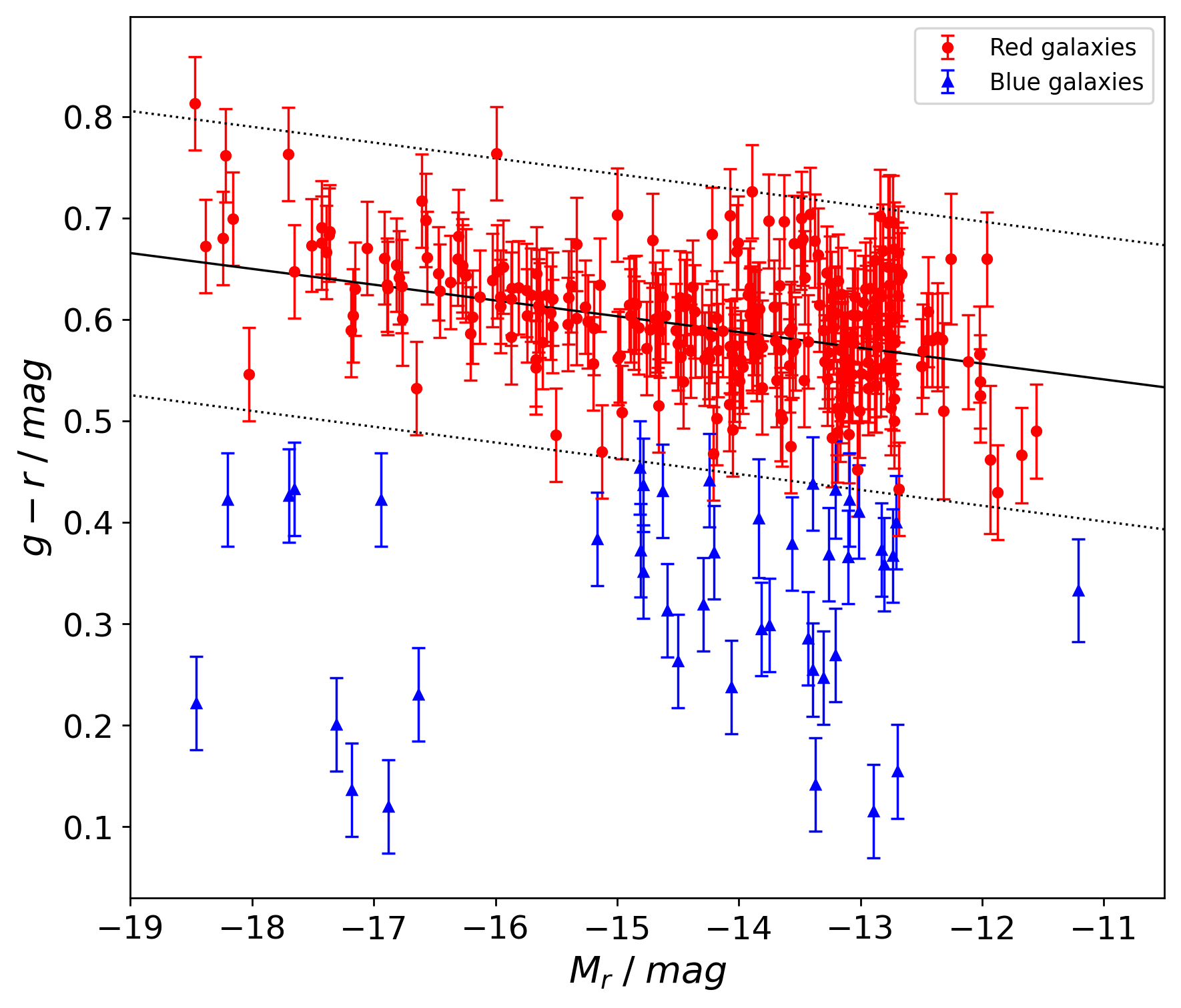}
    \caption{Color-magnitude relation for all Hydra I dwarf galaxies in our catalog. 
    The solid line is the linear fit of the red sequence (Eq. \ref{cmr}), while the dotted lines are the $2\sigma$ deviations from the fit.
    The two galaxy populations: blue triangles form the blue dwarf population, while red dots are the red dwarfs, with their color errors.}
    \label{fig:CMD}
\end{figure}

In Fig. \ref{fig:CMD} we show the color-magnitude relation (CMR) of the dwarf galaxies in our sample, from  $M_r=-18.5\;mag$, down to $M_r\simeq-11.5\;mag$. 
A well-defined CMR exists for early-type dwarf galaxies in our sample brighter than $M_r\simeq-14\;mag$: on average the brighter the galaxy, the redder its color is.
Fainter than that a wide scatter emerges, and nothing clear can be said.
We carry out a linear fit to all dwarf data points that were visually classified as ETGs, namely the early-type with structure, and \emph{Smooth} early-type categories. 
We obtain:
\begin{equation}
    (g-r) = -0.016\cdot  M_r + 0.37
    \label{cmr}
\end{equation}
with a rms of $0.07$. 
The fitted red sequence (RS) is overlaid in the color-magnitude diagram (CMD, Fig. \ref{fig:CMD}).
Except for two galaxies, all dwarfs are not redder than $2\sigma$ from the fitted RS. 
As expected, the scatter becomes larger going to fainter magnitudes.

In order to separate late-type from early-type galaxies in a more reliable way than the visual classification, we use the fitted RS to split the dwarf sample into two subpopulations. 
We refer to the dwarfs which are at least $2\sigma=0.14\;mag$ bluer than the RS as blue dwarfs. 
The remaining objects are named as red dwarfs.
A visual representation of this splitting is given in Figure \ref{fig:CMD}, where we color-code the objects according to this classification scheme. 
The blue population consists of $43$ dwarf galaxies ($\simeq14\%$), while the red one of $274$ dwarfs. 
Blue dwarfs all have colors $g-r<0.5$~mag, and do not follow a clear trend as the ETGs do along the RS, rather they are spread in the bottom part of the CMD.
We also notice that there are no blue dwarf galaxies in the range between $M_r\sim-16.5\;mag$ and $M_r\sim-15\;mag$.
In the next sections we analyze other scaling relations between photometric parameters, considering separately the two different dwarf populations. 

\subsection{Scaling relations for structural parameters}

\begin{figure*}[ht]
    \centering
    \includegraphics[width=0.9\textwidth]{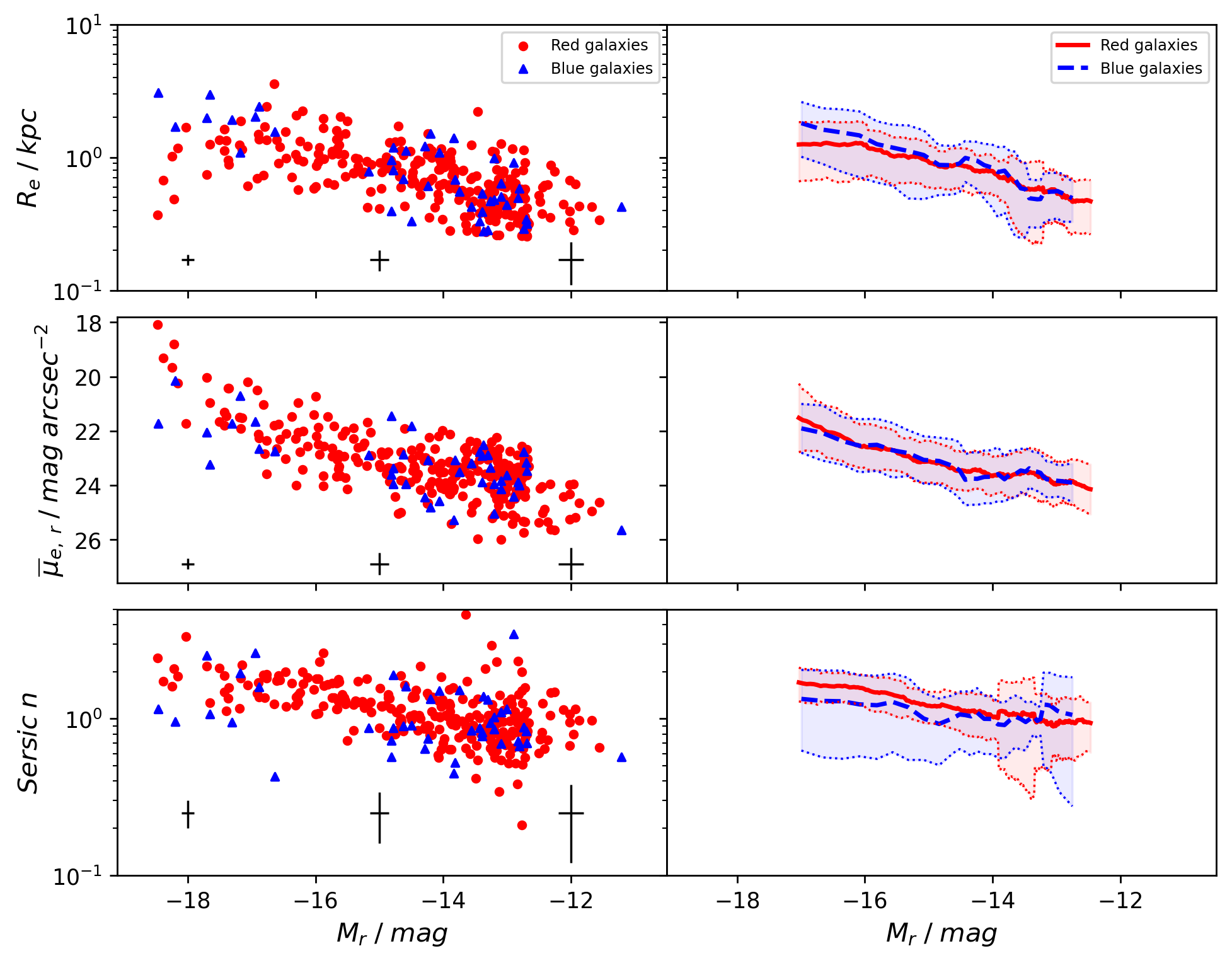}
    \caption{From the top to the bottom row: effective radius $R_e$, mean effective surface brightness in \textit{r}-band $\overline{\mu}_{e,r}$, and Sérsic index $n$ of our dwarf sample, as a function of total $r$-band absolute magnitude $M_r$. The left panels show the individual galaxies, while the right panels show the running means, for both populations.
    We use a fixed number of elements in each bin to get the local mean, $N=40$ for red dwarf population, and $N=10$ for blue dwarfs. 
    Blue triangles and dashed lines correspond to the blue population of galaxies, while the red dots and solid line are the red dwarfs. The shaded areas correspond to two standard deviations of the points using the same running mean. Effective radius and Sérsic index panels are in logarithmic scale.
    Average error bars are shown for different magnitudes.}
    \label{fig:scaling_rel}
\end{figure*}

In Figure \ref{fig:scaling_rel} we plot the structural parameters ($R_e,\;\overline{\mu}_{e,r}, n$) measured for all the dwarf galaxies in our catalog, as function of their total $r$-band magnitude. 
Both blue and red galaxy parameters show correlations with $M_r$.
These are highlighted by the running mean trends, shown in the right panels.

Within the errors, in all diagrams the distribution of red and blue dwarfs are very similar, especially up to $M_r=-16\;mag$.
At brightest magnitudes, where also errors are smaller, there are more appreciable differences, although they are statistically relevant only for the effective radii. 
Indeed, the red dwarf galaxies show a flattening in the $R_e$ vs. $M_r$ relation toward the bright end ($M_r<-16\; mag$), while blue dwarfs do not.
Both size and surface brightness clearly depend on total magnitude, and we see an almost linear correlation. 
Sérsic index seems to depend too, but given the larger errors it is less clear to confirm. 
This parameter describes the shape of the galaxy's radial light profile, where $n=1$ represents an exponential profile, whereas larger values indicates more concentrated profiles, and values smaller than 1 represent a flatter profile.
The red dwarfs show, on average, a mild increase in the Sérsic index from $n\lesssim1$ to $n\simeq2$, in the range from $M_r\simeq-14\;mag$ to $\;M_r\simeq-18\;mag$, which means that the brighter the red galaxies become, the more centrally concentrated they are.
For blue dwarfs it is hard to tell if there is the same trend, given the wide scatter.
However, brightest blue dwarf galaxies have on average smaller $n$ than red dwarfs of comparable luminosity.
A similar $log(n)$ vs. $M_r$ trend has also been found in other clusters \citep[for the Centaurus and Fornax clusters, respectively]{Misgeld2009cent,venhola2019optical}.
\cite{carlsten2021elves} also found a correlation between the Sérsic index and the stellar mass of early-type dwarfs, analyzing dwarf satellites
around Milky Way like hosts within Local Volume loose groups ($D<12\;Mpc$).
A more detailed comparison is reported in Section \ref{discussion-rel}.

\subsection{Axis ratio}
 
\begin{figure}[ht]
    \centering
    \includegraphics[width=0.49\textwidth]{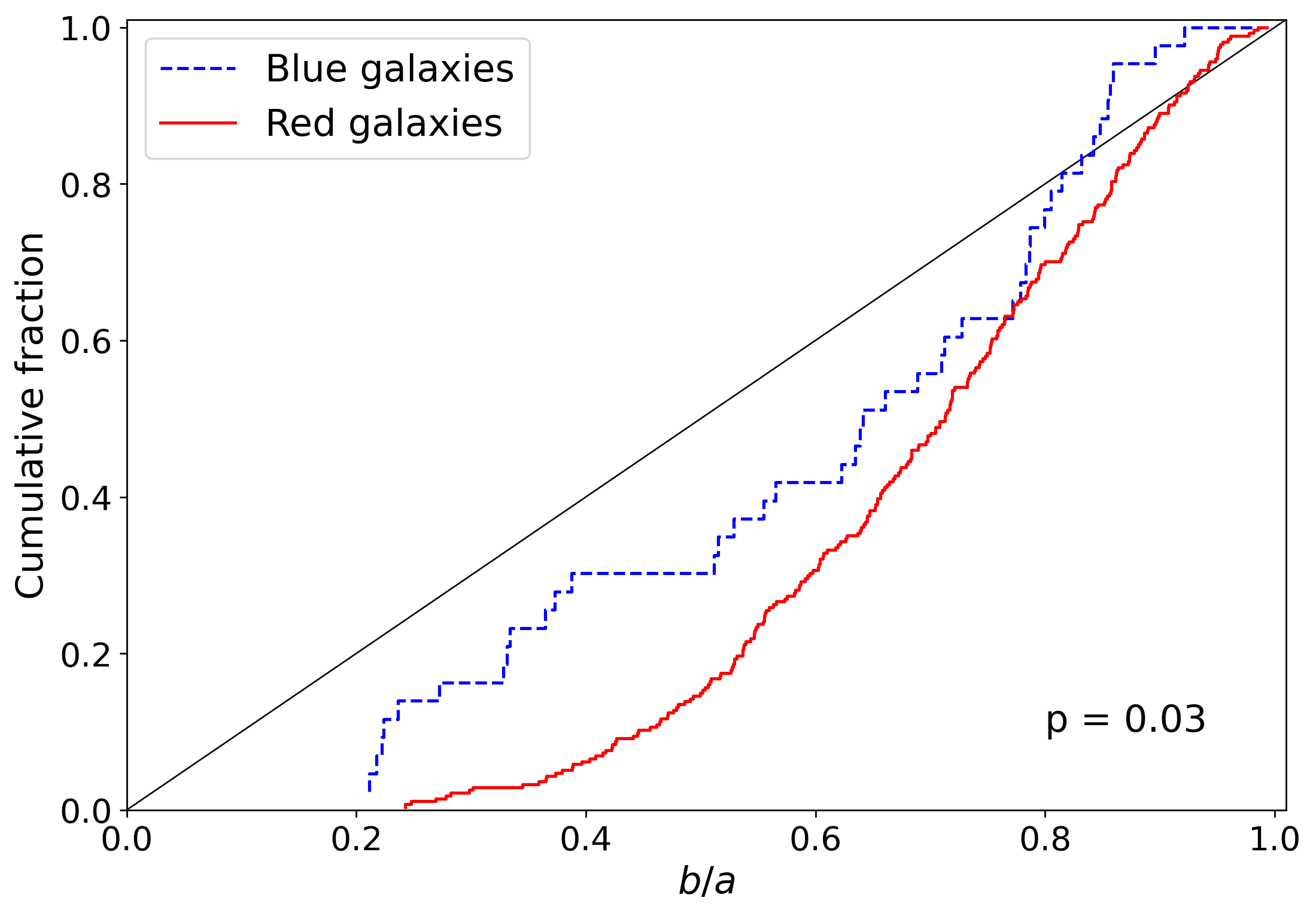}
    \includegraphics[width=0.49\textwidth]{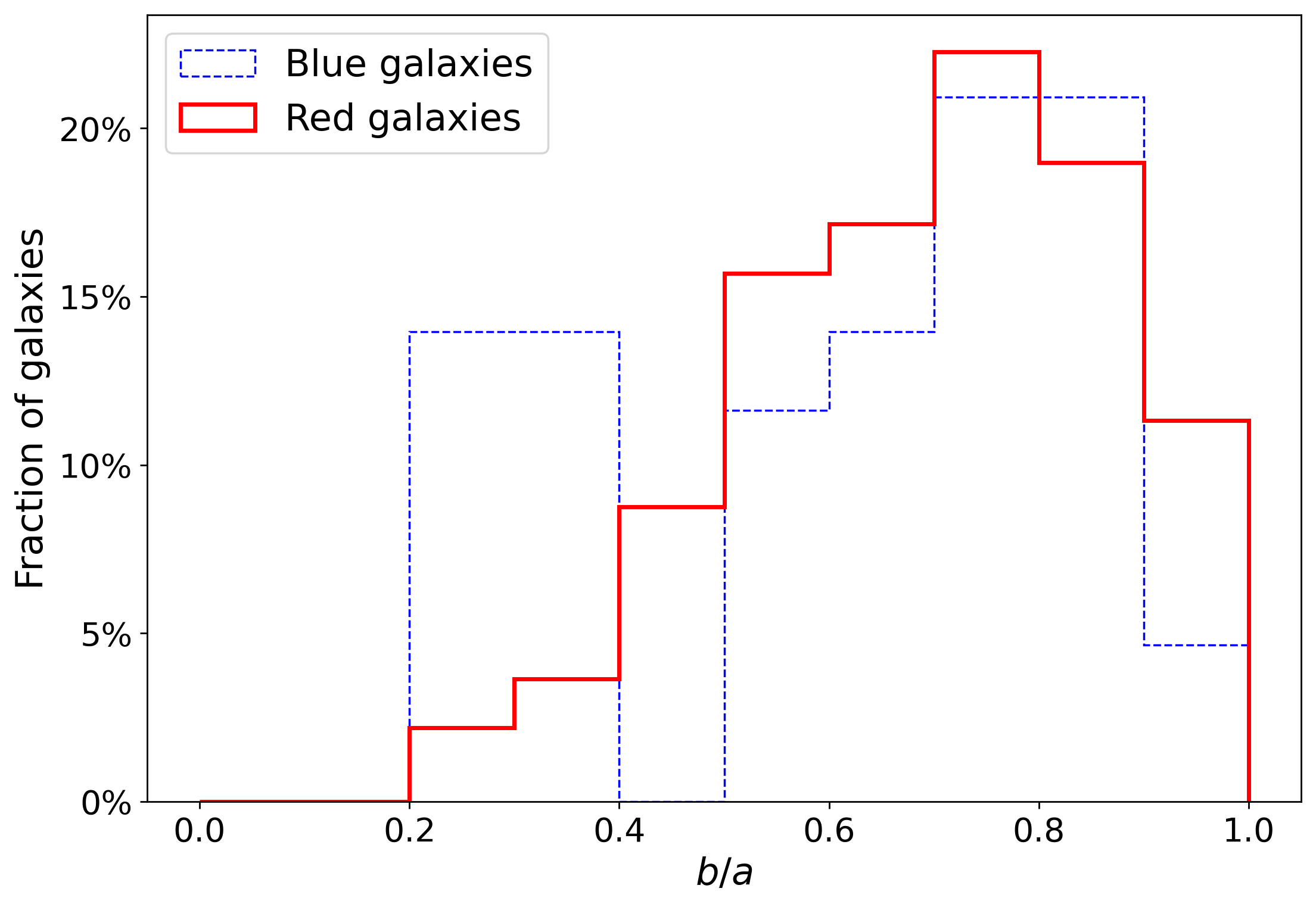}
    \caption{Axis ratio distributions for the red and blue galaxies, represented by the red solid line, and by the blue dashed line, respectively.
    \emph{Top panel-} Cumulative distributions of the two populations. The black diagonal line shows a flat $b/a$ distribution. The reported value is the p-value of the performed KS test.
    \emph{Bottom panel-} Fraction of red and blue dwarfs in $b/a=0.1$ wide bins.
    }
    \label{fig:AR}
\end{figure}

In Figure \ref{fig:AR} (top panel) we show the cumulative distribution of the axis ratio, $b/a$, for both dwarf populations.
The blue distribution grows steeper than the red population for low $b/a$ values, then flattens at $b/a\sim0.4$, and then continues to grow toward unity, reaching it faster than the red galaxies distribution. 
We compare blue and red dwarf axis ratios performing a KS test between the two cumulative distributions. 
The p-value is equal to $0.03$, indicating that the two distributions are not taken from the same parent distribution (see subsection 6.1). 

Bottom panel of Fig. \ref{fig:AR} shows the histogram distributions of the two populations. 
Both red and blue distributions peak at about $b/a=0.8$.
A considerable fraction of blue dwarfs ($\sim30\%$) have axis ratio between 0.2 and 0.4, while there are few red galaxies with axis ratio lower than 0.4.
This is qualitatively consistent with the idea that the early-type galaxies are rounder compared to the late-types, which show more disky features, and therefore prefer lower axis-ratios.
However, one has to take into account the fact that we observe a 2D projection of the intrinsic 3D shape distribution of the galaxies, but statistically the 2D observed distribution should reflect a behavior that is linked to the intrinsic 3D shape.

The $b/a$ distribution of dwarf galaxies in the Fornax cluster \citep{venhola2019optical}, also considering the separation between late-types and early-types, has the same trend found for the dwarf galaxies in our Hydra I sample. 
The distribution of non-nucleated dEs in Fornax is very close to the axis-ratio distribution of the red galaxies in Hydra I;
indeed the respective cumulative distributions grow with the same rate.
Fornax LTGs have qualitatively the same trend of Hydra I blue dwarfs, with a second peak in the distribution between 0.2 and 0.4.

\subsection{Clustercentric properties}\label{clust-centric-sec}

\begin{figure*}
    \centering
    \includegraphics[width=0.9\textwidth]{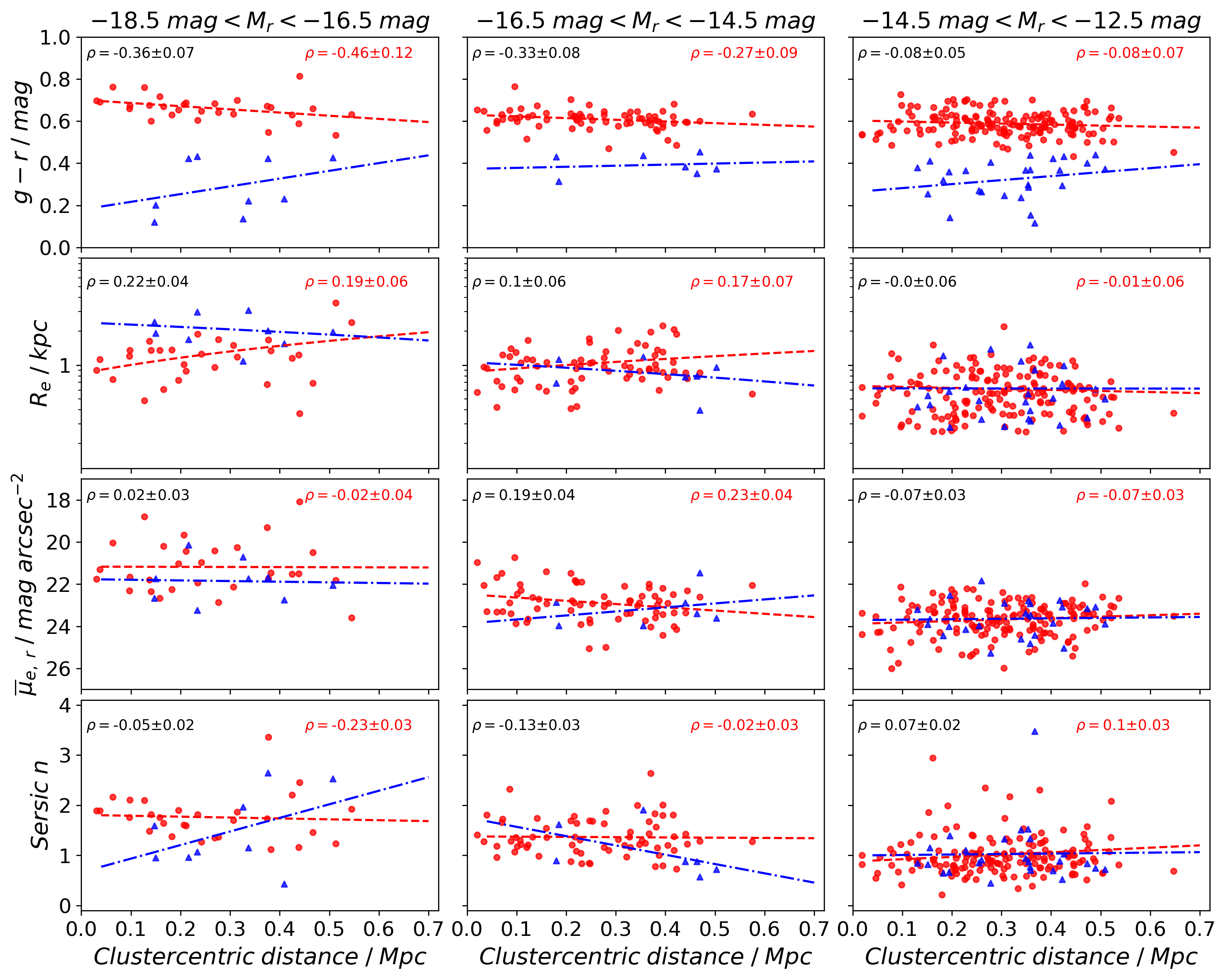}
    \caption{Structural parameters of Hydra's dwarf galaxies as a function of their projected clustercentric distance. Panel rows from top to bottom show: $g-r$ color, effective radius $R_e$, mean effective surface brightness $\overline{\mu}_{e,r}$, and the Sérsic index $n$. 
    Each column corresponds to a different luminosity bin. 
    Linear trend lines are shown in each panel, for both dwarf populations.
    Red dots and dashed lines are used for red dwarfs, while blue triangles and dash-dotted lines represent the blue dwarf population. 
    The $\rho$ parameter in each panel expresses Spearman's rank correlation coefficient for the correlation between the parameter and the distance from the cluster center (black indicates the whole dwarf sample, red only the red dwarfs).}
    \label{fig:cdist-prop}
\end{figure*}

\begin{figure}
    \centering
    \includegraphics[width=0.45\textwidth]{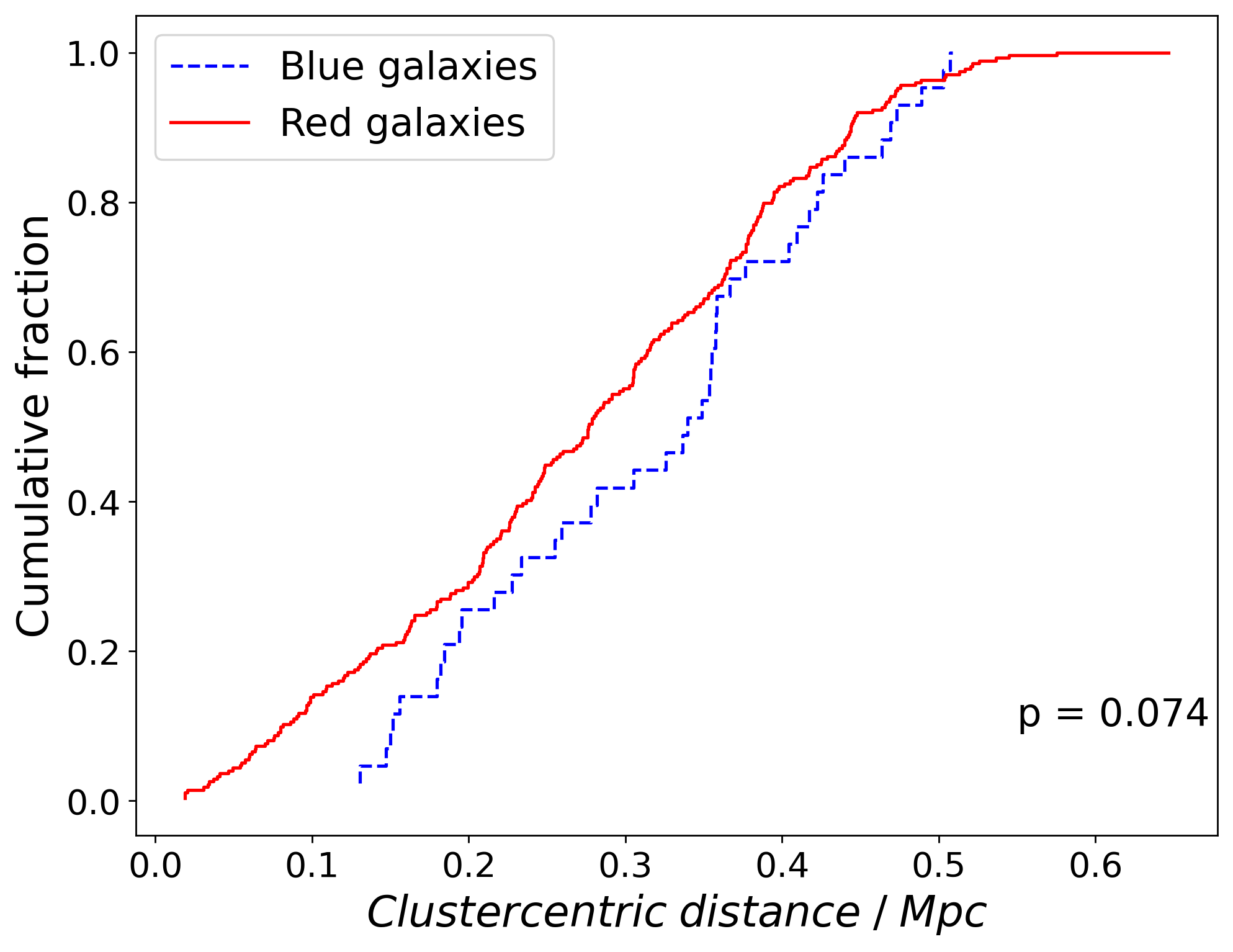}
    \caption{Cumulative clustercentric distributions for both red and blue dwarfs, represented by a red solid line and a blue dashed line, respectively. The number on the bottom is the p-value of the performed KS test.}
    \label{fig:cdist_cumul}
\end{figure}

In order to understand how the environment affects the dwarf galaxies, we analyze how their properties vary depending on their location in the Hydra I cluster.
If the role played by the environment is important, we expect to see galaxies with similar luminosity having different properties according to their projected clustercentric distance.
However it is important to remind that distances are projected, and hence galaxies might be located further out.
We binned the dwarf sample into $2\;mag$ wide magnitude bins.
The faintest bin (that is $M_r>-12.5\;mag$) covers a smaller magnitude range, due to our detection limit. 
Therefore, to make a statistical robust analysis, and a fair comparison with literature, we focus on the first $3$ bins.
Fig. \ref{fig:cdist-prop} shows the structural parameters as a function of the clustercentric distance for both galaxy populations, divided into the three magnitude bins.
To quantitatively test the correlations between the different parameters and the clustercentric distance we use the Spearman's rank correlation test
\footnote{The Spearman's correlation test assesses how well the relationship between two variables can be described using a monotonic function. A perfect correlation is expressed by $+1$ and $-1$. Intuitively, negative values indicate decreasing monotonic correlation, while positive values indicate increasing monotonic correlation \citep{kokoska2000crc}. Generally correlation above $0.4$ are considered relatively strong; between $0.2$ and $0.4$ are moderate, and below $0.2$ are consistent with no-correlation at all.}
, showing the correlation coefficients $\rho$ in each panel of Fig. \ref{fig:cdist-prop}.
The correlation coefficients are obtained considering both the whole dwarf population ($blue\;+\;red$), and only the red dwarfs.
We find that some parameters behave similarly with respect to the clustercentric distance in all luminosity bins, while some others have different trends for the high luminosity galaxies.
More in detail:

\renewcommand{\labelitemi}{\textendash}
\begin{itemize}
    \item Red dwarfs become redder moving inward in the bright $(-18.5,-16.5\;mag)$ bin, and show the same trend also in the mid-luminosity bin $(-16.5,-14.5\;mag)$, but with a weaker correlation. 
    For red dwarfs in the faintest bin, a rather flat trend is observed.
    Blue dwarfs show the opposite trend, which means they become bluer at larger distances. 
    The trend is much weaker in the intermediate luminosity bin. 
    
    \item On average, the effective radius increases with increasing clustercentric distance for red dwarfs in high- and mid-luminosity bins. 
    However, the correlation is weak, and galaxies show a wide scatter.
    By contrast, both blue and red dwarfs in the range $(-14.5,-12.5\;mag)$ show no correlation with distance, and an even wider scatter.
    
    \item A rather flat trend is observed for the mean effective surface brightness $\overline{\mu}_{e,r}$ in each luminosity bin. 
    Only red dwarfs in the mid-magnitude bin exhibit a weak negative correlation, i.e. $\overline{\mu}_{e,r}$ becomes fainter going outward from the center.
    The opposite weak trend is reported for blue dwarfs in the same bin.
    
    \item Dwarfs in each magnitude bin do not show any clear trend between the Sérsic index and clustercentric distance. 
    Only bright red dwarfs ($M_r<-16.5\;mag$) have on average slightly increasing $n$ toward the center, with a mild correlation. 
    
\end{itemize}
\renewcommand{\labelitemi}{\textbullet}
In addition, we study how the two dwarf populations are spatially distributed, producing a clustercentric cumulative distribution, which is shown in Fig. \ref{fig:cdist_cumul}. 
We perform a two-sided KS test to assess the validity of the null hypothesis, namely whether the two samples are drawn from the same distribution. 
The corresponding p-value is $0.074$, or $7.4\%$.
Interestingly, we do not find blue dwarf galaxies, according to our definition, closer than $\sim0.15\;deg$ to the cluster center.
This is in agreement with the fact that galaxies located at small clustercentric distances have on average entered the cluster earlier, and have already been quenched, explaining the lack of blue dwarfs in the very central region.
The red dwarf galaxy distribution grows regularly and approaches unity smoothly. 

\section{Discussion}\label{Discussion}

In this paper we present a new catalog of dwarf galaxies in the Hydra I cluster, that more than doubles the previously known samples, based on VEGAS deep imaging data that cover the cluster out to $0.4 R_{vir}$.
In the following sections we discuss the two main outcomes of this work: the 2D projected distribution of the galaxies in the cluster and the clustercentric properties of the dwarf galaxy population.
Moreover, we extend the comparison of Hydra's dwarf galaxies properties with the dwarf populations inhabiting other clusters.

\subsection{Dwarfs versus Giants}

\begin{figure}
    \centering
    \includegraphics[width=0.45\textwidth]{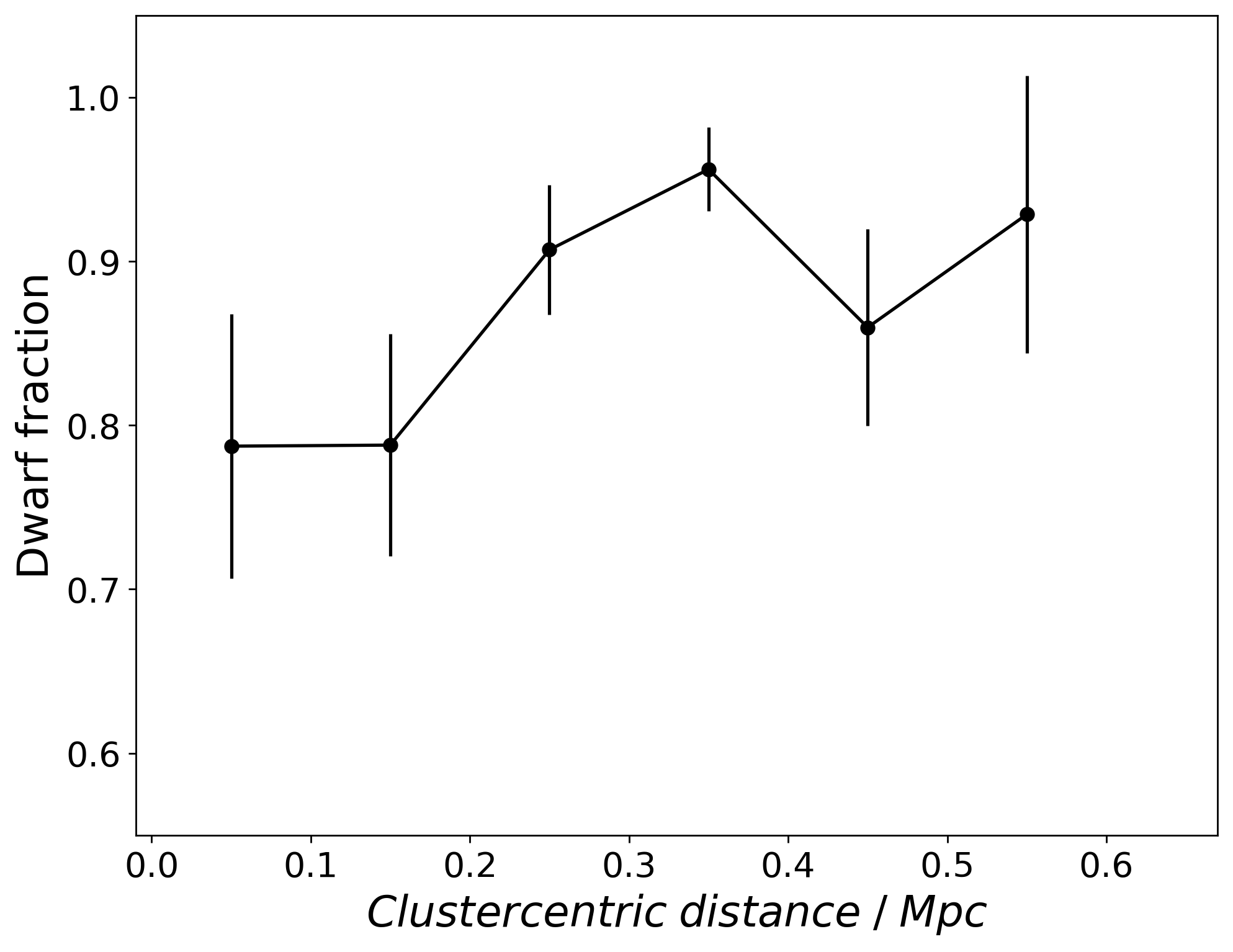}
    \caption{Dwarf fraction as a function of the projected clustercentric distance. 
    }
    \label{fig:dw_frac-dist}
\end{figure}

We find that the dwarf fraction in the central region of the Hydra I cluster is $\sim0.88$, according to the definition $M_r>-18.5\;mag$, and to our faint magnitude  limit.
In Fig. \ref{fig:cdist_cum} we compare dwarf and giant radial distributions for the Hydra I cluster. 
The Kolmogorov-Smirnov test gives as result that the two populations have statistically different radial distributions, with $p-value=0.006$.
Moreover, Fig. \ref{fig:cdist_cum} shows that bright galaxies are distributed closer to the center than the dwarfs.
We derived the dwarf fraction in bins of different projected clustercentric distances, with a step of $0.1\;Mpc$. 
Results are displayed in Fig. \ref{fig:dw_frac-dist}.
The dwarf fraction increases toward the outer part of Hydra I, so we can deduce that the cluster inner region seems to be more hostile to dwarfs, than the cluster outskirts.
This can be explained by tidal disruption of the dwarfs at smaller distances \citep[and references therein]{popesso2006}.

\citet{choque-challapa2021} analyze the dwarf fraction in 10 galaxy clusters ($0.015\lesssim z \lesssim 0.03$) from the KIWICS survey (PIs R. Peletier and A. Aguerri). 
They consider as dwarf galaxies the individuals in the range $-19.0<M_r<-15.5\;mag$.
The authors find dwarf fractions in all clusters $\gtrsim 0.7$.
Moreover, they calculate the dwarf fraction in the same way for the well-studied Fornax and Virgo clusters, using the catalogs from \cite{venhola2018} and \cite{kim2014extended}, finding it to be $\sim 0.7$ for both.
If we adopt the same definition as them, complementing again our catalog with the one by \cite{Christlein2003}, we infer a dwarf fraction of $0.71$, thus in prefect agreement with the fraction of the other nearby clusters, and consistent with Fornax and Virgo's fractions.
Table \ref{tab:dwarf_fraction} summarizes the comparison.

\begin{table}[ht]
    \caption{Dwarf fractions in different clusters. }
    \label{tab:dwarf_fraction} 
    \centering
    \begin{tabular}{cc}
    \hline \hline
        Cluster & Dwarf fraction \\
        \hline
         Hydra I & 0.71 \\
         Fornax & 0.70 \\
         Virgo & 0.72 \\
         Average KIWICS clusters & 0.84 \\
        \hline
    \end{tabular}

\textbf{Notes} - The dwarf fraction is calculated as the number of dwarf galaxies in the range $-19.0<M_r<-15.5\;mag$, divided by the total number of galaxies brighter than $M_r=-15.5\;mag$.

\end{table}

\citet{choque-challapa2021} find that in 5 out of 10 studied clusters the radial distributions of dwarf and giant are different.
These cases correspond to the more massive clusters in their sample, and in these cases the giant galaxies are centrally more concentrated, a similar result to what we observe for Hydra I.

\subsection{Colors of Hydra's dwarf galaxies}

In this section we aim at comparing the dwarf galaxy colors we computed for the Hydra I cluster, with the analog population in other local clusters, that is the Fornax and the Virgo clusters.
The most complete dwarf galaxy sample to date in the Virgo cluster core, in terms of luminosity, is that of the NGVS \citep[][]{Ferrarese2012}, presented in \cite{ferrarese2020next}.
For the Fornax cluster, we take as reference sample the dwarf catalog produced by the Fornax Deep Survey \citep[][]{Peletier2020}, presented by \cite{venhola2018}.

\begin{figure}[ht]
    \centering
    \includegraphics[width=0.49\textwidth]{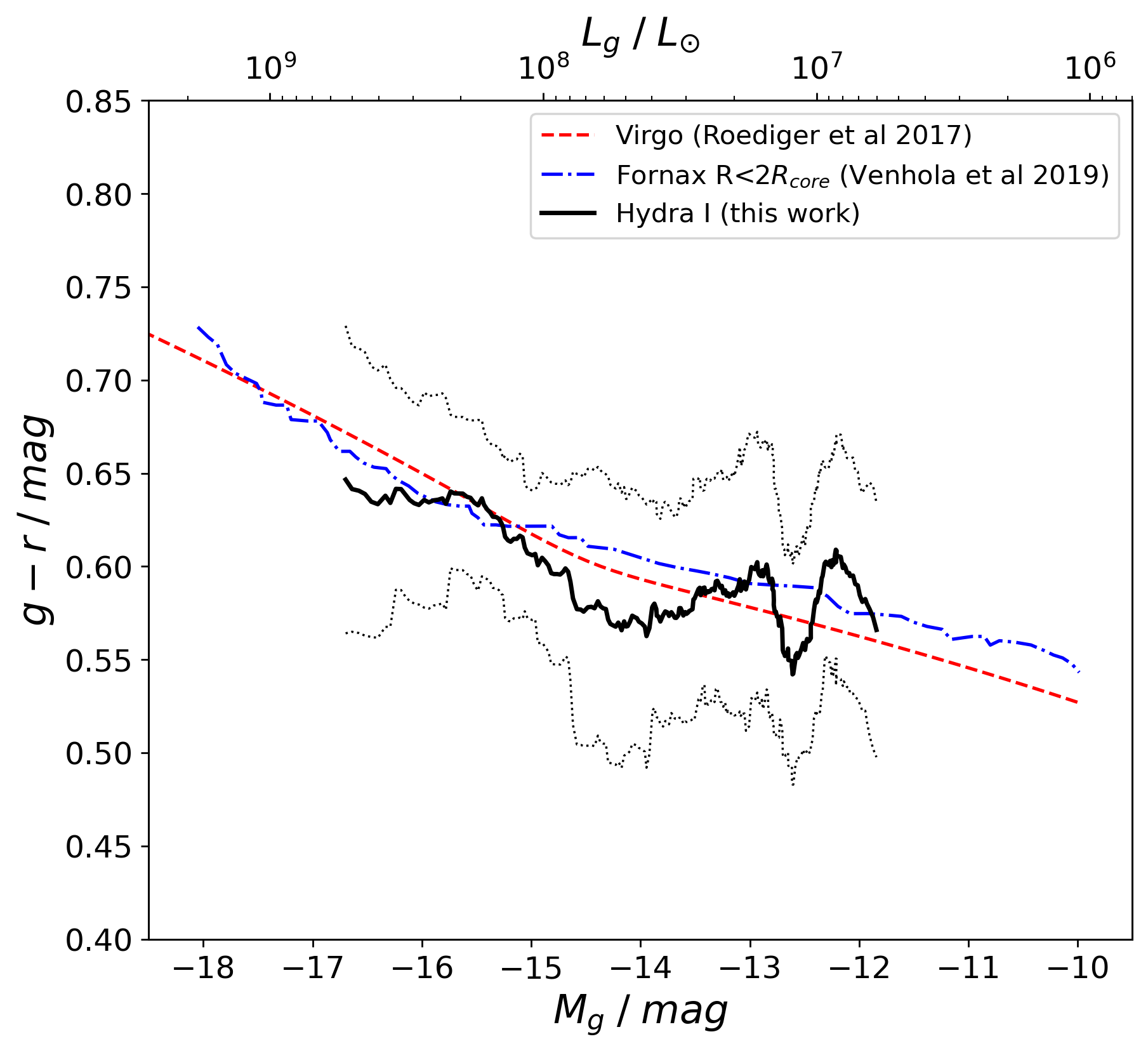}
    \caption{Comparison between Hydra I color-magnitude relation in this work and the red sequence for the Virgo and Fornax clusters. The black line represents the running mean of the CMR for the Hydra I's dwarfs in our sample, and the dotted lines show $1\sigma$ deviation (we use a fixed bin wide $N=30$ individuals). The red dashed line is the CMR in the core of the Virgo cluster \citep{roediger2017ngvs}. The blue dash-dotted line shows the Fornax CMR within $2R_{core}$ \citep{venhola2019optical}.}
    \label{fig:RScomp}
\end{figure}

In the color-magnitude diagram shown in Figure \ref{fig:RScomp}, we compare the CMR of dwarf galaxies in Hydra I, with the red sequences (RS) of the Virgo and Fornax cluster cores \citep[][respectively]{roediger2017ngvs,venhola2019optical}. 
We use the $g$-band absolute magnitudes on the x-axis instead of the $r$-band, which is the one used in the analysis of the Virgo cluster by \cite{roediger2017ngvs}.
For Hydra I's CMR, we use the running mean of the color-magnitude relation for the early-type dwarf galaxies in our sample, selected as done for the $r$-band in Section \ref{cmr_sec}. 
\cite{roediger2017ngvs} analyzed the RS of the Virgo cluster core galaxies, up to $\sim 10^6 L_{\odot}$ ($M_g\simeq-10\;mag$), finding the first evidence for a moderate flattening of the RS at the faint end, in different colors. 
Therefore, their RS has a shape which is better described by a double power law, with two different slopes, with a knee at $M_g\simeq-14\;mag$.
\cite{venhola2019optical} did not derive the RS analytically, but rather showed the running means of the CMR  within different radii from the center.
To make a fair comparison, we consider their $g-r$ vs. $M_g$ relation only for dwarfs within $2R_{core}$ \citep[i.e. $\sim0.5\;Mpc$,][]{Ferguson1989}.

The RS we present is, within the errors, in good agreement with the color-magnitude relations of both Fornax and Virgo clusters.
The Fornax cluster RS matches the change in slope toward the faint luminosity of the Virgo cluster RS, while the Hydra I slope is relatively noisy for faint magnitudes, but consistent with such a flattening.
Except for this point, we infer that the CMRs of the early-type galaxies are similar within the uncertainties in the Virgo, Fornax and Hydra I cluster cores.

LSB galaxies in the Coma cluster also show a range of color, and a color-magnitude relation, consistent with the CMR of Virgo, Fornax and Hydra I \citep{Alabi2020coma}.
\cite{bower1992precisioncmr} evidenced that there are no intrinsic differences between the CMRs of galaxies in Virgo and Coma. 
Detailed CMR analysis of three local clusters (namely A85, A496, A754) also confirm close agreement with that of the Coma cluster, and thus with Virgo, Fornax and Hydra I, \citep{Mcintosh2005}.
More in general, other observations suggested that such a red sequence exists in clusters over a wide redshift range \citep[for instance $0.2\leq z \leq 1.1$,][]{Bell2004cmr}, and our result provides further proof of its universality.

Comparing the cumulative radial distributions for Hydra I red and blue dwarfs (see Fig. \ref{fig:cdist_cumul}), we can rule out with a good confidence degree the hypothesis that they are statistically similar, given the KS test $p-value=0.074$.
However, we note that red dwarf galaxies are located closer to the cluster center, with no blue dwarf galaxies within $0.15\;Mpc$ projected distance. 
For what concerns the blue dwarfs fraction, we observed it to be $\sim0.14$. 
The fraction in Fornax, estimated in a similar way is 0.15, hence in prefect agreement with that of Hydra I.
In the Virgo cluster, the fraction of blue dwarfs is instead larger, $\sim0.3$ \citep[]{venhola2019optical,kim2014extended,choque-challapa2021}.
The similar dwarf fraction in Fornax and Hydra I clusters is intriguing, considering that Hydra I is more than twice as massive virial mass, and has a velocity dispersion almost two times higher \citep[see][and therein references for a summary of Fornax properties]{venhola2018}.

\subsection{Galaxy scaling relations in Local galaxy clusters}\label{discussion-rel}

We show in Fig. \ref{fig:scaling_rel} how dwarfs properties scale as a function of the total luminosity, finding that size, surface brightness and Sérsic index correlate with the dwarfs total magnitude.

\begin{figure*}
    \centering
    \includegraphics[width=0.49\textwidth]{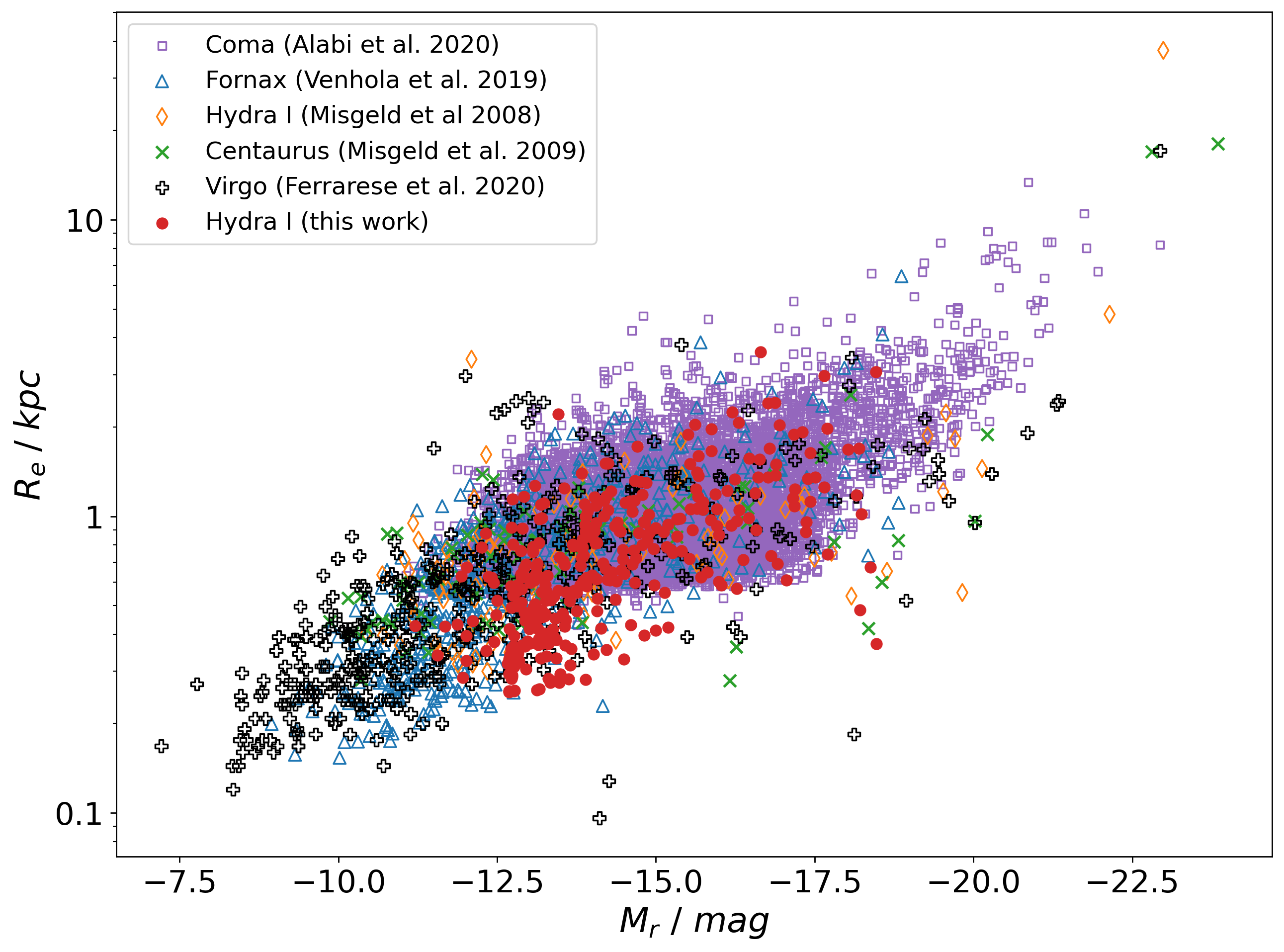}
    \includegraphics[width=0.49\textwidth]{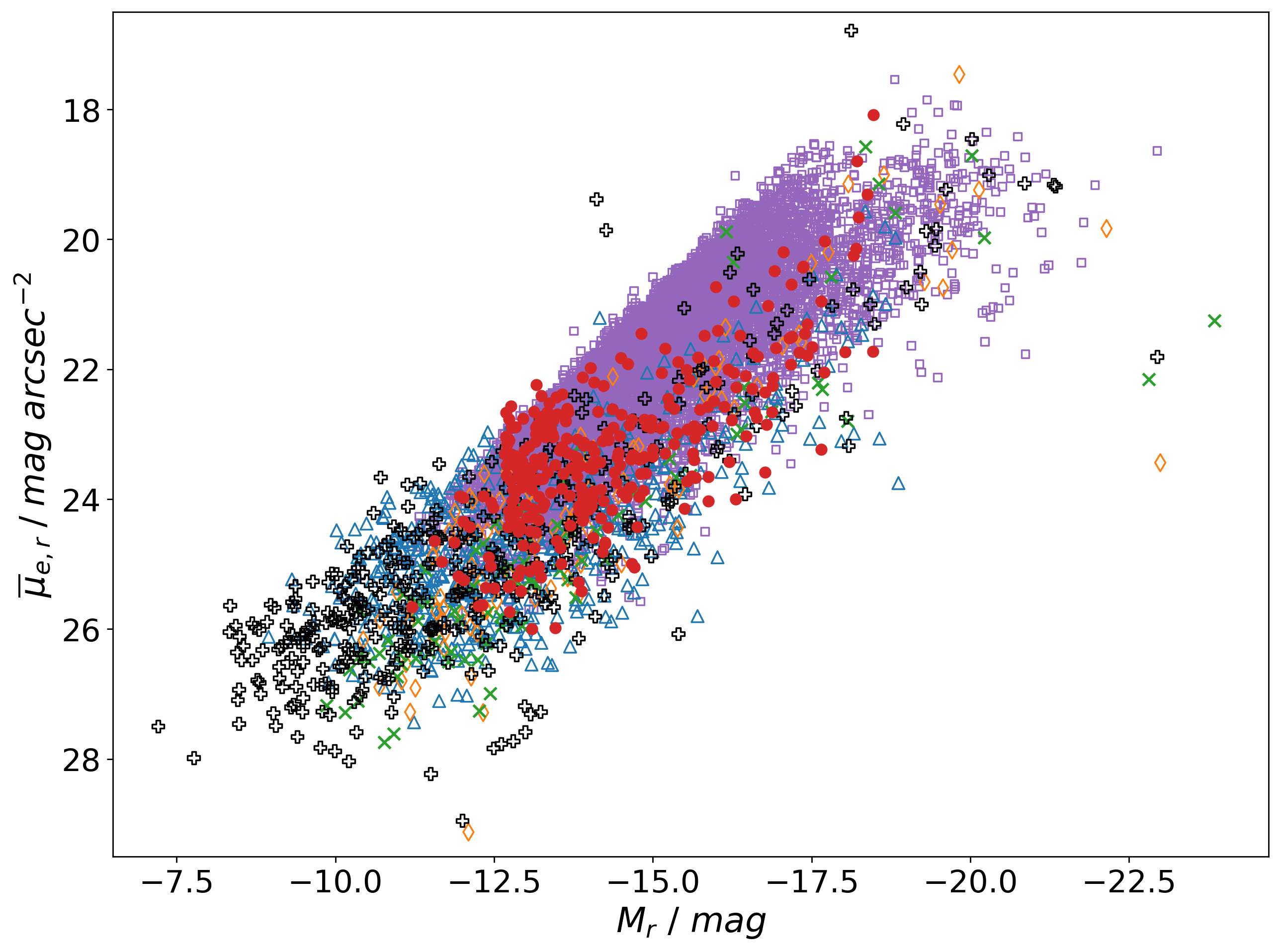}
    \includegraphics[width=0.49\textwidth]{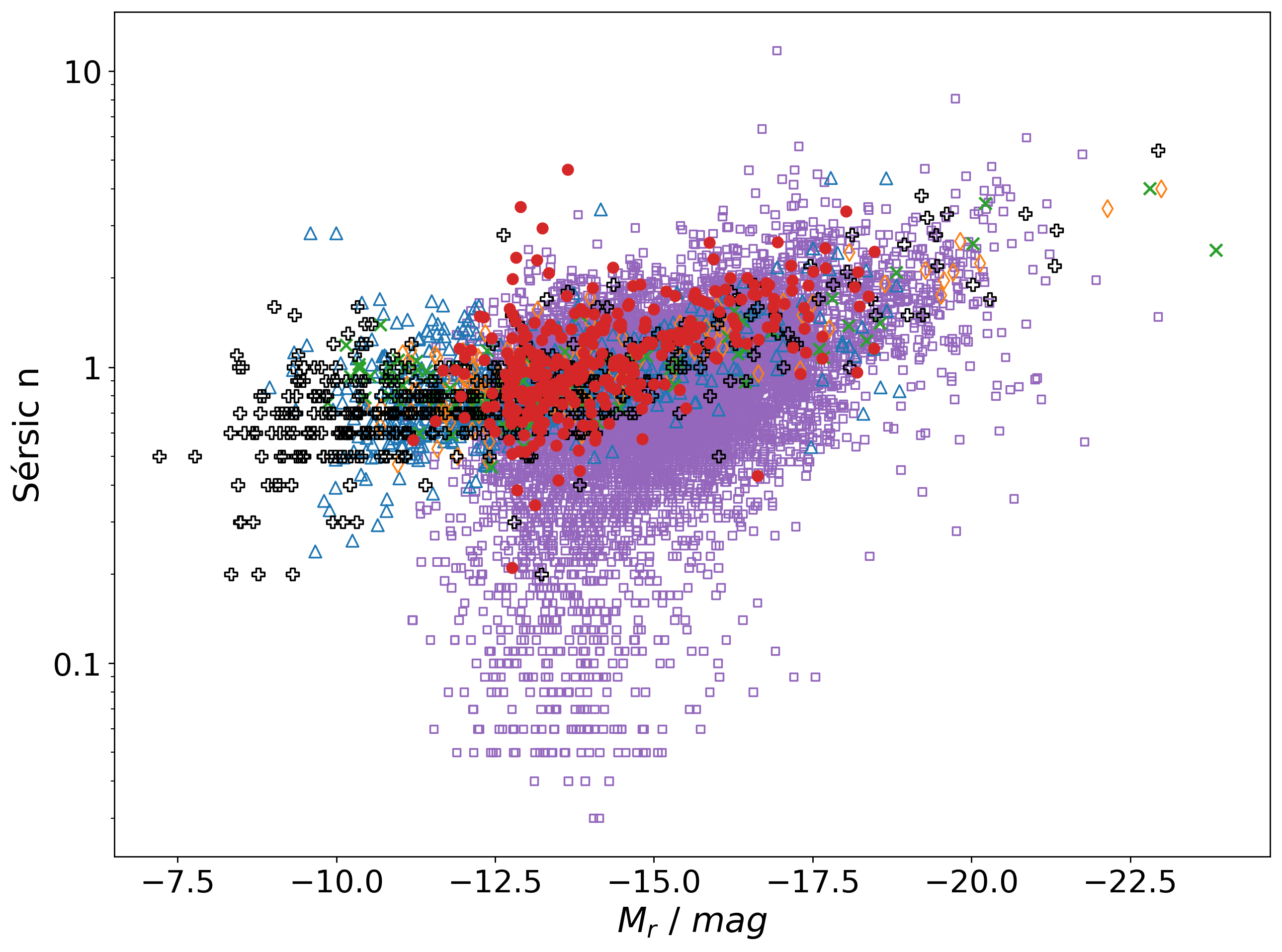}
    \caption{Comparison of galaxies scaling relations among different Local Universe clusters. 
    \emph{Top-left panel}: Effective radius vs. $r$-band magnitude ($M_r$).
    \emph{Top-right panel}: Mean effective surface brightness ($r$-band) as a function of the absolute $r$-band magnitude. 
    \emph{Bottom panel}: Sérsic index $n$ as a function of $M_r$.
    The red circles represents Hydra I dwarf galaxies from our sample. The open blue triangles are the Fornax dwarfs \citep{venhola2019optical}. The open orange diamonds indicate Hydra I early-types galaxies by \cite{misgeld2008early}, while the green crosses represents Centaurus cluster galaxies \citep{Misgeld2009cent}. Empty black plus are Virgo galaxies \citep[NGVS sample][]{ferrarese2020next}; open purple squares are Coma galaxies \citep{Alabi2020coma}.}
    \label{fig:comparison}
\end{figure*}

In Fig. \ref{fig:comparison} we compare in detail the scaling relations with the analogous relations from other clusters. 
We compare the following clusters in the Local Universe: Virgo \citep[sample from ][]{ferrarese2020next}, Fornax \citep[FDS dwarfs sample ][]{venhola2019optical}, Centaurus \citep{Misgeld2009cent}, Hydra I \citep[in addition to our sample we included also][]{misgeld2008early}, and Coma \citep{Alabi2020coma}. 
There is general agreement between galaxies from different clusters. 
Considering the $log(R_e)$ vs. $M_r$ plane, a linear relation is observed for dwarfs almost over every magnitude range.
Starting from $M_r\simeq-8\;mag$ galaxies have increasing size with increasing luminosity up to $M_r\simeq -13\;mag$. 
Then, in the range $-13\lesssim M_r \lesssim -17.5\;mag$ there is a change in the slope of the relation, with $R_e$ growing slower as a function of the luminosity over this magnitude range. 
Brighter than $-17.5\;mag$ the slope changes again, and the size increases with higher luminosity.
Also in less dense environments, a similar linear relation between $log R_e$ and total luminosity holds: \cite{carlsten2021elves} report a similar relation for dwarf satellites in the Local Volume ($D<12\;Mpc$), that extends down to $R_e \simeq 100\;pc$ and $M_V \simeq -7\;mag$. 
\citet{Poulain2021matlas} reveal analogous behavior of dwarf galaxies located in the low to moderate density environments of the MATLAS fields.

Looking at the $\overline{\mu}_{e,r}$ vs. $M_r$ diagram (top-right panel of Fig. \ref{fig:comparison}), galaxies occupy a well defined band which stretches from $M_r\simeq-20\;mag$ down to $\simeq-8\;mag$, with few outliers. 
The average scatter increases moderately toward fainter magnitudes.
Again, this relationship holds also for dwarfs in less dense environments \citep{carlsten2021elves}. 

For the last scaling relation analyzed in Fig. \ref{fig:scaling_rel}, that is Sérsic index vs magnitude, we already mentioned the similarity between Hydra I's galaxies and dwarfs in other local environments, both clusters and loose groups.
In the bottom panel of Fig. \ref{fig:comparison} a detailed comparison of the $n$ vs. total $r$-band magnitude is shown for galaxies of different clusters.
The Sérsic index $n$ varies with magnitude, on average increasing for brighter magnitudes.
A mild flattening is observed only toward faintest luminosities ($M_r<-11\;mag$).
However, a much larger scatter and possibly a different slope is observed for galaxies in the Coma cluster, especially at $-11<M_r<-16\;mag$.
Moreover, this relation also holds for early-type dwarf satellites around Milky Way-like galaxies, and for dwarfs in loose groups in the Local Universe.
Indeed, \cite{carlsten2021elves} observed in their sample the same trend for the early-type satellites in the stellar mass range $10^{5.5}<M_*<10^{8.5}M_{\odot}$.
On average, the faintest dwarfs in the different environments are well represented by Sérsic profiles with $0.5\lesssim n\lesssim 1$, while bright ones are better represented by radial profiles with indices $n>1$.
This means that the most luminous dwarfs are more centrally concentrated, while less luminous ones have a flatter profile in the center. 

From our analysis we conclude that photometric scaling relations are very similar across clusters of different characteristics. 
Indeed, Fornax, Virgo, Centaurus, Hydra I and Coma span a wide range of virial masses \citep[from $10^{13}M_{\odot}$ to $10^{15}M_{\odot}$,][]{Girardi1998}, and radii \citep[from $0.7\;Mpc$ to $4\;Mpc$,][]{Girardi1995}, and they also have very different velocity dispersion (from $\sim 350\;km/s$ to $\sim 1000\;km/s$). 
Moreover, these four clusters show different structures, and some of them have a major segmentation in subgroups.
However, from a photometric point of view, there is no observational evidence that the dwarf galaxies in these local clusters are different.
Moreover, the same trends are observed also for dwarf satellites around Milky-Way like galaxies \citep{Carlsten2020,carlsten2021elves}.
These similarities may suggest that the evolution of the dwarf galaxies is somehow independent from the properties of the environment, and therefore that this is playing perhaps a marginal role.

\subsection{Substructures in the Hydra I cluster}

The smoothed map of the projected galaxy distribution shown in Fig. \ref{fig:smooth_map} 
suggests that the dwarf galaxies in the cluster are not uniformly distributed.
The dwarf galaxies projected number density peaks close to the cluster core, off-center on the NW side to the region where the BCGs are located.
The peak offset is of $\simeq5\arcmin$ from the cluster center (represented by \object{NGC 3311}).
In this region, also the X-ray emission is a bit more extended.
Two more over-densities are found with respect to the cluster core, one on the N-NW and the other on the SE side, respectively (see Fig.~\ref{fig:cdist_cum}). 
All these regions are dominated by the light from the brightest cluster members.
\citet{arnaboldi2012} pointed out that an entire group of small galaxies is falling through the cluster core, and they have already been partially dissolved by tidal interactions. 
\citet{barbosa2018} has observed the sloshing of the central part of NGC~3311 with respect to the stellar halo. 
The velocity displacement of the cD vs. the baricenter of the cluster core indicate subcluster merging.
This subcluster merging is consistent with the elongated loop-sided number distribution of dwarfs in the Hydra I core, as shown in Fig. \ref{fig:smooth_map}.

The other two subgroups have different distributions. 
In the north side of the cluster core, cluster members are arranged in a filament-like structure, with galaxies extending along the SN direction. 
The projected distribution of the dwarf galaxies seems to be coherent with this structure. 
In the southern part of this structure, a dwarf galaxy (HCC~087)  currently being  tidally disrupted by the galaxy cluster’s potential itself was discovered, or, alternatively, by the nearby elliptical HCC~005 \citep{Koch2012}.
On the SE side of the cluster, the brightest galaxy is \object{NGC 3316}, with a systemic velocity comparable to the galaxies in the core (that is $cz\sim4000$ km/s) and a foreground group of galaxies, dominated by \object{NGC 3312} and \object{NGC 3314AB}, which is falling into the cluster potential.
The over-density found in the dwarf galaxy distribution is also consistent with these two substructures.

The presence of several substructures in the Hydra I cluster was also observed by \cite{Lima-dias2021}.
Using two different approaches, they found possible subgroups of galaxies within the virial radius. 
Among these, two of the possible substructures are at a clustercentric distances of $\sim0.2\; Mpc$ and $\sim 0.3\;Mpc$, and within the errors they coincide with the two over-densities we report in the right panel of Fig. \ref{fig:smooth_map}. 

It was also observed that the cluster does not show a Gaussian velocity distribution, therefore substructures could be important in Hydra I.
\cite{fitchett1988dynamics} show that the velocity distribution within $40\arcmin$ ($\sim 600\;kpc$) from the cluster center has different peaks, one of which corresponds to the radial velocity of \object{NGC 3312}. 
The presence of small-scale substructures in dwarfs distribution, large $\sim 100-150\;kpc$, have also been observed in the Fornax cluster \citep{ordenes2018}.
As for Hydra I, the Fornax galaxy number density does not to peak in the cluster center.
Rather, the Fornax cD galaxy  NGC~1399 occupies an apparent saddle-point between two main dwarf galaxy over-densities \citep[see also ][]{Munoz2015ngfs}.

In conclusion, the 2D projected density distribution derived for the dwarf galaxies in the Hydra I cluster supports the idea that this environment is still in an active assembly phase, where several subgroups of galaxies are merging into the cluster potential.

To test whether or not the dwarfs close in projection to the cluster substructures have the same properties as those in the general cluster environment, we divided them into four groups, based on the smoothed map distribution.
Specifically, we divided the dwarf galaxies in our catalog into the three mentioned substructures (core, north, and southeast groups), and in the remaining part of the studied field. 
As limit for the core group we considered the yellow contour in Fig. \ref{fig:smooth_map}, which corresponds to the 0.1 probability density level (i.e. 10\% of the probability mass will lie above the contour). 
We label galaxies as core galaxies if they are located, in projection, within a circle which circumscribes the yellow contour. 
For the N- and SE- group we choose the purple contour as density limit, which correspond to 0.25 probability density level.
Given the less regular shape of these two contours, we choose box shape regions which completely include them. 
Again, projected location of galaxies is considered as group membership criterion.
The three regions are chosen such that they do not overlap each other. 
Galaxies outside these three substructures are included in a fourth group, named "outside".

\begin{figure*}
    \centering
    \includegraphics[width=.8\textwidth]{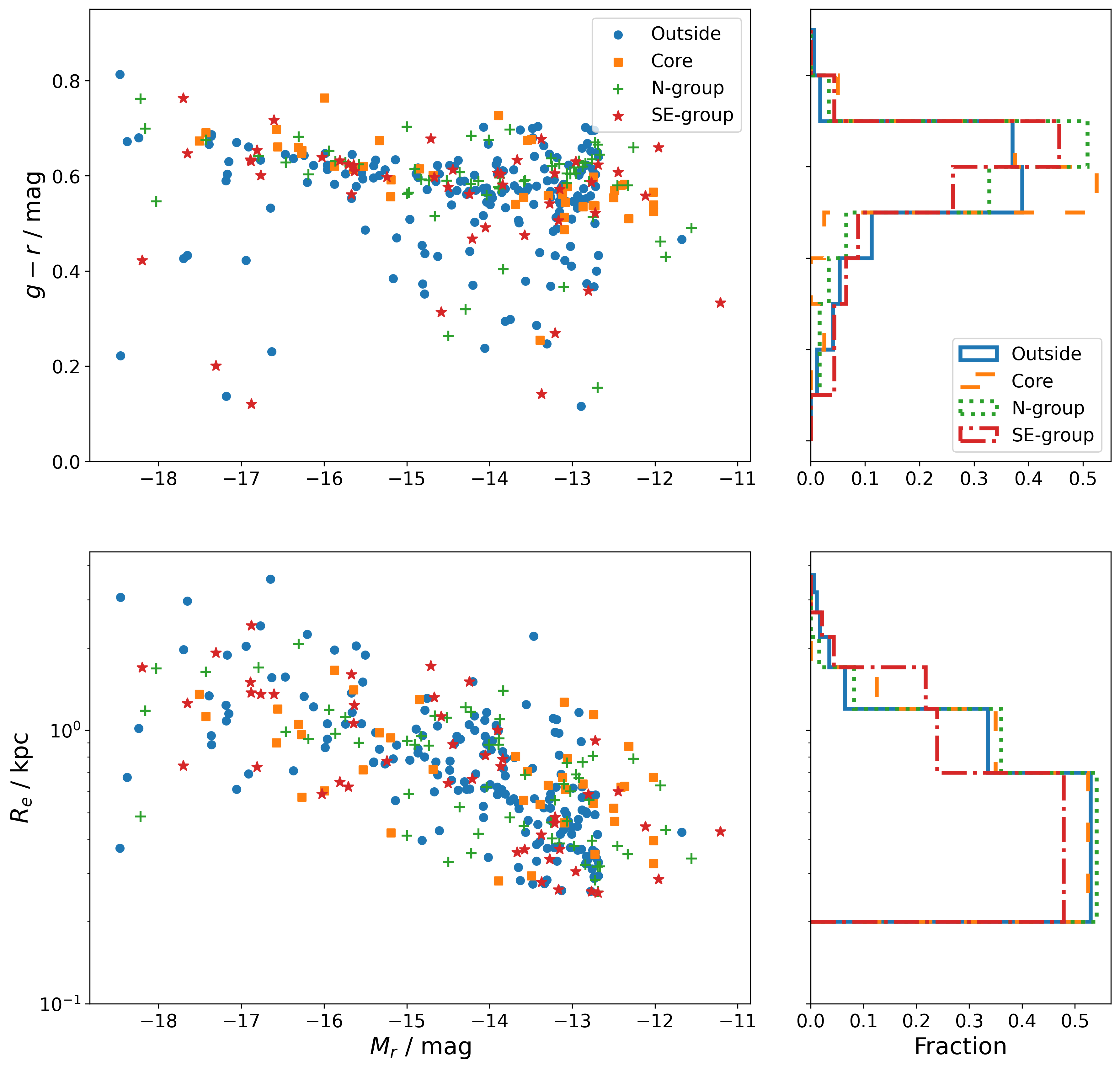}
    \caption{Comparison between dwarf galaxies in different cluster substructures. Dwarfs not included in the three main subgroups are labeled as "Outside".
    \emph{Top panels}. The color-magnitude diagram is shown on the left, while the $g{-}r$ color distributions are shown as histograms on the right panel. Symbol definitions are given in the legends. 
    \emph{Bottom panels}. On the left is reported the $R_e$ vs. $M_r$ plane, while on the right are the $R_e$ distributions for dwarfs in the four substructures.}
    \label{fig:group_comparison}
\end{figure*}

We compare galaxy $g{-}r$ color and $R_e$ in the four different structures. 
This choice is guided by the fact that the color is intrinsically linked to the underlying stellar population of the galaxy, while the $R_e$ indicates the galaxy's size and light concentration.
Thus, both color and $R_e$ are associated to the evolutionary processes acting on the dwarfs.
The analysis is presented in Fig. \ref{fig:group_comparison}, where we show $g{-}r$ and $R_e$ as a function of the total $r$-band magnitude, and their distribution, for the different groups.

The CMD in Fig. \ref{fig:group_comparison} shows that majority of the dwarfs located in projection within the cluster core follow the RS, and all but one have $g{-}r\gtrsim 0.5\;mag$.
Interestingly, we notice that the SE-group has the highest fraction of blue dwarfs ($g{-}r\leq0.4\;mag$).
However, no other significant difference emerges between dwarfs in the groups, and galaxies in projection outside the three main substructures.
Also for what concerns the effective radius no clear difference is found between dwarfs belonging to different substructures. 

Bearing in mind that the separation in groups is based on the projected location in the cluster, this result confirms what is pointed out in Sec.\,\ref{discussion-rel}. 
Indeed, no major difference is found among galaxies in different substructures in our sample.

\subsection{Origin of the clustercentric trends}

In Fig. \ref{fig:cdist-prop} we examine how the structural and physical properties of the dwarf galaxies in the Hydra I cluster core change as a function of the clustercentric distance.
Details of the results are given in Section \ref{clust-centric-sec} but here it is important to observe that these clustercentric properties depend on the galaxy luminosity, i.e. on the galaxy stellar mass.
This suggests that massive and low-mass dwarfs might be affected by different environmental processes, or that the relative importance of the environmental mechanisms depends on the dwarf galaxy mass.

We observe that Hydra I red dwarfs have, on average, $g-r$ color getting redder with decreasing clustercentric distance, with a stronger correlation for brighter bins.
\citet{venhola2019optical} analyzed the same clustercentric trends for dwarf galaxies in the Fornax cluster, splitting dwarfs in $2\;mag$ wide magnitude bins as we do.
The authors report color's trends which are consistent with our findings in the different magnitude bins. 
For the effective radii, we notice a weak positive correlation between with clustercentric distances for the high- and mid-luminosity dwarfs, but no clear trend is observed for the faintest dwarfs. 
Interestingly, \citet{venhola2019optical} found that for the most luminous dwarfs $R_e$ decreases toward the inner parts of Fornax, in agreement with Hydra I dwarfs of comparable luminosity;
for the faintest Fornax dwarfs the size increases going inward with a weak correlation.
The latter result is not in agreement with the Hydra I faint dwarf galaxies.

Thanks to our in-depth spatially resolved photometrical analysis, we are now in the position to discuss what evolution mechanisms can be responsible for such clustercentric trends, and hence what is the role of the environment on cluster galaxies. 
The main processes involved in the transformation of galaxies from star forming to quiescent ETGs in clusters are galaxy harassment and ram-pressure stripping, introduced in Sect. \ref{sec:Intro}.
A direct observational consequence of ram pressure should be that a galaxy, with a given mass, would become redder with fainter $\overline{\mu}_{e,r}$.
From a structural point of view, the galaxy should not change its concentration, therefore Sérsic $n$ and $R_e$ should not be affected.
In the case of partial gas stripping, the process leads to a higher central surface brightness and increased central concentration of the light profile, if star formation results in the central region. 
As consequence of galaxy harassment, the light concentration of a galaxy increases (that is its central surface brightness), and because the star formation activity is generally quenched, it becomes redder.
Thus, from an observational point of view, at a given mass, harassment slowly increases the Sérsic index $n$ of galaxies, which become redder with brighter $\overline{\mu}_{e,r}$.
Unfortunately, harassment and partial ram-pressure stripping have similar photometric effects on galaxies, making it very hard to distinguish between them.

\begin{figure*}[ht]
    \centering
    \includegraphics[width=0.9\textwidth]{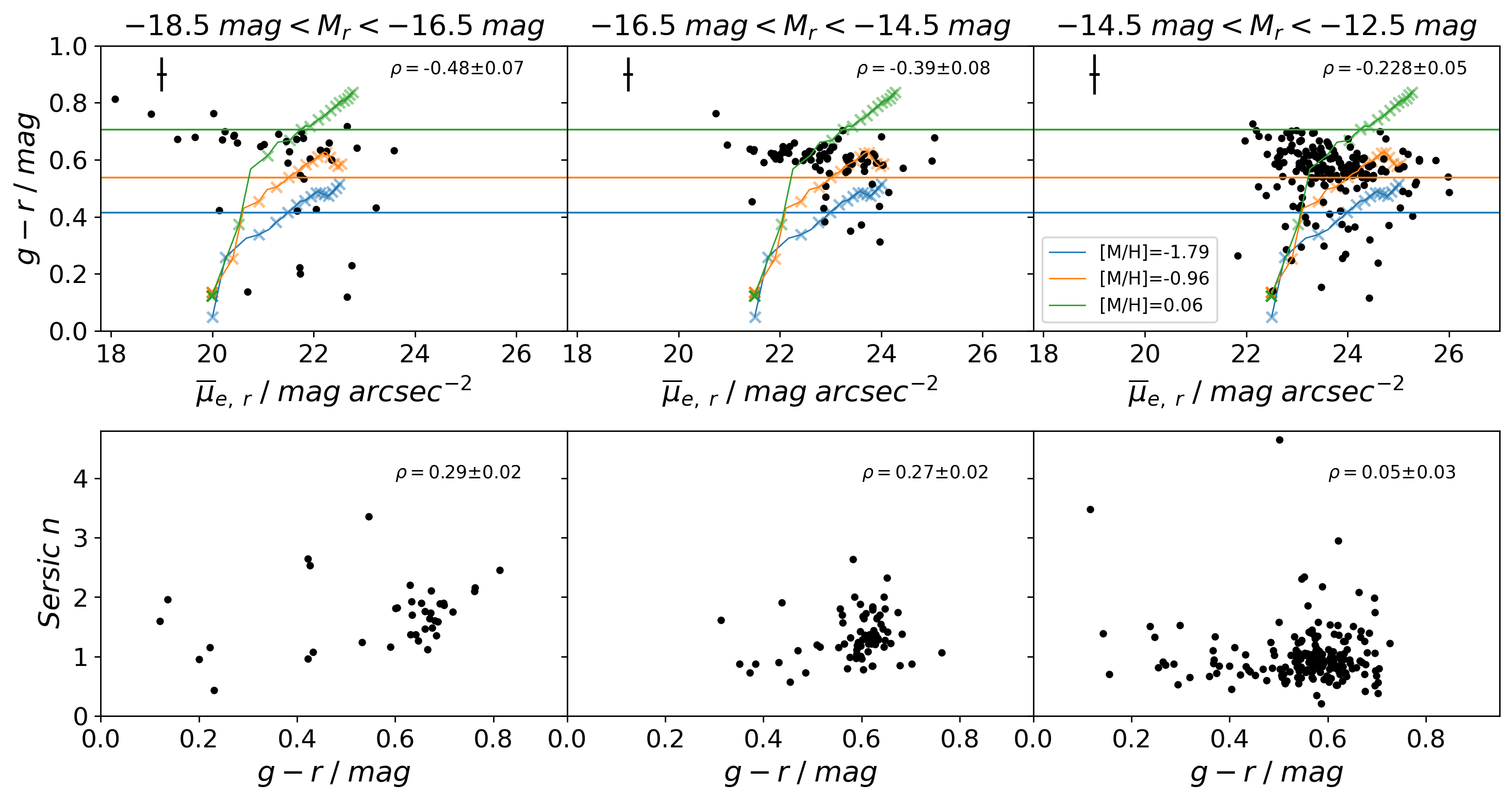}
    \caption{Top panels show dwarf galaxies $g-r$ colors as a function of the $r$-band mean effective surface brightness. 
    Evolutionary tracks of single stellar population for a Kroupa-like IMF for different metallicities are shown. The crosses on each track correspond to 0.5, 1, 2, 3,...,13, 14 $Gyr$. 
    Horizontal lines show the $4\;Gyr$ age level of each track.
    Lower panels show galaxies Sérsic indices $n$ as a function of their colors. Galaxies are split in three luminosity bins, represented by the different columns. In each panel is reported the $\rho$ value of the corresponding Spearman's correlation test, and its error.}
    \label{fig:env_prop}
\end{figure*}

To test whether Hydra I dwarf population show signs of ram-pressure and/or harassment, as predicted by theory, we analyze the correlation between $g-r$ color and surface brightness, and concentration, for galaxies of different luminosities. 
In Figure \ref{fig:env_prop} we present this comparison, using the Sérsic index as indicator of light concentration. 
To better interpret the results we test statistically the correlation using the Spearman's rank correlation test, which is reported in each panel of Fig. \ref{fig:env_prop}.
Dwarf galaxies are binned in magnitude bins, as done for the clustercentric trends. 
We find that in each luminosity bin there is a negative correlation between color and $\overline{\mu}_{e,r}$, which means galaxies become redder with increasing surface brightness.
However, the correlation has increasing strength with increasing luminosity bins: it is robust for dwarfs with $M_r<-16\;mag$, moderate for mid-luminosity dwarfs, and mild for low-luminosity dwarfs.
In addition, to compare the trend between the surface brightness and color with the passive evolution of a single stellar population, we overlay the surface brightness vs. color evolutionary tracks on each panel of Fig. \ref{fig:env_prop}, using the models of \citet{Vazdekis2010}. 
We use a Kroupa-like initial mass function (IMF), and show tracks for the following metallicities: $[Z/H]=-1.79,-0.96,0.06\;dex$. 
The ages range from 0.5 to 14 $Gyr$.
The colors are predicted by the models, while initial surface brightness is set manually. 
Low-luminosity dwarfs show a $\overline{\mu}_{e,r}$ vs. $g-r$ relation that approximately follows the passive evolutionary tracks.
Similarities can be found also for the mid-luminosity bin ($-16.5<M_r<-14.5\;mag$).

Concerning the Sérsic index $n$, there is a moderate positive correlation in the first two magnitude bins ($M_r<-14.5\;mag$), therefore the redder the galaxy, the higher the Sérsic index on average. 
By contrast, dwarfs in the faintest bin are consistent with no correlation, and dwarfs span over a wide range of Sérsic indices.

Interpreting these results in light of the theory predictions, we can conclude that mid- and high-luminosity Hydra I dwarfs ($M_r<-14.5\;mag$) are mildly consistent with the harassment and/or partial ram-pressure scenarios. 

\cite{venhola2019optical} tested environmental mechanisms for Fornax dwarfs of different luminosity, as we do for Hydra I. 
They concluded that the most luminous galaxies ($M_r<-16\;mag$) are consistent with being 
quenched via harassment and/or partial ram-pressure stripping, because they become redder with increasing concentration and surface brightness.
Therefore, Hydra I and Fornax brighter dwarfs ($M_r<-16\;mag$) are consistent with being affected by the same environmental processes. 
\cite{janz2021signatures} compared color vs mean effective surface brightness for Virgo and Fornax galaxies, binning them in magnitude bins too. 
They found that the color–surface brightness diagrams are remarkably similar for the two clusters; galaxies with $M_*>10^8M_{\odot}$ have constant or increasing $g-r$ color with increasing surface brightness, as happens for the Hydra I most luminous dwarfs.
On the other hand, they report that low-mass dwarfs ($M_*<10^8M_{\odot}$) show the opposite trend.
However, when splitting the population in LTGs and ETGs, the authors found that only the low-mass LTGs are consistent with a fading and reddening following the quenching of star formation.
For the ETGs there are no (or only weak) correlations between $g-r$ and surface brightness in all mass bins.
Discussing possible explanations for this discrepancy, \citet{janz2021signatures} concluded that the most likely scenario has to do with an early quenching (for example via gas removal), for which the current fading and reddening becomes less significant and less defined. 
This could mean that when the brightest dwarfs fall in clusters and loose their gas, they are able to keep some of it, which accumulates in their centers, forms stars, of higher metallicity, and then it makes the galaxy redder, and with an higher surface brightness.

The low-luminosity Hydra I dwarfs, while having on average redder colors with brighter surface brightness, do not show any increase in light concentration, as expected for harassment and partial gas stripping. 
However, they show a good agreement with passive evolutionary tracks of stellar population.
\citet{venhola2019optical} observed that the faintest dwarfs in Fornax ($M_r>-14.5\;mag$) show a positive correlation between $g-r$ color and $\bar{\mu}_{e,r}$, while for the concentration vs. color relation there is a rather flat trend. 
Therefore, the quenching of the Fornax low-mass galaxies is consistent with complete gas-stripping and subsequent fading, as happening also for Virgo low-mass dwarfs \citep{janz2021signatures}.
In this case, faint Hydra's dwarfs seem to differ from Fornax ones, with the latter being more consistent with the full ram-pressure scenario.
However, we cannot rule out the complete ram-pressure stripping for low-luminosity Hydra I dwarfs, and a more detailed analysis is required.


\section{Summary and conclusions}\label{Summary}

This work generated a new catalog of resolved dwarf galaxies for the $56.7\times46.6\;arcmin^2$ central area in the Hydra I cluster, covering almost half its virial radius, from the deep multiband imaging data of the VEGAS survey. 
The catalog contains 317 galaxies fainter than $M_r=-18.5\;mag$ and with semi-major axis larger than $a>0.84\;arcsec$ ($\sim 200\;pc$ at Hydra I's distance), of which $61$ were already present in \cite{misgeld2008early} catalog, and 70 in the work of \cite{Christlein2003}. 
The detection efficiency reaches $50\%$ at the limiting magnitude $M_r=-11.5\;mag$ , and at the mean effective surface
brightness $\overline{\mu}_{e,r}=26.5\;mag/arcsec^2$.
Using the new photometric catalog we analyzed the principal scaling relations and compared the results with the corresponding populations in other nearby galaxy clusters. 
Moreover, we investigated how these properties are affected by the surrounding environment, discussing the most likely mechanisms responsible for galaxy transformation. 
To this aim we studied how the different properties change as a function of the clustercentric distance, dividing the dwarf sample in bins of different magnitude.

The main results of this analysis are the following:

\renewcommand{\labelitemi}{\textendash}
\begin{itemize}
    \item The dwarf galaxies are not uniformly distributed in the cluster, rather they are grouped in substructures in different locations across the covered area (Fig. \ref{fig:smooth_map}). We recognized three over-densities of galaxies, a peak NW of the cluster core, a group north of the core, and another subgroup of galaxies in the SE region of the field. All these areas also overlap with the bright galaxy over-densities.
    Additionally, dwarfs and giants have radial distribution statistically different, with the former being less concentrated tan giants in the core.
    
    \item The CMR derived from our observations is consistent with the previous CMR of \cite{misgeld2008early}, and with the red sequence of other galaxy clusters in the local Universe (see Fig. \ref{fig:RScomp}). 
    
    \item Galaxies photometric scaling relations are very similar across Local Universe clusters with different properties (see Fig. \ref{fig:comparison}). There is no observational evidence for dwarfs inhabiting Virgo, Fornax, Centaurus, Hydra I and Coma to be different from each other, from a photometrical and structural point of view.

    \item From the clustercentric trends, we concluded that dwarf galaxies with $M_r<-14.5\;mag$ become redder toward the center of the cluster (Fig. \ref{fig:cdist-prop}), and become slightly larger with increasing distance from the center. 

\end{itemize}
\renewcommand{\labelitemi}{\textbullet}

Based on these results we draw two main conclusions about the structure of the Hydra I cluster.
The existence of subgroups of galaxies (dwarfs and giants) in the cluster is in agreement with previous findings by \citet{Lima-dias2021},\citet{arnaboldi2012}, and \citet{barbosa2018}, and this suggests that the assembly of the cluster is still ongoing.
The cluster environment transforms and influences the galaxies evolution, with the brightest and mid-luminosity dwarfs ($M_r<-14.5\;mag$) being consistent with the effects of harassment and/or partial gas stripping.


\begin{acknowledgements}
We thank the anonymous referee for his/her comments and constructive suggestions.
This work is based on visitor mode observations collected at the European Southern Observatory (ESO) La Silla Paranal Observatory within the VST Guaranteed Time Observations, Programme ID: 099.B-0560(A).
ALM acknowledges financial support from the INAF-OAC.
INAF authors acknowledge financial support for the VST project (P.I. P. Schipani).
DAF thanks the ARC for supoort via DP200102574.
GD acknowledges support from CONICYT project Basal AFB-170002, and FONDECYT REGULAR 1200495.
CS is supported by an `Hintze Fellow' at the Oxford Centre for Astrophysical Surveys, which is funded through generous support from the Hintze Family Charitable  Foundation. 
This research made use of Astropy,\footnote{\url{http://www.astropy.org}} a community-developed core Python package for Astronomy \citep[][]{astropy:2013,astropy:2018}.

\end{acknowledgements}

\bibliographystyle{aa} 
\bibliography{Bibliography.bib}


\begin{appendix}
\section{Estimates of uncertainties}\label{App:A}

\begin{figure*}
    \centering
    \includegraphics[width=0.95\textwidth]{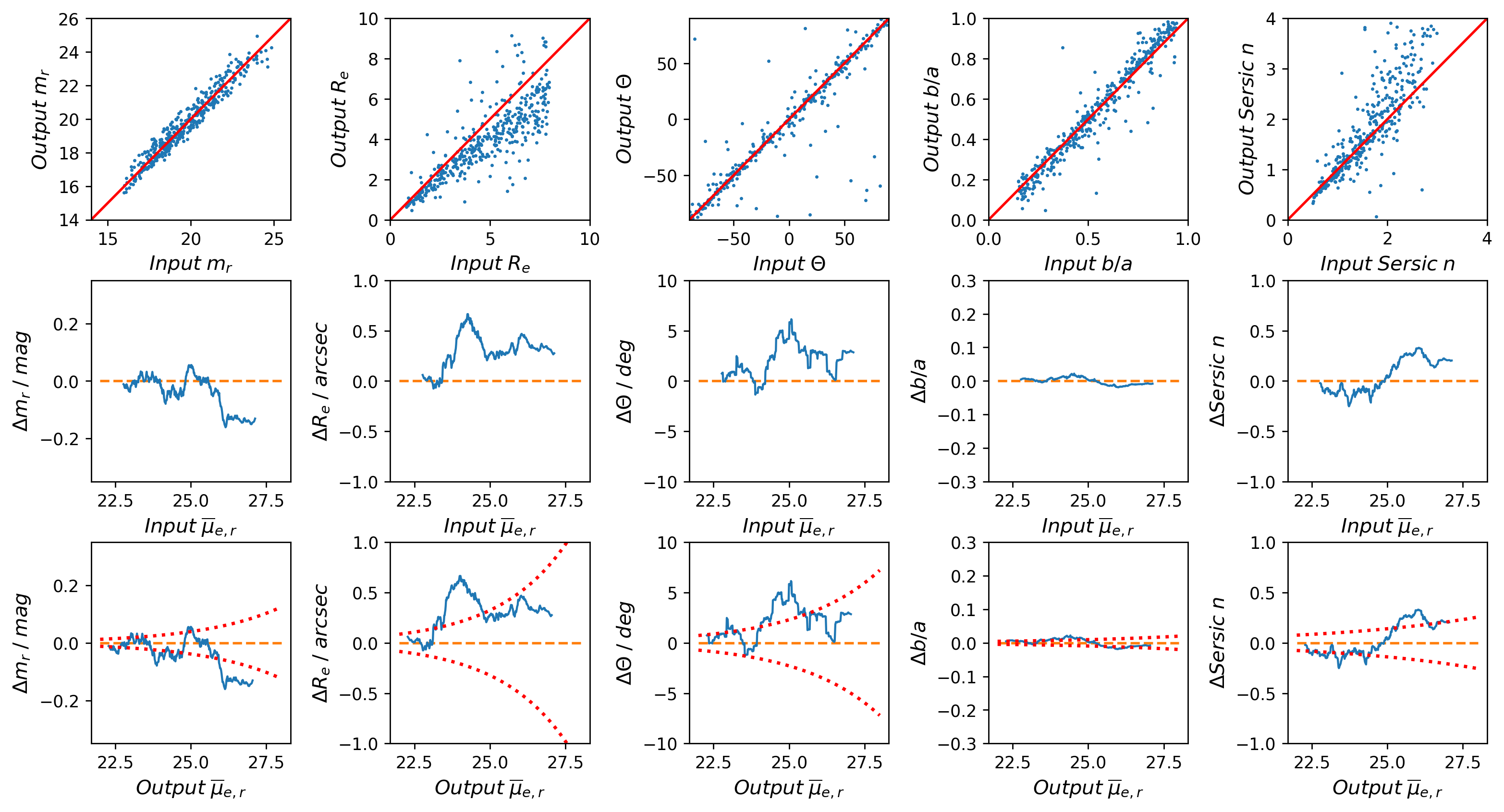}
    \caption{Comparison between \emph{True} and measured properties of $400$ mock galaxies.
    \emph{Top row panels}: Comparison between the input structural parameters of the mock galaxies, and the values measured by the photometric pipeline. The parameters compared are: apparent magnitude ($m_r$), effective radius ($R_e$), position angle ($\Theta$), axis ratio ($b/a$), and Sérsic index ($n$). The red solid lines represent the 1:1 ideal ratio.
    \emph{Mid and bottom row panels}: The blue lines are the running means of the differences between the input and output parameters, as a function of their input and output $\overline{\mu}_{e,r}$, respectively. The differences are calculated as input - output. The red dotted lines in the bottom row panels show the fits to the standard deviations of residuals, as defined in Eq. (\ref{error_eq}).}
    \label{fig:error_tot}
\end{figure*}

To estimate more realistic errors on the fitted parameters we use an approach based on mock galaxies, following the works of \citet{venhola2017fornax}, \citet{venhola2018}, and \cite{hoyos2011hst}. 
The photometric measurements we carried out have uncertainties arising from two different sources: the brightest galaxies usually show more structures than the simple single Sérsic profile assumed by the models we used, while at the low surface brightness end we are limited by the signal-to-noise. 
In order to characterize the typical uncertainties of the fit parameters, we make photometric measurements for a bunch of detected mock galaxies, generated as described in Sec. \ref{mock-gal}.
Specifically, $400$ detected mock galaxies, having a wide range of structural properties, went through the same photometric pipeline as the real galaxies.
In Figure \ref{fig:error_tot} we show a comprehensive comparison between the original mock galaxies parameters, and the measured ones. 
We also plot there the average differences between input and output parameters.
As expected, for all parameters the deviations of the measured values from the original ones increase toward lower surface brightness galaxies.
In the case of the effective radius this is more evident, indeed for larger $R_e$ values {\sc GALFIT} tends to underestimate considerably the size of the galaxies.

To estimate the uncertainties, we fit the standard deviations of the input-output residuals, defined as
\begin{equation}
    \sigma = \sqrt{\frac{\sum_{j=1}^N{(input_j-output_j)^2}}{N-1}}
    \label{std_eq}
\end{equation}
where $N$ is the number of mock galaxies in a given $\overline{\mu}_{e,r}$ bin.
Similar to \citet{venhola2017fornax}, \citet{venhola2018}, and \citet{hoyos2011hst}, we fit the $\sigma$ with the function
\begin{equation}
    log(\sigma)=\alpha \times \overline{\mu}_{e,r} + \beta
    \label{error_eq}
\end{equation}
where $\alpha$ and $\beta$ are the free parameters, and $\overline{\mu}_{e,r}$ is the mean effective surface brightness obtained by the photometric pipeline.
Fitting results are listed in Table \ref{tab:error_parameters}, and are represented by the red dotted lines in the last row of Fig. \ref{fig:error_tot}.
We point out that these uncertainties measured empirically are larger than the standard uncertainties given by {\sc GALFIT}.
Our results are in good agreement with the ones obtained with same procedure by \cite{venhola2017fornax} and \citet{venhola2018}.
The errors derived in this way for individual galaxies are given with their photometric parameters in the catalog.  

\begin{table}[ht]
    \begin{center}
    \caption{Fit parameters of Eq. (\ref{error_eq}).}
        \begin{tabular}{ccc}
        \hline \hline
         Parameter & $\alpha$ & $\beta$  \\
         \hline \hline
         $m_r$ & $0.168$ & $-5.612$ \\
         $R_e$ & $0.192$ & $-5.301$ \\
         $\Theta$ & $0.167$ & $-3.787$ \\
         $b/a$ & $0.108$ & $-4.718$ \\
         Sérsic $n$ & $0.0876$ & $-3.046$ \\
        \end{tabular}
    \end{center}
    
    {\bf Notes} - The first column indicates the fitted structural parameter, while the second and the third columns show the slope ($\alpha$) and the constant coefficient ($\beta$) of the equation.
    \label{tab:error_parameters}
\end{table}

\section{Band conversion}\label{App:B}

In this paper we worked with SDSS $g$ and $r$ photometric bands, but we extensively compared with works that used different photometric systems.
To convert quantities from the Johnson $V$, $R$ and $I$ filters, to SDSS $r$ and $g$ we used the following equation listed in \citet{kostov2018}:

\begin{equation}
    V = r -0.017 + 0.492 \cdot (g-r)
\end{equation}
\begin{equation}
    V-I = 0.27 + 1.26\cdot(g-r)
\end{equation}
\begin{equation}
    r = V - 0.14 - 0.748\cdot(V-R)
\end{equation}
\begin{equation}
    g-r = -0.19 +1.52\cdot(V-R)
\end{equation}

\section{The Hydra I galaxy cluster dwarf catalog}

We show in Table \ref{table} an example table from the Hydra I galaxy cluster dwarf catalog (HCDC). 
The full catalog is available at \url{}.
In the table, we provide also the parameters' errors. 

\begin{sidewaystable*}
\caption{An example table of our Hydra I dwarf galaxies catalog.}
\label{table}

{\small

\begin{tabular}{lcccccccccc}%

\hline
\hline
Target    & R.A.      & Dec.       & $M_r$           & $R_e$        & $\overline{\mu}_e$ & b/a           & $\theta$    & n           & r            & g-r         \\
 & [deg] & [deg] & [mag] & [arcsec] & [mag/arcsec$^2$] & & [deg] & & [mag] & [mag] \\
 
\hline

HCDC\_001 & 159.45055 & -27.11779  & -18.469 $\pm$ 0.003 & 1.50 $\pm$ 0.01  & 18.08   & 0.900 $\pm$ 0.002 & 57.5 $\pm$ 0.2  & 2.46 $\pm$ 0.03 & 16.08 $\pm$ 0.05 & 0.81 $\pm$ 0.09 \\
HCDC\_002 & 158.8406  & -27.695715 & -18.462 $\pm$ 0.011 & 12.36 $\pm$ 0.07 & 21.73   & 0.236 $\pm$ 0.004 & -86.8 $\pm$ 0.7 & 1.16 $\pm$ 0.07 & 16.59 $\pm$ 0.05 & 0.22 $\pm$ 0.09 \\
HCDC\_003 & 159.59425 & -27.582651 & -18.381 $\pm$ 0.004 & 2.72 $\pm$ 0.03  & 19.31   & 0.874 $\pm$ 0.002 & 23.2 $\pm$ 0.3  & 1.74 $\pm$ 0.04 & 16.04 $\pm$ 0.05 & 0.67 $\pm$ 0.09 \\
HCDC\_004 & 159.0078  & -27.685272 & -18.241 $\pm$ 0.005 & 4.10 $\pm$ 0.03  & 19.66   & 0.415 $\pm$ 0.003 & 28.9 $\pm$ 0.3  & 1.61 $\pm$ 0.05 & 16.31 $\pm$ 0.05 & 0.68 $\pm$ 0.09 \\
HCDC\_005 & 159.20451 & -27.388655 & -18.218 $\pm$ 0.004 & 1.95 $\pm$ 0.02  & 18.79   & 0.862 $\pm$ 0.002 & -49.2 $\pm$ 0.2 & 2.10 $\pm$ 0.04 & 16.26 $\pm$ 0.05 & 0.76 $\pm$ 0.09 \\
HCDC\_006 & 159.42046 & -27.53289  & -18.199 $\pm$ 0.006 & 6.84 $\pm$ 0.04  & 20.14   & 0.217 $\pm$ 0.003 & -6.9 $\pm$ 0.4  & 0.96 $\pm$ 0.05 & 16.38 $\pm$ 0.05 & 0.42 $\pm$ 0.09 \\
HCDC\_007 & 158.98225 & -27.23673  & -18.161 $\pm$ 0.006 & 4.75 $\pm$ 0.04  & 20.25   & 0.464 $\pm$ 0.003 & 77.0 $\pm$ 0.4  & 1.87 $\pm$ 0.05 & 16.46 $\pm$ 0.05 & 0.70 $\pm$ 0.09 \\
HCDC\_008 & 158.95192 & -27.172167 & -18.028 $\pm$ 0.011 & 6.79 $\pm$ 0.07  & 21.74   & 0.959 $\pm$ 0.004 & 11.8 $\pm$ 0.7  & 3.36 $\pm$ 0.07 & 16.38 $\pm$ 0.05 & 0.55 $\pm$ 0.09 \\
HCDC\_009 & 159.23766 & -27.567867 & -17.703 $\pm$ 0.006 & 3.00 $\pm$ 0.04  & 20.03   & 0.753 $\pm$ 0.003 & -81.2 $\pm$ 0.4 & 2.16 $\pm$ 0.05 & 16.71 $\pm$ 0.05 & 0.76 $\pm$ 0.09 \\
HCDC\_010 & 158.61127 & -27.50125  & -17.698 $\pm$ 0.012 & 7.96 $\pm$ 0.09  & 22.05   & 0.661 $\pm$ 0.005 & 35.8 $\pm$ 0.8  & 2.53 $\pm$ 0.08 & 16.75 $\pm$ 0.05 & 0.43 $\pm$ 0.09 \\
HCDC\_011 & 158.9451  & -27.646912 & -17.654 $\pm$ 0.020 & 11.98 $\pm$ 0.14 & 23.23   & 0.848 $\pm$ 0.006 & -5.5 $\pm$ 1.2  & 1.07 $\pm$ 0.10 & 16.78 $\pm$ 0.05 & 0.43 $\pm$ 0.09 \\
HCDC\_012 & 159.44597 & -27.572983 & -17.653 $\pm$ 0.008 & 5.06 $\pm$ 0.05  & 20.95   & 0.547 $\pm$ 0.004 & -50.8 $\pm$ 0.5 & 1.27 $\pm$ 0.06 & 16.84 $\pm$ 0.05 & 0.65 $\pm$ 0.09 \\
HCDC\_013 & 159.08057 & -27.479763 & -17.509 $\pm$ 0.011 & 5.47 $\pm$ 0.07  & 21.66   & 0.886 $\pm$ 0.004 & -65.2 $\pm$ 0.7 & 2.10 $\pm$ 0.07 & 16.86 $\pm$ 0.05 & 0.67 $\pm$ 0.09 \\
HCDC\_014 & 159.2191  & -27.537567 & -17.431 $\pm$ 0.009 & 4.53 $\pm$ 0.06  & 21.31   & 0.898 $\pm$ 0.004 & -87.4 $\pm$ 0.6 & 1.89 $\pm$ 0.07 & 16.90 $\pm$ 0.05 & 0.69 $\pm$ 0.09 \\
HCDC\_015 & 159.1116  & -27.390396 & -17.429 $\pm$ 0.011 & 6.60 $\pm$ 0.08  & 21.79   & 0.627 $\pm$ 0.004 & -42.8 $\pm$ 0.7 & 1.48 $\pm$ 0.07 & 16.96 $\pm$ 0.05 & 0.68 $\pm$ 0.09 \\
HCDC\_016 & 159.49373 & -27.239534 & -17.392 $\pm$ 0.010 & 5.40 $\pm$ 0.07  & 21.45   & 0.636 $\pm$ 0.004 & -58.1 $\pm$ 0.6 & 1.12 $\pm$ 0.07 & 17.04 $\pm$ 0.05 & 0.67 $\pm$ 0.09 \\
HCDC\_017 & 158.98024 & -27.754667 & -17.366 $\pm$ 0.007 & 3.86 $\pm$ 0.04  & 20.42   & 0.379 $\pm$ 0.003 & -44.6 $\pm$ 0.4 & 1.35 $\pm$ 0.06 & 17.29 $\pm$ 0.05 & 0.68 $\pm$ 0.09 \\
HCDC\_018 & 159.12296 & -27.75818  & -17.361 $\pm$ 0.007 & 3.57 $\pm$ 0.04  & 20.43   & 0.517 $\pm$ 0.003 & -35.6 $\pm$ 0.4 & 1.59 $\pm$ 0.06 & 17.14 $\pm$ 0.05 & 0.69 $\pm$ 0.09 \\
HCDC\_019 & 159.34555 & -27.545156 & -17.308 $\pm$ 0.011 & 7.75 $\pm$ 0.07  & 21.74   & 0.328 $\pm$ 0.004 & 31.4 $\pm$ 0.7  & 0.95 $\pm$ 0.07 & 17.26 $\pm$ 0.05 & 0.20 $\pm$ 0.09 \\
HCDC\_020 & 159.50995 & -27.166685 & -17.187 $\pm$ 0.010 & 4.99 $\pm$ 0.07  & 21.49   & 0.610 $\pm$ 0.004 & 76.2 $\pm$ 0.6  & 1.16 $\pm$ 0.07 & 17.30 $\pm$ 0.05 & 0.59 $\pm$ 0.09 \\
HCDC\_021 & 158.84036 & -27.390701 & -17.184 $\pm$ 0.007 & 4.37 $\pm$ 0.05  & 20.7    & 0.373 $\pm$ 0.003 & 1.9 $\pm$ 0.5   & 1.96 $\pm$ 0.06 & 17.32 $\pm$ 0.05 & 0.14 $\pm$ 0.09 \\
HCDC\_022 & 159.0893  & -27.7753   & -17.171 $\pm$ 0.012 & 7.60 $\pm$ 0.08  & 21.92   & 0.396 $\pm$ 0.004 & 33.2 $\pm$ 0.7  & 1.82 $\pm$ 0.07 & 17.28 $\pm$ 0.05 & 0.60 $\pm$ 0.09 \\
HCDC\_023 & 159.2929  & -27.066227 & -17.154 $\pm$ 0.010 & 4.64 $\pm$ 0.07  & 21.52   & 0.746 $\pm$ 0.004 & -81.6 $\pm$ 0.6 & 2.20 $\pm$ 0.07 & 17.26 $\pm$ 0.05 & 0.63 $\pm$ 0.09 \\
HCDC\_024 & 159.057   & -27.668182 & -17.057 $\pm$ 0.006 & 2.46 $\pm$ 0.04  & 20.2    & 0.654 $\pm$ 0.003 & 82.5 $\pm$ 0.4  & 1.64 $\pm$ 0.05 & 17.46 $\pm$ 0.05 & 0.67 $\pm$ 0.09 \\
HCDC\_025 & 159.49046 & -27.244827 & -16.942 $\pm$ 0.011 & 8.19 $\pm$ 0.07  & 21.67   & 0.211 $\pm$ 0.004 & 17.1 $\pm$ 0.7  & 2.64 $\pm$ 0.07 & 17.55 $\pm$ 0.05 & 0.42 $\pm$ 0.09 \\
HCDC\_026 & 159.48895 & -27.10833  & -16.913 $\pm$ 0.007 & 2.79 $\pm$ 0.04  & 20.49   & 0.595 $\pm$ 0.003 & 57.0 $\pm$ 0.4  & 1.46 $\pm$ 0.06 & 17.58 $\pm$ 0.05 & 0.66 $\pm$ 0.09 \\
HCDC\_027 & 159.42653 & -27.765316 & -16.890 $\pm$ 0.013 & 6.04 $\pm$ 0.09  & 22.13   & 0.647 $\pm$ 0.005 & 57.8 $\pm$ 0.8  & 1.70 $\pm$ 0.08 & 17.45 $\pm$ 0.05 & 0.63 $\pm$ 0.09 \\
HCDC\_028 & 159.22984 & -27.725649 & -16.884 $\pm$ 0.013 & 5.54 $\pm$ 0.09  & 22.26   & 0.883 $\pm$ 0.005 & -10.4 $\pm$ 0.9 & 1.37 $\pm$ 0.08 & 17.43 $\pm$ 0.05 & 0.63 $\pm$ 0.09 \\
HCDC\_029 & 159.34265 & -27.545198 & -16.879 $\pm$ 0.016 & 9.76 $\pm$ 0.11  & 22.66   & 0.388 $\pm$ 0.005 & 10.6 $\pm$ 1.0  & 1.59 $\pm$ 0.09 & 17.50 $\pm$ 0.05 & 0.12 $\pm$ 0.09 \\
HCDC\_030 & 159.37286 & -27.630747 & -16.812 $\pm$ 0.008 & 2.96 $\pm$ 0.05  & 21.02   & 0.858 $\pm$ 0.004 & 51.7 $\pm$ 0.5  & 1.90 $\pm$ 0.06 & 17.59 $\pm$ 0.05 & 0.65 $\pm$ 0.09 \\
HCDC\_031 & 159.03699 & -27.253405 & -16.791 $\pm$ 0.017 & 6.84 $\pm$ 0.12  & 22.85   & 0.856 $\pm$ 0.006 & -66.3 $\pm$ 1.1 & 1.37 $\pm$ 0.09 & 17.61 $\pm$ 0.05 & 0.64 $\pm$ 0.09 \\
HCDC\_032 & 159.5429  & -27.039248 & -16.770 $\pm$ 0.022 & 9.71 $\pm$ 0.17  & 23.59   & 0.844 $\pm$ 0.007 & -31.0 $\pm$ 1.4 & 1.92 $\pm$ 0.10 & 17.59 $\pm$ 0.05 & 0.63 $\pm$ 0.09 \\
HCDC\_033 & 159.32234 & -27.592731 & -16.766 $\pm$ 0.014 & 5.47 $\pm$ 0.10  & 22.35   & 0.857 $\pm$ 0.005 & -34.8 $\pm$ 0.9 & 1.81 $\pm$ 0.08 & 17.59 $\pm$ 0.05 & 0.60 $\pm$ 0.09 \\
HCDC\_034 & 158.63142 & -27.35273  & -16.648 $\pm$ 0.011 & 14.42 $\pm$ 0.08 & 21.81   & 0.248 $\pm$ 0.004 & -20.8 $\pm$ 0.7 & 1.24 $\pm$ 0.07 & 16.28 $\pm$ 0.05 & 0.53 $\pm$ 0.09 \\
HCDC\_035 & 158.73383 & -27.639797 & -16.634 $\pm$ 0.016 & 6.28 $\pm$ 0.12  & 22.75   & 0.799 $\pm$ 0.005 & 81.6 $\pm$ 1.0  & 0.43 $\pm$ 0.09 & 17.76 $\pm$ 0.05 & 0.23 $\pm$ 0.09 \\

\hline
\end{tabular}}
\\
\\
\textbf{Notes} - Column 1 reports the name of the Hydra's dwarf galaxy candidate. 
In columns 2 and 3 we list the coordinates of the galaxies. 
In columns 4 and 5 the total absolute $r$-band magnitude and the effective radius $R_e$, in arcsecond are reported, respectively. 
Column 6 gives the mean effective surface brightness in $r$-band. 
Columns 7 to 9 list dwarfs axis-ratio, position angle, and Sérsic index. 
In column 10 we give the aperture magnitude within the effective radius in the $r$-band, and in the last column the $g-r$ aperture color of the galaxies. 
The parameters listed here are derived through the 2D fits preformed with {\sc GALFIT} \citep{Peng2002,Peng2010}.
All magnitudes and colors are corrected for Galactic extinction using correcting values from \citet{schlegel1998maps} and \citet{Schlafly2011}.

\end{sidewaystable*}

\end{appendix}

\end{document}